%% file: main.tex
\documentclass[aps, prl, reprint, amsfonts, amssymb, amsmath, showkeys, nofootinbib, twoside, superscriptaddress, longbibliography]{revtex4-2}

\usepackage{acronym}
\usepackage{graphicx}% Include figure files
\usepackage{dcolumn}% Align table columns on decimal point
\usepackage{bm}% bold math
\usepackage{hyperref}% add hypertext capabilities
\usepackage[mathlines]{lineno}% Enable numbering of text and display math
\usepackage{graphicx}
\usepackage{xcolor}
\usepackage[normalem]{ulem}

\newcommand{\blue}[1]{\textcolor{black}{#1}}

\include{macros}

\begin{document}

\preprint{APS/123-QED}

\title{Evidence for Three Subpopulations of Merging Binary Black Holes at Different Primary Masses }

\author{Sharan Banagiri}
\email{sharan.banagiri@monash.edu}
\affiliation{\SPA}
\affiliation{\OzGravMonash}

\author{Eric Thrane}
\affiliation{\SPA}
\affiliation{\OzGravMonash}

\author{Paul D. Lasky}
\affiliation{\SPA}
\affiliation{\OzGravMonash}

%\author{TBD}

\date{\today}% It is always \today, today,

\begin{abstract}
With the release of the fourth LIGO--Virgo--KAGRA gravitational-wave catalog (GWTC-4), we are starting to gain a detailed view of the population of merging binary black holes. The formation channels of these black holes is not clearly understood, but different formation mechanisms may lead to subpopulations with different properties visible in gravitational-wave data. Adopting a phenomenological approach, we find GWTC-4.0 data supports the presence of at least three subpopulations, each associated with a different range of black hole mass and with sharp transition boundaries between them. Each subpopulation is characterized by different distributions for either the mass ratios, the black-hole spin magnitudes or both. \PopOneLabel~ with primary mass $m_1 \leq \mtmid$ ($90 \%$ credibility), is characterized by a nearly flat mass ratio distribution $q=m_2/m_1$, and by small spin magnitudes ($\chi \leq \PopOneSpinUL$). \PopTwoLabel, with $\mtmid \leq m_1 \leq \mthigh$, has a much sharper preference for mass ratio $q \approx 1$. \PopThreeLabel, with $m_1 \geq \mthigh$, has support for large spin magnitudes, and tentative support for mass ratios $q\approx0.5$. We interpret these transitions as evidence for multiple subpopulations, each potentially associated with a different formation pathways. We suggest {putative formation scenarios for each subpopulations, and explore chemically homogeneous evolution, population III stars and dynamical formation channels as an explanation for \PopTwoLabel }. Our findings for \PopThreeLabel~are largely consistent with recent claims of hierarchical mergers but with some curious differences in properties. 
\end{abstract}

\maketitle

\acrodef{GW}[GW]{gravitational wave}
\acrodef{BBH}[BBH]{binary black hole}
\acrodef{CHE}[CHE]{chemically homogeneous evolution}

\textit{\textbf{Introduction}.---} The sensitivity of \ac{GW} detectors has increased massively in the last ten years~\cite{LIGOScientific:2016aoc, KAGRA:2025oiz} as a result of technological upgrades~\cite{Capote:2024rmo, Ganapathy:2023, Wenxuan:2024elc, Buikema:2020} and an improved understanding of instrumental noise~\cite{LIGO:2024kkz}. With the recent GWTC-4.0 catalog~\cite{LIGOScientific:2025hdt, LIGOScientific:2025yae, GWTC4:Catalog_results} by LIGO-Virgo-KAGRA (LVK)~\cite{LIGOScientific:2014pky, VIRGO:2014yos, KAGRA:2020tym}, the number of confidently-detected \ac{BBH} mergers has now surpassed 150, offering an unprecedented view into the nature and formation of these systems and their populations~\cite{GWTC4:astrodist}.  In the early days of \ac{GW} astronomy, it seemed plausible that just a small number of detections might enable us to determine the primary formation channel of of \acp{BBH} mergers (e.g.~\cite{Vitale:2015tea, Stevenson:2017dlk, Farr:2017gtv}). However, this hope has since dissipated as \ac{GW} data has unveiled a rich and complex picture of the population of merging binaries, not easily explained with a single formation channel. Furthermore, uncertainties in the physics underlying black hole formation and merger among the various possible channels make it hard to obtain robust predictions (see Ref.~\cite{Mandel:2021smh} for a review). The relative contributions of various \ac{BBH} formation channels is still a topic of active study (e.g.~\cite{Zevin:2020gbd, Afroz:2024fzp, Banagiri:2025dxo, Colloms:2025hib}).

Several analyses have claimed evidence of \ac{BBH} subpopulations. References~\cite{Vitale:2015tea,  Stevenson:2017dlk, Talbot:2017yur, Baibhav:2022qxm, Vitale:2022dpa, Pierra:2024fbl, Li:2023yyt, GWTC4:astrodist} used \ac{BBH} spin tilts to discriminate between different subpopulations. While promising from a theoretical perspective~\cite{Kalogera:1999tq, Rodriguez:2016vmx, Farr:2017uvj, Gerosa:2018wbw}, using this as a clean discriminator has proven difficult. First, precise measurements of \ac{BBH} spin tilts with \acp{GW} is difficult. Second, recent studies have challenged commonly held assumptions and shown that these models can be subject to significant theoretical uncertainties~\cite{Wang:2021yjf, Banerjee:2023ycw, Baibhav:2024rkn, Kiroglu:2025bbp}. There is also evidence for global correlations between various parameters~\cite{Callister:2021fpo, Biscoveanu:2022qac, Adamcewicz:2022hce, Adamcewicz:2023mov, GWTC4:astrodist} but interpreting them astrophysically is not straightforward.

\begin{figure*}[ht]
\centering
    \includegraphics[width=0.9\textwidth]{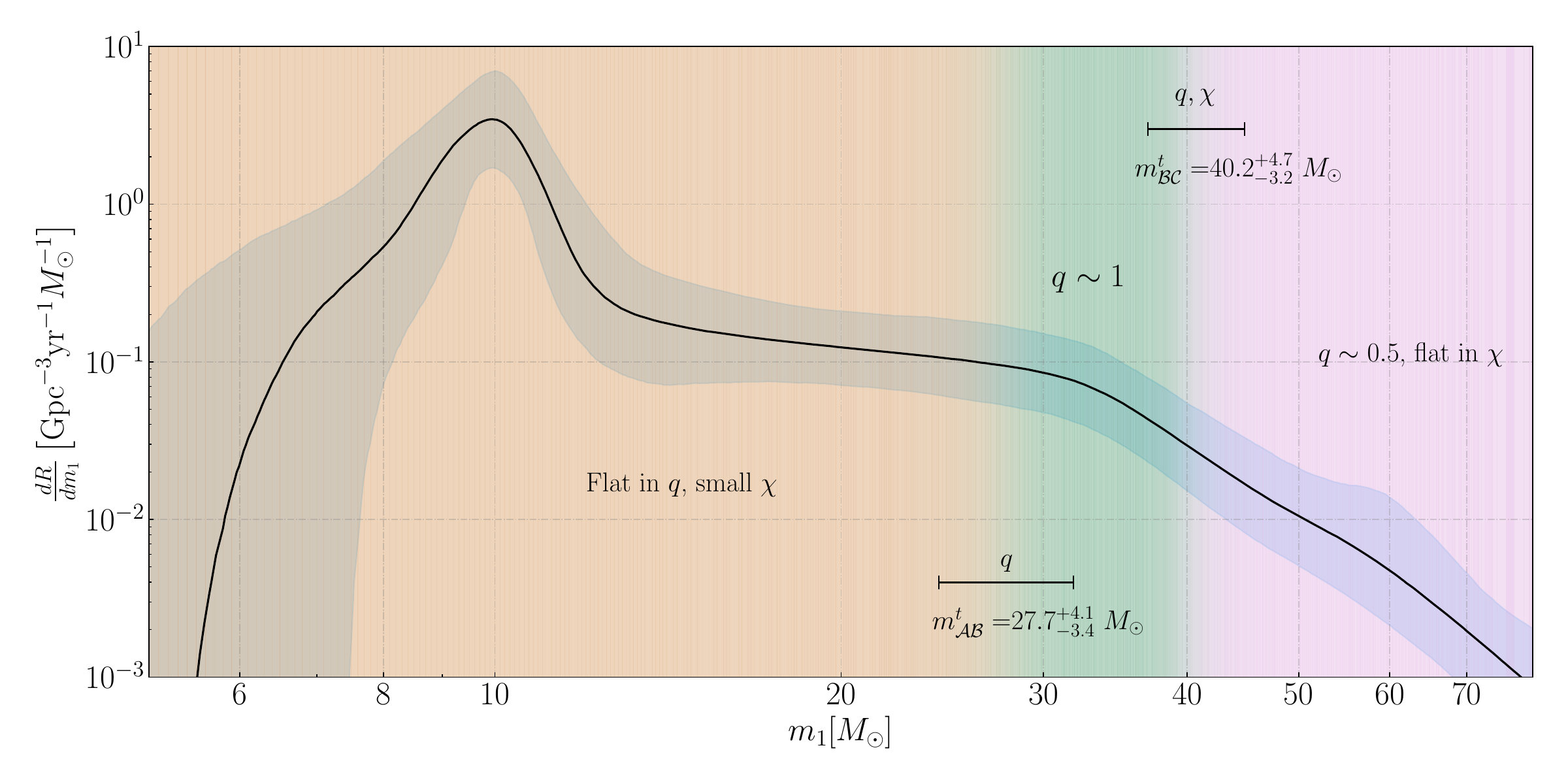}
    \caption{\label{fig:main_overiew} The three subpopulations inferred with the \textsc{multipop} model are shown as three colored regions along the primary mass spectrum. The transition regions between the colors correspond to the inferred transition masses. \PopOneLabel, characterized by $m_1 \lesssim 28 M_{\odot}$ has an almost flat mass ratio distribution and a $\chi\lesssim 0.5$. \PopTwoLabel, characterized by $ 28 M_{\odot} \lesssim m_1 \lesssim 40 M_{\odot}$ has a mass-ratio distribution that is more sharply peaked at $q=1$ while retaining the same spin distribution as \PopOneLabel. 
    \PopThreeLabel, at $m_1 \gtrsim 40 M_\odot$ shows support for a spin-magnitude distribution with significant support at $\chi > 0.5$ and tentative preference for $q \sim 0.5$. The solid black curve and the shaded region around it corresponds to the median population predictive distribution and the 90\% credible levels of the primary mass mass spectrum.
    % Eric: no need to mention what _doesn't_ change... just focus on what does change.
    }
\end{figure*}

Another potentially fruitful way is to adopt a phenomenological approach: to look for subpopulations that have clear and distinguishable properties. If two groups of binaries are found to have different distributions for some parameter like mass ratio, spin or redshift, one can then infer that the relevant physical processing behind their formation are different.\footnote{This does not necessarily mean that the two subpopulations originate from entirely different formation scenarios. For example, they can both be formed through isolated binary channel but go through different mass transfer phases.} Several studies have pursued this strategy to search for subpopulations with interesting properties~\cite{Li:2022jge, Li:2023yyt, Godfrey:2023oxb, Hussain:2024qzl, Ray:2024hos, Antonini:2024het, Antonini:2025zzw,Galaudage:2024meo, GWTC4:astrodist,Tong2025, Antonini:2025ilj, Li:2024jzi, Wang:2022gnx}.

It can be particularly informative to find subpopulations separated by clear transition masses. Knowing both the \ac{BBH} mass scales and their properties can enable us to constrain and infer their formation physics. This is the approach we take in this \textit{Letter} by searching for subpopulations with distinct primary black hole masses. Using cumulative GWTC-4.0 data~\cite{GWTC4:Catalog_results}, we find evidence for at least three such subpopulations, each characterized by different mass ratio pairing or different spin distributions. Crucially, the data strongly prefers the interpretation that all of these subpopulations exist in conjunction. 

\textit{\textbf{Methods \& Results.}}---We model the \ac{BBH} population with simple phenomenological components. 
We use the strongly-parameterized fiducial models from Ref.~\cite{GWTC4:astrodist}. The primary mass distribution $p(m_1)$ is a broken power law with two Gaussian peaks. The spin magnitude distributions $p(\chi)$ are (independently and identically distributed) truncated normals, and the spin tilt distributions $p(\cos\theta)$ are mixtures between an isotropic component, and an aligned component modeled with a truncated Gaussian distribution~\cite{spin,Vitale:2022dpa}. The redshift distributions $p(z)$ are modeled as  a power law in $(1+z)$~\cite{Fishbach:2018edt}. Finally, we model the mass-ratio distributions, $p(q)$, where $q = m_2 / m_1$, either as a power law or a truncated normal.

 A key feature of our work is that the distributions of $z, q, \chi$ and $\cos \theta$ are allowed to change shape at up to three transition masses, corresponding to four subpopulations.\footnote{We stop at three transition points because we do not find support for more than two.}
The functional form of the distribution may be the same on either side of the transition, but the hyperparameters controlling the shape are allowed to have distinct values. For example, our model can accommodate a scenario where binaries with $m_1< 30 M_\odot$ are characterized by a mass ratio distribution of $p(q)\propto q^2$, but binaries with $m_1>30$ are characterized by a mass ratio distribution $p(q) = \text{const}$. More details about model parameterization and the priors used can be found in the Supplementary Materials.

We use standard hierarchical Bayesian inference techniques~\cite{Mandel:2018mve, Thrane:2019, Vitale:2020aaz} for inferring the \ac{BBH} population properties. We analyze data cumulative to GWTC-4.0~\cite{LIGOScientific:2025snk, GWTC4:Catalog_results, LIGOScientific:2025gwtc4zenodo, gwtc3_pe:zenodo, gwtc2_pe:zenodo} including all \acp{BBH} mergers with a false alarm rate less than $1 {\rm yr}^{-1}$. We do not include GW190814~\cite{LIGOScientific:2020zkf} because of the ambiguous nature of the secondary. We also do not include GW231123~\cite{LIGOScientific:2025rsn} in our analysis, driven by concerns that the seemingly extremal spins makes it susceptible to poorly-understood waveform systematics. Moreover, if its measured properties are real, the most parsimonious explanation would likely put it in a class of its own highly spinning, ultra-massive \ac{BBH} systems~\footnote{{However, see Refs.~\cite{Hussain:2024qzl, Adamcewicz:2025phm} which suggests that in binaries with rapid spins, both black holes spin rapidly. If real this might bring GW231123 more in line with other rapidly-spinning binaries although its spins are still likely too high to be explained as, for example, 2G + 2G mergers}}. 

We then tested several models allowing upto three transition masses. These yield a more-or-less consistent picture of the \ac{BBH} population, wherein we find subpopulations with different mass-ratio or spin magnitude distributions, but no statistically significant evidence that the distribution of other parameters have a transition. {We relied on Bayes factors for model selection and to avoid over fitting the data. Bayes factors carry an Occam penalty for more complex models and ensure that a more flexible model is selected only when its flexibility is required by the data. We direct the reader to Sec. 3 Ref.~\cite{Thrane:2019} for an overview on Bayes factors and Bayesian model selection in gravitational-wave astronomy}.

The rest of this \textit{Letter} focuses on the model with the highest Bayes factor. All quoted uncertainties correspond to $90\%$ credible levels unless stated otherwise. 

We find that the data strongly prefers three distinct subpopulations with two transition masses. The three subpopulations differ in either their mass-ratio distributions, their spin magnitude-distributions or both. For brevity we refer to this model as \textsc{multipop}. 
An overview of our main results with \textsc{multipop} is shown in Fig.~\ref{fig:main_overiew}. Formally, the mass-ratio distribution of \textsc{multipop} is modeled as:
\begin{equation}
    p(q) \propto 
\begin{cases}
    q^{\beta_0}, & \text{if } m_1 \leq m^t_{\mathcal{AB}},\\
    q^{\beta_1}, & \text{if } m^t_{\mathcal{AB}} \leq m_1 \leq m^t_{\mathcal{BC}},\\
    \mathcal{N}(q | \mu_q, \sigma_q),              &\text{if } m_1 > m^t_{\mathcal{BC}},\\
\end{cases}
\label{Eq:Multipop_pq}
\end{equation}
and the spin-magnitude distribution as:
\begin{equation}
    p(\chi) \propto 
\begin{cases}
    \mathcal{N}(\chi | \mu_{\chi_0}, \sigma_{\chi_0}),& \text{if } m_1 \leq m^t_{\mathcal{BC}},\\
    \mathcal{N}(\chi | \mu_{\chi_1}, \sigma_{\chi_1}),              &\text{if } m_1 > m^t_{\mathcal{BC}}.\\
\end{cases}
\label{Eq:Multipop_pchi}
\end{equation}
{Redshift and spin tilts are found to need just a single global distribution.}

We measure the transition masses separating these subpopulations at $m^t_{\mathcal{ AB}} = \mtmid$ and $m^t_{\mathcal{BC}} = \mthigh$. These are statistically consistent with transition values individually inferred elsewhere in the literature. In particular, Ref.~\cite{Li:2022jge} finds a transition in $p(q)$ at $29.3^{+5.7}_{-4.0} M_{\odot}$ with \acp{BBH} above this mass have a stronger preference for equal-mass pairing~\footnote{This preference for equal mass pairing for \acp{BBH} at these masses has also been confirmed with GWTC-4.0 data~\cite{GWTC4:astrodist}.}. Meanwhile Ref.~\cite{Li:2023yyt} and Ref.~\cite{Antonini:2025zzw} measure a transition in spins at $m_1 = 41.56^{+20.64}_{-8.55} (1 \sigma \text{ credible interval})$ and $m_1 = 46^{+7}_{-5} M_{\odot}$, respectively. More recently, Ref.~\cite{Tong2025} also found a transition at $45^{+5}_{-4} M_{\odot}$ in the secondary mass distribution. {Since this Letter was made public, newer work in the literature also found evidence to support such transitions; in particular Ref.~\cite{Plunkett:2026pxt, Wang:2025nhf} finds a transition at $\sim 45 M_{\odot}$ to a population of binaries with more rapid spins, while Ref.~\cite{Ray:2025xti} finds evidence for a gentler transition in the secondary mass that is still consistent with our result.}

The \textsc{multipop} model is preferred over the fiducial strongly-parameterized model from Ref.~\cite{GWTC4:astrodist} (which we refer to as \textsc{gwtc4-fiducial}) by a natural log Bayes factor of $\MultipopLogB$. We direct the reader to the Supplementary materials and to Ref.~\cite{GWTC4:astrodist} for more details about \textsc{gwtc4-fiducial}. We now describe the characteristic properties of each of the subpopulations. 

\begin{figure}[t]
    \centering
    \includegraphics[width=0.5\textwidth]{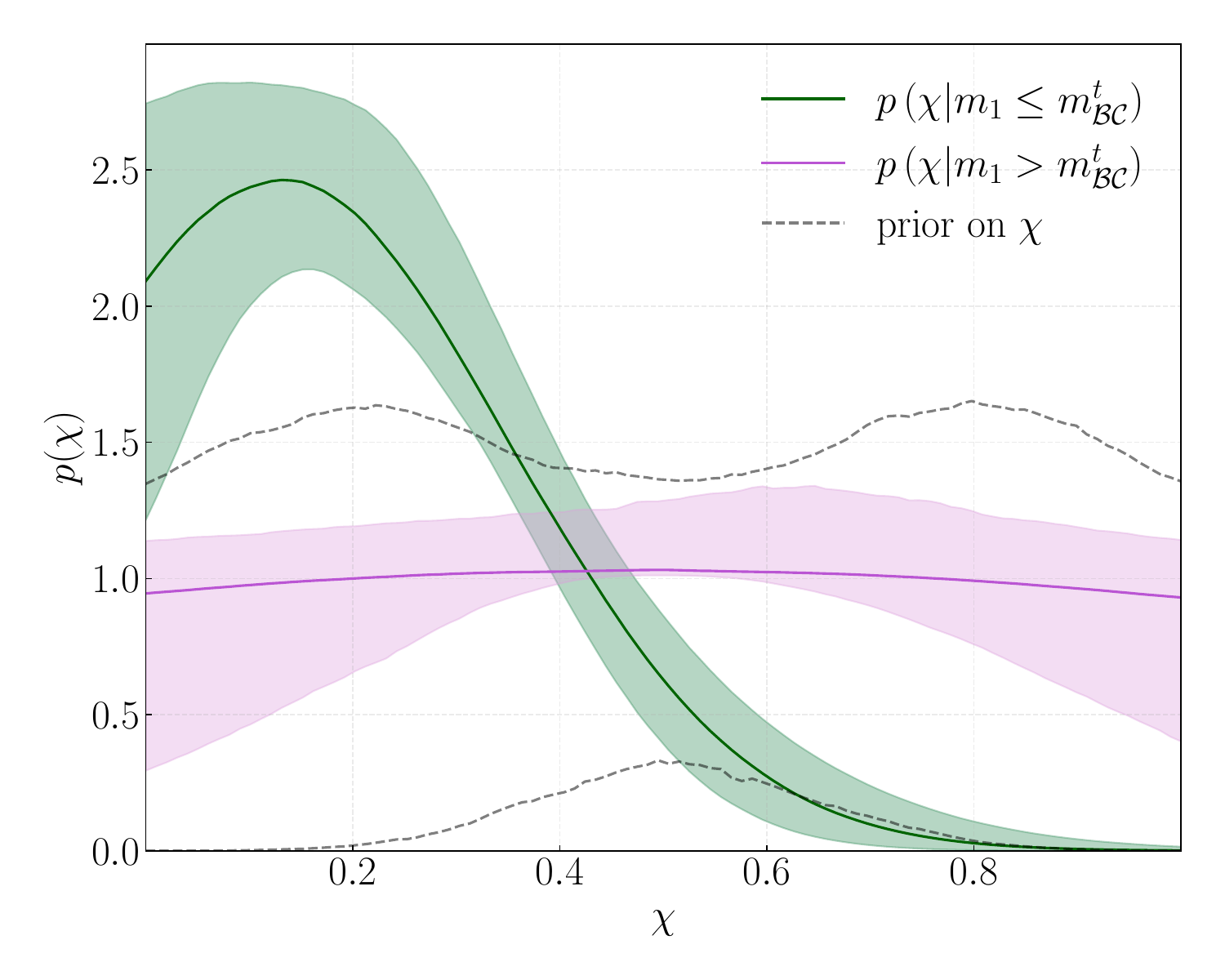}
    \caption{The spin magnitude distributions for different subpopulations of binary black holes. 
    In this and all subsequent figures, the solid curves are the population predictive distributions and the shaded regions are the 90\% credible interval. The $90\%$ credible range of the population prior on the spin magnitudes (see Supplementary Tab.~\ref{tab:priors}) is shown with the dashed lines. \PopOneLabel~ and \PopTwoLabel~are both characterized by comparatively small spins (green) with $\chi < \PopOneChiUpperLimit$. On the other hand, \PopThreeLabel~(magenta) is characterized by a comparatively broader distribution with significant support for $\chi > 0.5$.}
    \label{fig:p_chi}
\end{figure}

\begin{figure}[t!]
         \centering
     % \begin{minipage}{0.49\textwidth}
         \centering
         \includegraphics[width=0.5\textwidth]{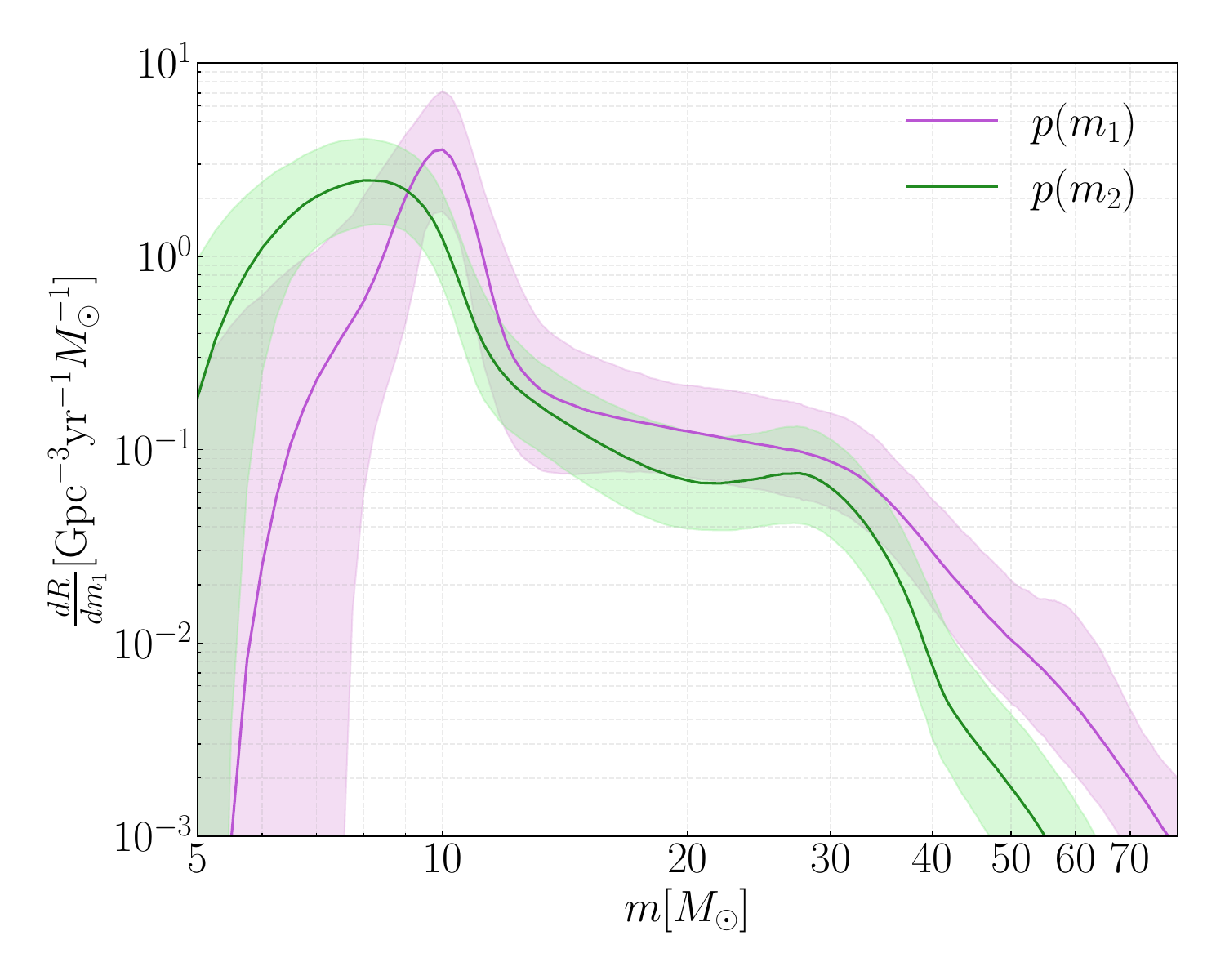}
     % \end{minipage}
     %  % \begin{minipage}{0.49\textwidth}
     %  %    \centering
     %  %    \includegraphics[width=\textwidth]{figures/pm1m2.pdf}
     % \end{minipage}
     % \caption{
     % Left: The reconstructed marginalized distributions of primary mass, $m_1$ and secondary mass, $m_2$. The solid curves are the population predictive distribution and the shaded regions are the 90\% credibility intervals.
     % The distribution of $m_2$ is characterized by a bump at $~30 M_{\odot}$, which is not immediately apparent in the distribution of $m_1$. 
     % Both distributions fall above $40 M_\odot$, but the $m_2$ distribution falls more rapidly than the $m_1$ distribution. Right: reconstructed merger rate density in the $m_1, m_2$ plane. There is a preference for nearly equal masses between $\sim 30 - 40 M_{\odot}$. 
     % At $m_1 \geq 40 M_\odot$, $q\lesssim1$ pairing becomes more common.
     % }
    \caption{ The reconstructed marginalized distributions of primary mass, $m_1$ and secondary mass, $m_2$. The distribution of $m_2$ is characterized by a bump at $~30 M_{\odot}$, which is not immediately apparent in the distribution of $m_1$. Both distributions fall above $40 M_\odot$, but the $m_2$ distribution falls more rapidly than the $m_1$ distribution. }
   \label{fig:p_m1_m2}
\end{figure}

\begin{figure}[t!]
    \centering
\includegraphics[width=0.5\textwidth]{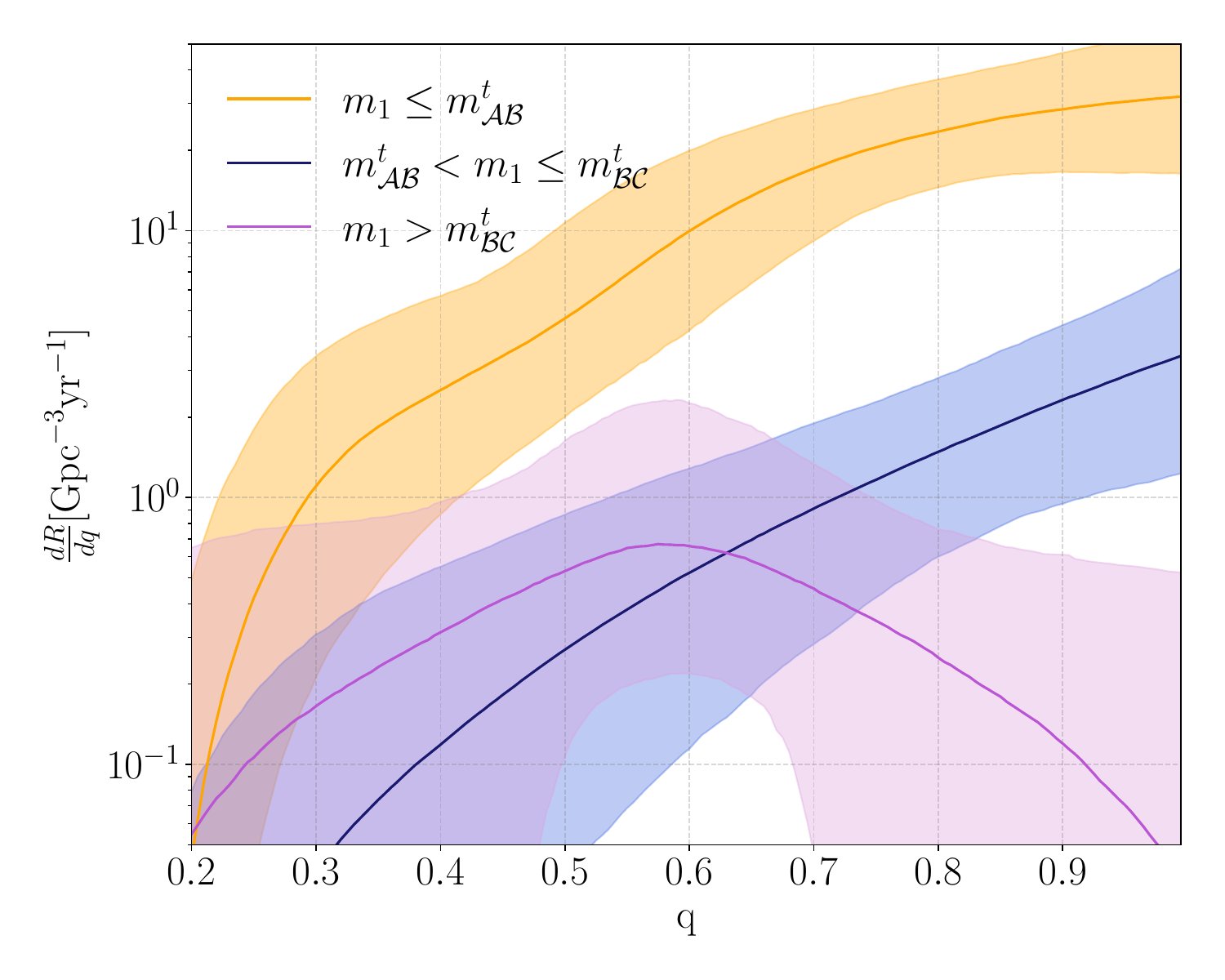}
    \caption{\blue{The reconstructed merger rate of the three subpopulations as a function of the mass ratio. The distribution for \PopThreeLabel~exhibits a tentative peak at $q \approx 0.5$. }}
    \label{fig:Component_q}
\end{figure}

\begin{figure*}[ht]
    \centering
    \includegraphics[width=0.9\textwidth]{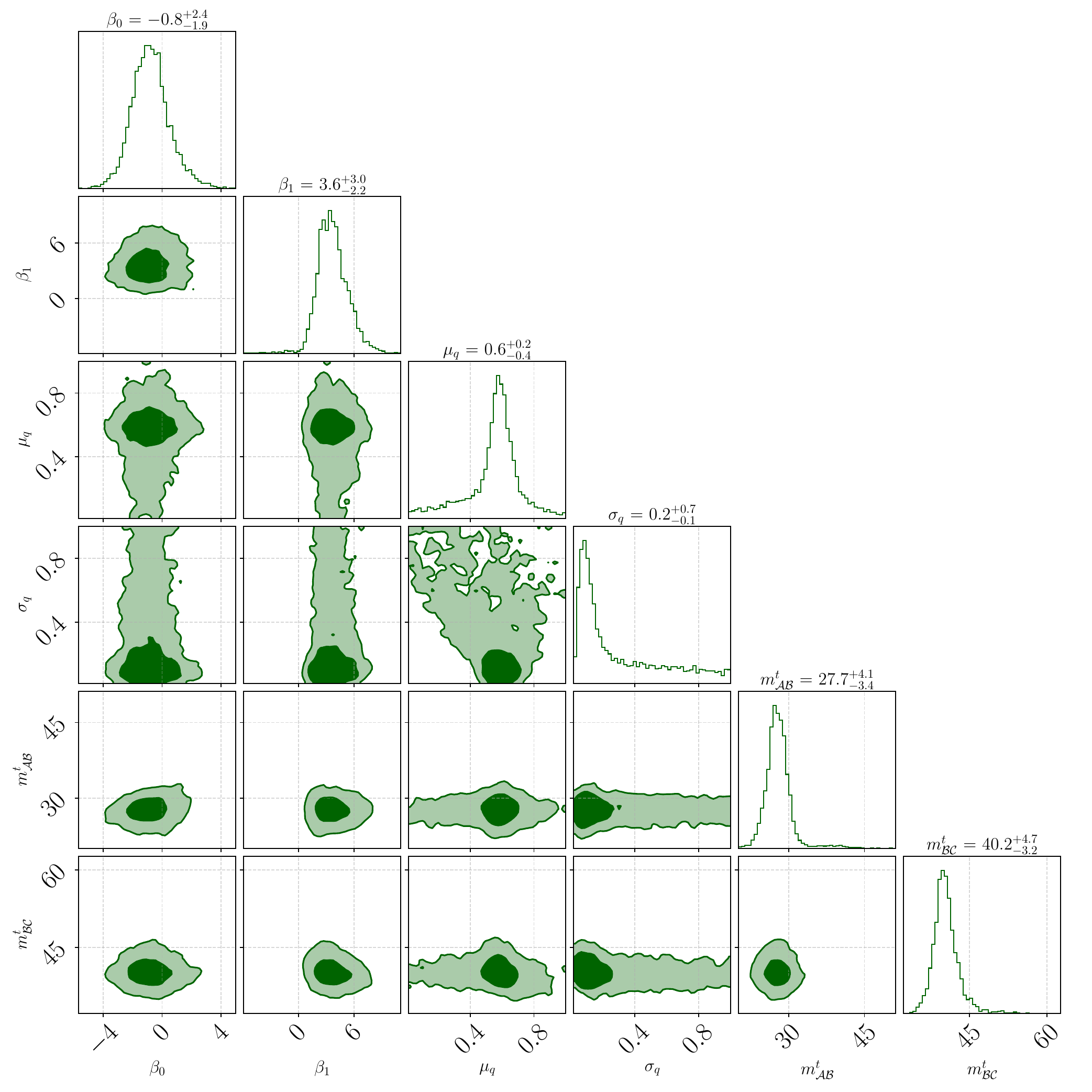}
    \caption{\label{fig:q_corner} Corner plot of the mass ratio and transition parameters for the \textsc{multipop} model.}
\end{figure*}

\begin{itemize}

\item \textit{\PopOneLabel}: $m_1 \leq \mtmid$. 
The mass ratio distribution is somewhat flat; we infer $\beta_0=\FirstComponentBeta$. The spin-magnitude distribution is mostly localized within $\chi \leq \PopOneSpinUL$ -- see Fig.~\ref{fig:p_chi}. Since this subpopulation includes a large range of black hole masses from $3-28 M_\odot$, we speculate that that it may actually include contributions from multiple formation channels.

\item \textit{\PopTwoLabel}: $\mtmid<m_1\leq\mthigh$. 
This subpopulation is characterized by a stronger preference for equal mass pairing; we infer $\beta_1=\SecondComponentBeta$ and $\beta_1 > \beta_0$ at $97\%$ credibility. The spin properties of this subpopulation are indistinguishable from those of \PopOneLabel. Interestingly, by allowing for these two transitions, evidence for a peak in $p(m_1)$ at $m_1 \approx 35 M_{\odot}$ is diminished. However, we see tentative signs of an peak in $p(m_2)$ at a slightly lower mass; see Fig.~\ref{fig:p_m1_m2}. 

\item \textit{\PopThreeLabel}: $m_1 > \mthigh$. The subpopulation is characterized by a wider spin-magnitude distribution that is incompatible with the first two; see Fig.~\ref{fig:p_chi}. The mass-ratio distribution also changes. We find tentative evidence that $q \approx 0.5$ for this subpopulation, see {Figs.~\ref{fig:Component_q} and \ref{fig:q_corner}}. But this is not conclusive and more data is required to confirm this possibility. 
\end{itemize}

We now turn to measuring the statistical significance of individual features and transitions. We define two different metrics for this purpose. The first is the relative Bayes Factor, $\mathcal{B}_{R} $, which is the gain in evidence when a new feature is added to \textsc{gwtc4-fiducial} model. For example, we can compare the evidence after adding a transition in $p(q)$ to \textsc{gwtc4-fiducial} with the evidence without any transitions. We also define the differential Bayes factor, $\mathcal{B}_{D} $ which compares the evidence of the \textsc{multipop} model (Eqs~\ref{Eq:Multipop_pq}, \ref{Eq:Multipop_pchi}) to the evidence when a feature is taken out. We direct the reader to Eqs~\ref{Eq:relative_bayes}, \ref{Eq:differential_bayes} in the Supplementary Materials for formal definitions. These two metrics allow us to gauge the significance of a feature both on its own, and in conjunction with other features that capture the presence of multiple subpopulations. Table.~\ref{tab:bayes_factors} shows the $\ln \mathcal{B}_R$ and $\ln \mathcal{B}_D$ values for the features identified by the \textsc{multipop} model. 

\begin{table}[ht]
    \begin{tabular}{|c|c|c|}
    \hline
    Feature / model & $\ln \mathcal{B}_{R}$ & $\ln \mathcal{B}_{D}$ \\
    \hline
    \textsc{multipop} model & $\MultipopLogB$ & -- \\
    $p(\chi)$ transition at $m_1 = m^t_{\mathcal{BC}}$ & $\ChiTransistionHighRelativeLogB$ & $\ChiTransistionHighDifferentialLogB$\\
    $p(q)$ and $p(\chi)$ both transition at $m_1 = m^t_{\mathcal{BC}}$ & $\MassRatioChiTransistionHighRelativeLogB$ & $\MassRatioChiTransistionHighDifferentialLogB$\\
    $p(q)$ transitions at both $m_1 = m^t_{\mathcal{AB}}$ and $m^t_{\mathcal{BC}}$ & $\MassRatioTransistionMidHighRelativeLogB$ & $\MassRatioTransistionMidHighDifferentialLogB$ \\
    $p(q)$ transition at $m_1 = m^t_{\mathcal{BC}}$ & $\MassRatioTransistionHighRelativeLogB$ & $\MassRatioTransistionHighDifferentialLogB$\\
    $p(q)$ transition at $m_1 = m^t_{\mathcal{AB}}$ & $\MassRatioTransistionMidRelativeLogB$ & $\MassRatioTransistionMidDifferentialLogB$
    \\
    \hline
    \end{tabular}
    \caption{Relative Bayes factor ${\cal B}_R$ and differential Bayes factors ${\cal B}_D$ for the features that define the \textsc{multipop} model. The relative Bayes factor measures the gain in evidence when a feature is added on top of the \textsc{gwtc4-fiducial}. The differential Bayes factor compares the evidence of the \textsc{Multipop} model with and without a feature. The transition points, $m^t_{\mathcal{AB}}$ and $m^t_{\mathcal{BC}}$ (see Fig.~\ref{fig:main_overiew}), are localized at $\mtmid$ and $\mthigh$, respectively. }
    \label{tab:bayes_factors}
\end{table}

The individual feature most strongly supported by data is that \PopThreeLabel~has a distinct spin-magnitude distribution with support for $\chi > 0.5$. When we allow the mass ratio distribution to simultaneously change for this subpopulation, this significantly bolsters evidence for a transition at $m_1 \geq m^t_{\mathcal{BC}}$. The hypothesis that there are three subpopulations with distinct mass-ratio distributions (when keeping the spin magnitude distribution fixed) is strongly supported with $\ln \mathcal{B}_R$ and $\ln \mathcal{B}_D$ of $6.0$ and $4.5$, respectively. {Finally, we note that when we tested a variant of \textsc{multipop} that models the mass-ratio distribution of \PopThreeLabel~as a power-law, it is disfavored with a $\log \mathcal{B}$ of 4.8 compared to the standard version described in Eq.~\ref{Eq:Multipop_pq}.}

\textit{\textbf{Astrophysical Interpretation.}}---We turn now to the astrophysical interpretation of these subpopulations, guided not just by the observed transitions in $p(q)$ and $p(\chi)$ but also the lack of any transitions in spin-tilt or redshift distributions.

\PopOneLabel: This subpopulation is characterized by a somewhat flat mass-ratio distribution and spin magnitudes constrained at $\chi \leq 0.5$. Astroseismological observations of low mass stars find highly effective angular momentum transport~\cite{Fuller:2019abc}, which extrapolated to more massive stars imply black hole spins with $\chi\lesssim0.01$~\cite{Ma:2019cpr, Fuller:2019sxi}. The fact that we see $\chi > 0.01$ for this subpopulation implies that they have been spun up or retained their angular momentum in some other way. \blue{Alternatively, such assumptions of extremely-efficient angular momentum transport might be incorrect for massive stars; models based on the Taylor-dynamo~\citep{Spruit:2001tz} suggest that the spin of isolated black holes can be as high as $\chi \approx 0.1$~\citep{Heger:2004qp, Qin:2018vaa, Belczynski:2017gds}}.

It is also possible that multiple formation channels contribute to the subpopulation below $m_1 \leq m^t_{\mathcal{AB}}$, especially around the $10 M_{\odot}$ peak which is the global maxima of the mass \ac{BBH} mass spectrum~\cite{Tiwari:2020otp, Edelman:2021zkw, GWTC4:astrodist}. There is tentative observational evidence that this is the case with some analyses finding substructure among the subpopulation that makes up the peak~\cite{Godfrey:2023oxb, Galaudage:2024meo, GWTC4:astrodist}. Our exploratory studies, when allowing for a transition around the  $10 M_{\odot}$ peak find broad posteriors for transition masses, perhaps indicating some substructure that is not well fit by our phenomenological models. Nevertheless, we find no clear evidence for multiple subpopulations there. We touch on this point again at the end of this \textit{Letter}. 

\PopTwoLabel: This subpopulation is characterized by a stronger preference for equal mass pairing. The preference for a  bump at $\sim 35 M_{\odot}$ in $p(m_1)$---first noted in Ref.~\cite{o3a_pop}---disappears with the more complicated pairing distribution of \textsc{multipop}; see Figs.~\ref{fig:main_overiew} \& \ref{fig:p_m1_m2}. We find that the absence of an $m_1$ bump is a consequence of \PopTwoLabel~having a distinct mass ratio distribution; when forced to share a common mass ratio distribution with either \PopOneLabel~or {\PopThreeLabel}, the peak reappears. However, we find hints of a bump in $p(m_2)$ at $\sim 30 M_{\odot}$; see Fig.~\ref{fig:p_m1_m2}. A similar result, albeit using the GWTC-3 catalog, was reported by Ref.~\cite{Farah:2023swu} with a model that independently models and pairs $p(m_1)$ and $p(m_2)$. Thus, we propose that the previously identified $m_1$ bump might be due to model misspecification \cite{wmf}. {Future data and analysis with different models will help to shed more light on the possibility of misspecification and the true nature of this feature.}

Recently, Ref.~\cite{Roy:2025ktr} conducted an investigation of \acp{BBH} at the $35 M_\odot$ bump. In effect; they studied binaries equivalent to our \PopTwoLabel. Comparing \ac{GW} inferences to a suite of different population-synthesis results, they conclude that none of the formation channels considered are consistent with the observed features of \PopTwoLabel. While we mostly agree with their assessment, we highlight two potentially promising formation scenarios. The first is \ac{CHE}~\cite{deMink:2009,deMink20092, deMink:2016vkw, Mandel:2015qlu, Marchant:2016wow}. Unlike Ref.~\cite{Roy:2025ktr}, we find a gentler and flatter $p(m_1)$, but still with a preference for equal-mass pairing, consistent with \ac{CHE}~\cite{Hastings:2020}. Perhaps a more difficult thing to reconcile with this formation scenario is that we do not find any difference in spin magnitude or tilt distributions between \PopTwoLabel~and \PopOneLabel. While \ac{CHE} typically involves stellar progenitors that are rotating at a significant fraction of their critical speed, the spins of the subsequent black holes are much more uncertain, with predictions ranging from $\chi \approx 0.4-1$~\cite{Marchant:2016wow, Riley:2020btf, Marchant:2023ncp}.  \PopTwoLabel~might be consistent with \ac{CHE} formation if the black holes are spun up to $\chi\approx 0.4$; see Fig.~\ref{fig:p_chi}. 

Another potential origin for this subpopulation is from population III stars. Multiple studies robustly identify a characteristic excess of \ac{BBH} mergers with masses around $30 M_{\odot}$~\cite{Kinugawa:2014zha, Kinugawa:2020ego, Iwaya:2023mse, Santoliquido:2023wzn} from population III stars. However, the star-formation rate (and hence the merger rate) of such systems should naturally peak at high redshift. While some fraction can experience long delay times and merge at low redshift, a degree of fine tuning is likely required to ensure the same star-formation history as the rest of the \acp{BBH} subpopulation. However, we caution that this conclusion can be susceptible to model misspecification as we only looked for (and failed to find) different redshift distributions at clearly separated $m_1$ ranges. 

Some studies have suggested a dynamical origin for this subpopulation~\cite{Ray:2024hos}. Canonically, such a population should have randomly aligned spins~\cite{Rodriguez:2016vmx} and -- assuming highly efficient angular momentum transport~\cite{Fuller:2019sxi} --  very small spin magnitudes. As we do not find evidence for a separate spin-magnitude or tilt distribution from \PopOneLabel, this interpretation is somewhat at odds with our results. \blue{However, the caveat about assumptions of extremely efficient angular momentum transport for massive stars also apply here and, as previously mentioned, models based on the Taylor-dynamo~\citep{Spruit:2001tz} suggest that the spin of isolated black holes can be as high as $\chi \approx 0.1$~\citep{Heger:2004qp, Qin:2018vaa, Belczynski:2017gds}}. Furthermore, recent suggestions of stellar collisions in dense clusters~\cite{Kiroglu:2025bbp} might be able to produce aligned spinning \acp{BBH} through dynamical scenarios.

Finally, it is unlikely that this subpopulation originates from pair-instability physics, a possibility that attracted much attention in the early days of \ac{GW} astronomy~\cite{Fishbach:2017zga, Talbot:2018cva}. However, multiple studies recently have argued that the putative pileup due to pair-instability physics should lie at least above $40 M_{\odot}$ (for e.g.~\cite{stevenson,Farmer:2019jed, Farmer:2020xne, Hendriks:2023yrw, Renzo:2020rzx}), and more likely above $45 m_{\odot}$---beyond the upper limit of this subpopulation. Although, for an alternative point of view see Refs.~\cite{Croon,Winch}. In any case, with the more complex pairing model introduced in this work, evidence of a pileup at $35 M_{\odot}$ is diminished. We find tentative support for a bump in $m_2$ at $\approx 30 M_\odot$, but this may be too low of a scale for pair instability physics to be relevant. 

\PopThreeLabel: The idea that massive black holes might be formed from hierarchical mergers of earlier generations of black holes has been explored with some of the first unexpectedly-massive \acp{BBH} detected by the LVK (for e.g.~\citep{LIGOScientific:2020iuh, hierarchical,gwtc2_hierarchical,fishbach2017, Chatziioannou:2019dsz, Doctor:2019ruh, Kimball:2019mfs}). Recently, this hypothesis has gained traction because of the discovery of an apparent transition in the effective spin distribution~\cite{Li:2023yyt, Antonini:2024het, Antonini:2025zzw} and an apparent gap in  the distribution of secondary mass~\cite{Tong2025}. 
 
Our inference of the mass-ratio distribution \PopThreeLabel~-- with a mean at $q \approx 0.5$ -- provides further tentative support for the hierarchical interpretation as this qualitatively resembles the distribution one might expect from the merger of a second-generation and a first-generation black holes. {While the inferred spin distribution is very broad, it does have support at at $\chi \sim 0.7$ (Fig.~\ref{fig:p_chi}), consistent with the interpretations of Refs~\cite{Li:2023yyt, Antonini:2024het, Antonini:2025zzw,Tong2025}} 
 
On the other hand, this distribution also has support at small $\chi$, indicating that a fraction of this subpopulation has small spins. {Furthermore, we found that a variant of \textsc{multipop} that allowed only $p(\chi_1)$ to transition at $m^t_{\mathcal{BC}}$ is disfavored by $\ln \mathcal{B} = 1.5$ compared to the version that lets both $p(\chi_1)$ and $p(\chi_2)$ transition}. Another potential inconsistency is that we find only a transition in the distribution of spin magnitudes but not of spin tilts between \PopThreeLabel~and the rest of \ac{BBH} population. These features, if confirmed, might suggest that hierarchical mergers in dense star clusters cannot fully explain \PopThreeLabel. Alternate scenarios such as \ac{CHE}~\cite{Marchant:2016wow, Riley:2020btf, Marchant:2023ncp} and accretion-based spin up in AGN disks~\cite{Bartos:2025pkv} might better explain some of these features. However, it is hard to see how either of these explanations would explain the preference for $q\approx 0.5$ if that is confirmed by future data. {Conversely, alternative accretion scenarios have been proposed that might be able to explain these properties better~\cite{Safarzadeh:2020vbv, vanSon:2020zbk}}. 

{A few other analyses in the literature also seem to suggest that in rapidly spinning subpopulations, both black holes posses rapid spins~\cite{Hussain:2024qzl, Adamcewicz:2025phm}}. Recently, Ref.~\cite{Galaudage:2024meo} also inferred a similar result when modeling component spins -- finding that the both $\chi_1$ and $\chi_2$ have support at high spin magnitudes for the high mass subpopulation (see their Fig. 6). However, they also infer a different spin tilt distribution, perhaps suggesting that any inference about this subpopulation can be particularly sensitive to model specification and misspecification.

\textit{\textbf{Conclusion.}}---In this \textit{Letter}, we investigate the possibility of multiple subpopulations of \acp{BBH} each associated with a different range of black hole masses. We find strong evidence that there are at least three such subpopulations with different mass-ratio and spin-magnitude distributions, with transitions between them inferred at $\mtmid$ and $\mthigh$. A caveat to these results is that we use strongly-parametrized models, potentially subject to model misspecification. {Model misspecification of the $p(q)$ distributions might be particularly important (Fig.~\ref{fig:p_m1_m2}) as we can resolve the features more precisely in the mass distributions than with spins or redshift. While our results are internally self-consistent when the power-law and normal distributions are swapped, testing with more models and future data can always help shed light on model misspecifications.}

However, the strength of the signal and that the fact that these transition points have been individually found in other analyses~\cite{Li:2023yyt, Tiwari:2021yvr, Antonini:2024het, Antonini:2025zzw, Tong2025, Antonini:2025ilj, Afroz:2025ikg} suggests that even with some degree of model misspecification, the presence of three subpopulations and their broad features are real. Furthermore, these results imply that usual assumptions of separability of the population distribution of \ac{BBH} parameters breaks down above $\sim 25 M_{\odot}$. 

Finally, our work focuses mostly on \acp{BBH} at higher masses as we do not find any clearly localized subpopulations at $m_1 < 25 M_{\odot}$. In particular, we do not find evidence for any transitions around the $10 M_{\odot}$ peak. There are reasons to expect substructure at this peak, both based on previous observations~\cite{Godfrey:2023oxb, Galaudage:2024meo, GWTC4:astrodist, Banerjee:2021xzp} and based on theoretical considerations. Isolated binary formation channels that undergo stable mass transfer often predict a peak near $10 M_{\odot}$(for e.g. Refs~\cite{Giacobbo:2018etu, Neijssel:2019, Tanikawa:2021qqi, vanSon:2022ylf, vanSon:2022myr}). Some studies of cluster formation channels also find a peak around $10 M_{\odot}$~\cite{Ye:2024ypm, Ye:2025ano} while others~\cite{Antonini:2022vib} place the peak at higher masses. We plan to revisit this subpopulation in a future study.

\textit{Acknowledgments}: We thank  Shanika Galaudage, Evgeni Grishin, Ryosuke Hirai, Simon Stevenson, Hui Tong and Mike Zevin for discussions and suggestions. We also thank Soumendra Roy for comments on the manuscript. S.B. acknowledges support from Australian Research Council grants CE230100016, LE210100002, and DP230103088. We thank Hui Tong for helpful discussions and suggestions. The authors are grateful for computational resources provided by the LIGO Laboratory and supported by NSF Grants No. PHY-0757058 and No. PHY-0823459. This material is based upon work supported by NSF's LIGO Laboratory which is a major facility fully funded by the National Science Foundation. This research has made use of data obtained from the Gravitational Wave Open Science Center (\href{https://gwosc.org}{gwosc.org}), a service of LIGO Laboratory, the LIGO Scientific Collaboration, the Virgo Collaboration, and KAGRA. LIGO Laboratory and Advanced LIGO are funded by the United States NSF as well as the STFC of the United Kingdom, the Max-Planck-Society (MPS), and the State of Niedersachsen/Germany for support of the construction of Advanced LIGO and construction and operation of the GEO\,600 detector. Additional support for Advanced LIGO was provided by the Australian Research Council. Virgo is funded, through the European Gravitational Observatory (EGO), by the French Centre National de Recherche Scientifique (CNRS), the Italian Istituto Nazionale di Fisica Nucleare (INFN) and the Dutch Nikhef, with contributions by institutions from Belgium, Germany, Greece, Hungary, Ireland, Japan, Monaco, Poland, Portugal, Spain. KAGRA is supported by Ministry of Education, Culture, Sports, Science and Technology (MEXT), Japan Society for the Promotion of Science (JSPS) in Japan; National Research Foundation (NRF) and Ministry of Science and ICT (MSIT) in Korea; Academia Sinica (AS) and National Science and Technology Council (NSTC) in Taiwan.

{\it Data and code availability:} The data and the code behind this is available as a Zenodo repository~\citep{banagiri_2026_20320700}.

\bibliography{short_author, lvk_refs}% Produces the bibliography via BibTeX.

\section{End Matter}

%\onecolumngrid
\setcounter{equation}{0}
\setcounter{figure}{0}
\setcounter{table}{0}
\setcounter{section}{0}

%% Prefixing the numbers with 'S'
\renewcommand{\theequation}{S\arabic{equation}}
\renewcommand{\thefigure}{S\arabic{figure}}
\renewcommand{\thetable}{S\arabic{table}}

\include{supplementary}

\end{document}

%% file: macros.tex
\newcommand{\SPA}{School of Physics and Astronomy, Monash University, Clayton VIC 3800, Australia}
\newcommand{\OzGravMonash}{OzGrav: The ARC Centre of Excellence for Gravitational Wave Discovery, Clayton VIC 3800, Australia}

\newcommand{\PopOneLabel}{ Subpopulation~${\cal A}$}
\newcommand{\PopTwoLabel}{Subpopulation~${\cal B}$}
\newcommand{\PopThreeLabel}{Subpopulation~${\cal C}$}

\newcommand{\MultipopLogB}{12.3}

\newcommand{\MassRatioTransistionMidRelativeLogB}{0.8}
\newcommand{\MassRatioTransistionMidDifferentialLogB}{2.2}

\newcommand{\MassRatioTransistionHighRelativeLogB}{2.7}
\newcommand{\MassRatioTransistionHighDifferentialLogB}{4.0}

\newcommand{\MassRatioTransistionMidHighRelativeLogB}{6.0}
\newcommand{\MassRatioTransistionMidHighDifferentialLogB}{4.5}

\newcommand{\ChiTransistionHighRelativeLogB}{7.9}
\newcommand{\ChiTransistionHighDifferentialLogB}{6.3}

\newcommand{\MassRatioChiTransistionHighRelativeLogB}{10.1}
\newcommand{\MassRatioChiTransistionHighDifferentialLogB}{11.5}

\newcommand{\mtmid}{27.7^{+4.1}_{-3.4} M_{\odot}}
\newcommand{\mthigh}{40.2^{+4.7}_{-3.2} M_{\odot}}

\newcommand{\FirstComponentBeta}{{-0.8}^{+2.4}_{-1.9}}
\newcommand{\SecondComponentBeta}{{3.6}^{+3.0}_{-2.2}}

\newcommand{\PopOneSpinUL}{0.5^{+0.1}_{-0.1}}

\newcommand{\PopOneChiUpperLimit}{{0.5}^{+0.1}_{-0.1}}

\newcommand{\RevOneEdit}[1]{{\color{black} #1}}

%% file: supplementary.tex
\section*{Supplementary Material}

\begin{table*}[ht]
\centering
 \input{priors}
 \caption{Descriptions and the priors used for the various parameters. Note that when a model uses both $m^t_1$ and $m^t_2$, we instead sample from uniformly from a 2-$d$ distribution U($m^t_1, m^t_2)$ subject to the constraint $m^t_2 \geq m^t_1$. }
\label{tab:priors}
\end{table*}

The analysis is built on the formalism of hierarchical Bayesian inference~\citep{Thrane:2019, Vitale:2020aaz}. This involves calculating hyperparameters $\Lambda$ that describe the population distribution of \ac{BBH} using discrete detections. The individual observations are constrained by the sensitivity of the detectors, and we usually apply some threshold $x_{\rm th}$ on the detection statistics $x$ that is stringent enough to neglect contamination from false alarms. One then obtains a posterior distribution for the population parameters as

\begin{equation}
    p(\Lambda | \{ d\}) \propto \pi(\Lambda) \prod_i  \frac{\int d \theta \mathcal{L}(d_i | \theta) \pi(\theta | \Lambda)}{\xi(x > x_{\rm th} | \Lambda)}.
    \label{Eq:hierarchical_inference}
\end{equation}

Here, $\theta$ are parameters that describe an individual detection (such as masses, spins, redshift etc), $\mathcal{L}(d_i | \theta)$ is the likelihood of the $i$-th detection and $\pi(\theta | \Lambda)$ describes the priors on those parameters conditioned on $\Lambda$. The denominator, $\xi(x > x_{\rm th} | \Lambda)$~\footnote{Assuming without loss of generality that a larger $x$ corresponds to a more statistically significant detection.} accounts for the selection effects that arise from the thresholding described above, and is calculated as the fraction of events from a population described by $\Lambda$ that will fall above this threshold when observed~\cite{Mandel:2018mve, Essick:2023upv}. 

For all models considered here, we retain the same strongly-parameterized primary mass distribution, \textsc{broken powerlaw + two peaks}~\cite{Callister:2023tgi, GWTC4:astrodist}. This models $p(m_1)$ as a broken powerlaw with two Gaussian peaks along with a low mass taper used as a smoothing function. For more details of the model, please see Appendix B.2 of Ref.~\cite{GWTC4:astrodist}. 

We model the other parameters using fidicual strongly-parametrized models adapted from Ref.~\cite{GWTC4:astrodist} (see their Table 1). We allow the distribution of all other parameters to have transitions along $m_1$ so that the distribution above and below each transition are independent of each other. 

The $i$-th subpopulation of $p(z)$ is therefore modeled as a powerlaw in $(1+z)$ with some index $\kappa_z^i$,
\begin{equation}
 p_i(z) \propto (1 + z)^{\kappa_z^i}. 
\end{equation}
The spin magnitudes of the $i$-th subpopulation are modeled as (independently and identically distributed) normals truncated between $[0, 1]$, with mean and standard deviation parameters of $\mu_{\chi}^i$ and $\sigma_{\chi}^i$,
\begin{equation}
    p_i(\chi) \propto \mathcal{N}(\chi | \mu_{\chi}^i, \sigma_{\chi}^i).
\end{equation}
The $i$-th subpopulation of $p(\cos \theta)$ is modeled as a normal distribution truncated between $[-1, 1]$, with mean and standard deviation parameters of $\mu_{\theta}^i$ and $\sigma_{\theta}^i$,
\begin{equation}
    p_i(\cos \theta) \propto \mathcal{N}(\cos \theta | \mu_{\chi}^i, \sigma_{\chi}^i)
\end{equation}
Finally, the  $i$-th subpopulation of $p(q)$ is modeled as a powerlaw with index $\beta_q^i$,
\begin{equation}
    p_i(q) \propto q^{\beta_q^i}.
\end{equation}
However, for \PopOneLabel~and \PopThreeLabel, we also explore an alternate parameterization of $p(q)$ as a truncated normal with $\mu_q$ and $\sigma_q$ as mean and standard deviation,
\begin{equation}
    p_i(q) \propto \mathcal{N}(q | \mu_q, \sigma_q).
\end{equation}
\RevOneEdit{This alternate parameterization for \PopOneLabel~is motivated by work in the literature that reports evidence for mergers with a mass ratio distribution peaking at $q \approx 0.7$ at around $10 M_{\odot}$~\cite{Godfrey:2023oxb, GWTC4:astrodist}. While for \PopThreeLabel~, it is motivated by literature that reports evidence of $2G + 1G$ hierarchical mergers at high masses~\cite{Tong2025}}.

We allow up to three transition points, thereby producing up to four subpopulations. The distribution of individual parameters can then have any number of these three transition points, independently of each other. The transition masses, $m^t_{\mathcal{ZA}}, m^t_{\mathcal{AB}}, \text{ and } m^t_{\mathcal{BC}}$, are themselves treated as independent parameters in the model to be inferred from the data. Table.~\ref{tab:priors} provides a brief description of all of these parameters and their priors. 

We tested these models using \acp{BBH} from the cumulative GWTC-4.0 catalog~\cite{LIGOScientific:2025snk, GWTC4:Catalog_results, LIGOScientific:2025gwtc4zenodo} comprising detections from O1, O2, O3 and O4a observing runs of LIGO and Virgo. We adopt the same threshold (false alarm rate less than $1 {\rm yr}^{-1}$) used for the dedicated \ac{BBH} population analysis by the LVK~\cite{GWTC4:astrodist}; however as described in the main text we do no include GW190418 and GW231123, yielding a total of 152 detections. The selection effect term ($\xi$ in Eq.~\ref{Eq:hierarchical_inference}) is calculated using the injection campaigns conducted by the LVK and released with their data release~\cite{Essick:2025zed, LVK:2025gwtc4sensitivityzenodo}.

We control for Monte Carlo errors when estimating Eq.~\ref{Eq:hierarchical_inference} by discarding draws from $\Lambda$ with large relative variances in the estimated log-likelihoods~\citep{Essick:2022ojx, Talbot:2023pex}. More specifically, we follow Ref.~\cite{GWTC4:astrodist} by adopting a threshold of 1 for the variance of the log of the estimator. All our analyses and models are implemented in \textsc{GWpopulation}~\citep{Talbot:2019} with \textsc{jax} for GPU-based Bayesian inference. We use \textsc{bilby}~\citep{Ashton:2019} and the nested sampler \textsc{dynesty}~\citep{Speagle:2020} for Bayesian inference and evidence estimation.

After testing all realistic variations of the models described, we found that \textsc{multipop}, described in Eqs~\ref{Eq:Multipop_pq} and \ref{Eq:Multipop_pq} has the highest Bayes factor and show clear evidence for multiple subpopulations at different masses. As described in the main text, the three subpopulations in this model have different $p(q)$ and $p(\chi)$ with transition masses at localized to $m^t_{\mathcal{AB}} = \mtmid$ and $m^t_{\mathcal{BC}} = \mthigh$. While we see hints of other structure along the mass spectrum, we do not find evidence for statistically significant, well-localized transitions either in redshift or in spin tilts. \RevOneEdit{Figure.~\ref{fig:Component_q}~plots the merger rate as a function of mass ratio for each component mass and Fig.~\ref{fig:q_corner} shows the corresponding corner plot for the mass ratio parameters along with the transition masses.}

The reference fiducial model \textsc{gwtc4-fiducial}, against which we compare \textsc{multipop}, has the same primary mass distribution. It models the mass ratio distribution as a power law, spin magnitudes as a truncated normal, spin tilts as a mixture between a uniform and an aligned component and the redshift distribution as a powerlaw in $(1+z)$. These are modeled globally in \textsc{gwtc4-fiducial} with no transitions or subpopulations and with their seperability strictly enforced, i.e. $p(m_1, q, \chi, \cos \theta, z) = p(m_1) \times p(q) \times p(\chi) \times p(\cos \theta) \times p(z)$. We direct the reader to the appendices of Ref.~\cite{GWTC4:astrodist} for more details of these models. 

We find that \textsc{multipop} is supported over the \textsc{gwtc4-fiducial} model by a natural log Bayes factor of $\MultipopLogB$. We define two metrics for measuring the significance of individual features $\lambda$ that relate to specific subpopulations. The first is the relative Bayes factor $\mathcal{B}_R$ defined as the ratio between the \textsc{gwtc4-fiducial} with just that feature $\lambda$ added, and the base \textsc{gwtc4-fiducial} model: 

\begin{equation}
 \ln \mathcal{B}_R (\lambda) = \ln \mathcal{Z}_{\textsc{gwtc4-fiducial} + \lambda} - \ln \mathcal{Z}_{\textsc{gwtc4-fiducial}}.
 \label{Eq:relative_bayes}
\end{equation}

The second metric is the differential Bayes factor defined as the drop in evidence when we take out the feature corresponding to $\lambda$ from \textsc{multipop}:

\begin{equation}
 \ln \mathcal{B}_D (\lambda)= \ln \mathcal{Z}_{\textsc{multipop}} - \ln \mathcal{Z}_{\textsc{multipop} - \lambda}.
  \label{Eq:differential_bayes}
\end{equation}

The values of the two Bayes factors for the various features characterizing the subpopulations are shown in Tab.~\ref{tab:bayes_factors}.

\begin{figure*}[ht]
    \centering
    \includegraphics[width=1.07\textwidth]{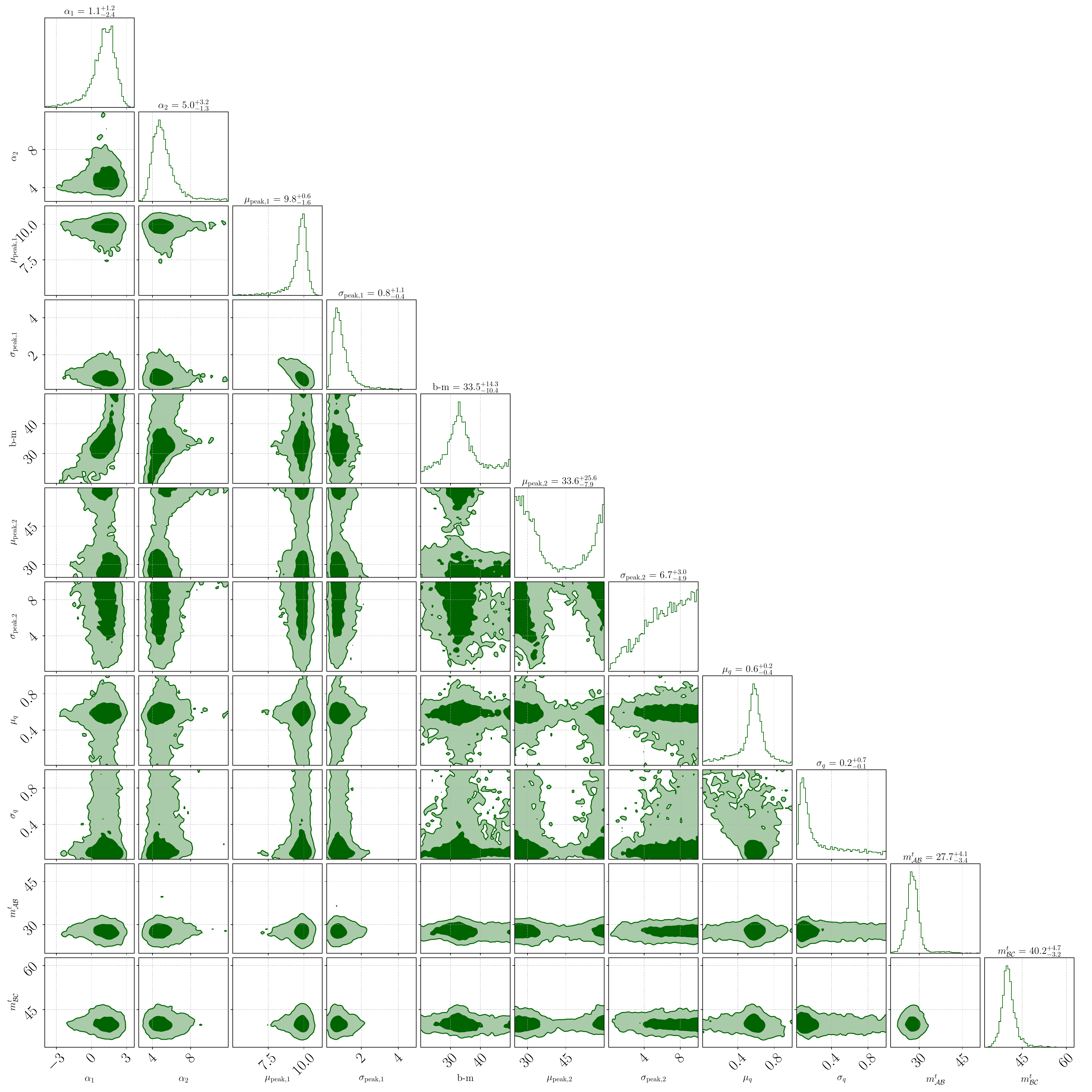}
    \caption{\label{fig:mass_model_corner} \textbf{Corner plot of the parameters of the primary mass model, and transition masses for the \textsc{multipop} model}.}
\end{figure*}

% \bibliography{short_author, lvk_refs}

%% file: priors.tex
\begin{tabular*}{0.95\textwidth}{c  @{\extracolsep{\fill}} c c } 
\hline 
Parameter & Description & Prior \\
\hline \hline 
\multicolumn{1}{c}{}  & \multicolumn{1}{c}{$p(m_1)$ priors} \\
\hline 
$\alpha_1$ & Power-law index of the first powerlaw   & U$(-4, 12)$ \\
$\alpha_2$ & Power-law index of the second powerlaw   & U$(-4, 12)$ \\
$m_{\rm low, 1}$ & Minimum primary blackhole mass & U$(3, 10) \, M_{\odot}$ \\
$b_m$  & Break mass of the broken powerlaw & U$(20, 50) \, M_{\odot}$ \\
$\delta_m^1$ & Taper size at low mass end & U$(0, 10) \, M_{\odot}$ \\
$\mu_{\rm peak, 1}$ & Location of the first Gaussian peak & U$(5, 20) \, M_{\odot}$ \\
$\mu_{\rm peak, 2}$ & Location of the second Gaussian peak & U$(25, 60) \, M_{\odot}$ \\
$\sigma_{\rm peak, 1}$ & Width of the first Gaussian peak & U$(0, 5) \, M_{\odot}$ \\
$\sigma_{\rm peak, 2}$ & Width of the first Gaussian peak & U$(0, 10) \, M_{\odot}$ \\
$\lambda_{1}, \lambda_{2}, \lambda_{3}$ & Mixing fractions for the broken powerlaw and the two Gaussian peaks & Dir$(\alpha=(1, 1, 1))$ \\
\hline
\multicolumn{1}{c}{}  & \multicolumn{1}{c}{Multipartite parameters} \\
\hline
$m^t_{\mathcal{ZA}}$ & First potential transition mass & U$(5, 20) \, M_{\odot}$ \\
$m^t_{\mathcal{AB}}$ & Second potential transition mass & U$(20, 65)\, M_{\odot}$ \\
$m^t_{\mathcal{BC}}$ & Third potential transition mass & U$(30, 75)\, M_{\odot}$ \\
$\kappa_z^i$ & Redshift powerlaw index of the $i$-th subpopulation & U$(-8, 8)$ \\
$\beta_q^i$ & Mass ratio powerlaw index of the $i$-th subpopulation & U$(-6, 11)$ \\
$\mu_q$ & Mean of the mass ratio distribution of the subpopulation & U$(0.01, 1)$ \\
$\sigma_q$ & \RevOneEdit{Width} of the mass ratio distribution of the subpopulation & U$(0.01, 1)$ \\
$\mu^i_{\chi}$ & Mean of the spin-magnitude distribution of the $i$-th subpopulation & U$(0, 1)$ \\
$\sigma^i_{\chi}$ & Width of the spin-magnitude distribution of the $i$-th subpopulation & U$(0.005, 2)$ \\
$\xi_{\theta}$ & Fraction of the aligned population (only used when tilts are modeled as seperable) &  U$(0, 1)$ \\
$\mu^i_{\theta}$ & Mean of the aligned part of the $i$-th subpopulation & U$(-1, 1)$ \\
$\sigma^i_{\theta}$ & Width of aligned part of the $i$-th subpopulation & U$(0.1, 4)$ \\
\hline
\multicolumn{1}{c}{}  & \multicolumn{1}{c}{Common parameters of the mass ratio distribution} \\
\hline
$m_{\rm low, 2}$ & Minimum primary blackhole mass & U$(3, 10) \, M_{\odot}$ \\
$\delta_m^2$ & Taper size at low mass end in the implicit distribution of $m_2$ & U$(0, 10) \, M_{\odot}$ \\
\hline
\end{tabular*}

%% file: main.bbl
\begin{thebibliography}{130}%
\makeatletter
\providecommand \@ifxundefined [1]{%
 \@ifx{#1\undefined}
}%
\providecommand \@ifnum [1]{%
 \ifnum #1\expandafter \@firstoftwo
 \else \expandafter \@secondoftwo
 \fi
}%
\providecommand \@ifx [1]{%
 \ifx #1\expandafter \@firstoftwo
 \else \expandafter \@secondoftwo
 \fi
}%
\providecommand \natexlab [1]{#1}%
\providecommand \enquote  [1]{``#1''}%
\providecommand \bibnamefont  [1]{#1}%
\providecommand \bibfnamefont [1]{#1}%
\providecommand \citenamefont [1]{#1}%
\providecommand \href@noop [0]{\@secondoftwo}%
\providecommand \href [0]{\begingroup \@sanitize@url \@href}%
\providecommand \@href[1]{\@@startlink{#1}\@@href}%
\providecommand \@@href[1]{\endgroup#1\@@endlink}%
\providecommand \@sanitize@url [0]{\catcode `\\12\catcode `\$12\catcode `\&12\catcode `\#12\catcode `\^12\catcode `\_12\catcode `\%12\relax}%
\providecommand \@@startlink[1]{}%
\providecommand \@@endlink[0]{}%
\providecommand \url  [0]{\begingroup\@sanitize@url \@url }%
\providecommand \@url [1]{\endgroup\@href {#1}{\urlprefix }}%
\providecommand \urlprefix  [0]{URL }%
\providecommand \Eprint [0]{\href }%
\providecommand \doibase [0]{https://doi.org/}%
\providecommand \selectlanguage [0]{\@gobble}%
\providecommand \bibinfo  [0]{\@secondoftwo}%
\providecommand \bibfield  [0]{\@secondoftwo}%
\providecommand \translation [1]{[#1]}%
\providecommand \BibitemOpen [0]{}%
\providecommand \bibitemStop [0]{}%
\providecommand \bibitemNoStop [0]{.\EOS\space}%
\providecommand \EOS [0]{\spacefactor3000\relax}%
\providecommand \BibitemShut  [1]{\csname bibitem#1\endcsname}%
\let\auto@bib@innerbib\@empty
%</preamble>
\bibitem [{\citenamefont {Abbott}\ \emph {et~al.}(2016)\citenamefont {Abbott} \emph {et~al.}}]{LIGOScientific:2016aoc}%
  \BibitemOpen
  \bibfield  {author} {\bibinfo {author} {\bibfnamefont {B.~P.}\ \bibnamefont {Abbott}} \emph {et~al.} (\bibinfo {collaboration} {LIGO Scientific, Virgo}),\ }\bibfield  {title} {\bibinfo {title} {{Observation of Gravitational Waves from a Binary Black Hole Merger}},\ }\href {https://doi.org/10.1103/PhysRevLett.116.061102} {\bibfield  {journal} {\bibinfo  {journal} {Phys. Rev. Lett.}\ }\textbf {\bibinfo {volume} {116}},\ \bibinfo {pages} {061102} (\bibinfo {year} {2016})},\ \Eprint {https://arxiv.org/abs/1602.03837} {arXiv:1602.03837 [gr-qc]} \BibitemShut {NoStop}%
\bibitem [{\citenamefont {Abac}\ \emph {et~al.}(2025{\natexlab{a}})\citenamefont {Abac} \emph {et~al.}}]{KAGRA:2025oiz}%
  \BibitemOpen
  \bibfield  {author} {\bibinfo {author} {\bibfnamefont {A.~G.}\ \bibnamefont {Abac}} \emph {et~al.} (\bibinfo {collaboration} {KAGRA, Virgo, LIGO Scientific}),\ }\bibfield  {title} {\bibinfo {title} {{GW250114: Testing Hawking{\textquoteright}s Area Law and the Kerr Nature of Black Holes}},\ }\href {https://doi.org/10.1103/kw5g-d732} {\bibfield  {journal} {\bibinfo  {journal} {Phys. Rev. Lett.}\ }\textbf {\bibinfo {volume} {135}},\ \bibinfo {pages} {111403} (\bibinfo {year} {2025}{\natexlab{a}})},\ \Eprint {https://arxiv.org/abs/2509.08054} {arXiv:2509.08054 [gr-qc]} \BibitemShut {NoStop}%
\bibitem [{\citenamefont {Capote}\ \emph {et~al.}(2025)\citenamefont {Capote} \emph {et~al.}}]{Capote:2024rmo}%
  \BibitemOpen
  \bibfield  {author} {\bibinfo {author} {\bibfnamefont {E.}~\bibnamefont {Capote}} \emph {et~al.},\ }\bibfield  {title} {\bibinfo {title} {{Advanced LIGO detector performance in the fourth observing run}},\ }\href {https://doi.org/10.1103/PhysRevD.111.062002} {\bibfield  {journal} {\bibinfo  {journal} {Phys. Rev. D}\ }\textbf {\bibinfo {volume} {111}},\ \bibinfo {pages} {062002} (\bibinfo {year} {2025})},\ \Eprint {https://arxiv.org/abs/2411.14607} {arXiv:2411.14607 [gr-qc]} \BibitemShut {NoStop}%
\bibitem [{\citenamefont {Ganapathy}\ \emph {et~al.}(2023)\citenamefont {Ganapathy} \emph {et~al.}}]{Ganapathy:2023}%
  \BibitemOpen
  \bibfield  {author} {\bibinfo {author} {\bibfnamefont {D.}~\bibnamefont {Ganapathy}} \emph {et~al.} (\bibinfo {collaboration} {LIGO O4 Detector Collaboration}),\ }\bibfield  {title} {\bibinfo {title} {Broadband quantum enhancement of the ligo detectors with frequency-dependent squeezing},\ }\href {https://doi.org/10.1103/PhysRevX.13.041021} {\bibfield  {journal} {\bibinfo  {journal} {Phys. Rev. X}\ }\textbf {\bibinfo {volume} {13}},\ \bibinfo {pages} {041021} (\bibinfo {year} {2023})}\BibitemShut {NoStop}%
\bibitem [{\citenamefont {Jia}\ \emph {et~al.}(2024)\citenamefont {Jia} \emph {et~al.}}]{Wenxuan:2024elc}%
  \BibitemOpen
  \bibfield  {author} {\bibinfo {author} {\bibfnamefont {W.}~\bibnamefont {Jia}} \emph {et~al.} (\bibinfo {collaboration} {members of the LIGO Scientific{\textdagger}}),\ }\bibfield  {title} {\bibinfo {title} {{Squeezing the quantum noise of a gravitational-wave detector below the standard quantum limit}},\ }\href {https://doi.org/10.1126/science.ado8069} {\bibfield  {journal} {\bibinfo  {journal} {Science}\ }\textbf {\bibinfo {volume} {385}},\ \bibinfo {pages} {1318} (\bibinfo {year} {2024})},\ \Eprint {https://arxiv.org/abs/2404.14569} {arXiv:2404.14569 [gr-qc]} \BibitemShut {NoStop}%
\bibitem [{\citenamefont {Buikema}\ \emph {et~al.}(2020)\citenamefont {Buikema} \emph {et~al.}}]{Buikema:2020}%
  \BibitemOpen
  \bibfield  {author} {\bibinfo {author} {\bibfnamefont {A.}~\bibnamefont {Buikema}} \emph {et~al.},\ }\bibfield  {title} {\bibinfo {title} {Sensitivity and performance of the advanced ligo detectors in the third observing run},\ }\bibfield  {journal} {\bibinfo  {journal} {Physical Review D}\ }\textbf {\bibinfo {volume} {102}},\ \href {https://doi.org/10.1103/physrevd.102.062003} {10.1103/physrevd.102.062003} (\bibinfo {year} {2020})\BibitemShut {NoStop}%
\bibitem [{\citenamefont {Soni}\ \emph {et~al.}(2025)\citenamefont {Soni} \emph {et~al.}}]{LIGO:2024kkz}%
  \BibitemOpen
  \bibfield  {author} {\bibinfo {author} {\bibfnamefont {S.}~\bibnamefont {Soni}} \emph {et~al.} (\bibinfo {collaboration} {LIGO}),\ }\bibfield  {title} {\bibinfo {title} {{LIGO Detector Characterization in the first half of the fourth Observing run}},\ }\href {https://doi.org/10.1088/1361-6382/adc4b6} {\bibfield  {journal} {\bibinfo  {journal} {Class. Quant. Grav.}\ }\textbf {\bibinfo {volume} {42}},\ \bibinfo {pages} {085016} (\bibinfo {year} {2025})},\ \Eprint {https://arxiv.org/abs/2409.02831} {arXiv:2409.02831 [astro-ph.IM]} \BibitemShut {NoStop}%
\bibitem [{\citenamefont {Abac}\ \emph {et~al.}(2025{\natexlab{b}})\citenamefont {Abac} \emph {et~al.}}]{LIGOScientific:2025hdt}%
  \BibitemOpen
  \bibfield  {author} {\bibinfo {author} {\bibfnamefont {A.~G.}\ \bibnamefont {Abac}} \emph {et~al.} (\bibinfo {collaboration} {LIGO Scientific, VIRGO, KAGRA}),\ }\bibfield  {title} {\bibinfo {title} {{GWTC-4.0: An Introduction to Version 4.0 of the Gravitational-Wave Transient Catalog}},\ }\href@noop {} {\  (\bibinfo {year} {2025}{\natexlab{b}})},\ \Eprint {https://arxiv.org/abs/2508.18080} {arXiv:2508.18080 [gr-qc]} \BibitemShut {NoStop}%
\bibitem [{\citenamefont {Abac}\ \emph {et~al.}(2025{\natexlab{c}})\citenamefont {Abac} \emph {et~al.}}]{LIGOScientific:2025yae}%
  \BibitemOpen
  \bibfield  {author} {\bibinfo {author} {\bibfnamefont {A.~G.}\ \bibnamefont {Abac}} \emph {et~al.} (\bibinfo {collaboration} {LIGO Scientific, VIRGO, KAGRA}),\ }\bibfield  {title} {\bibinfo {title} {{GWTC-4.0: Methods for Identifying and Characterizing Gravitational-wave Transients}},\ }\href@noop {} {\  (\bibinfo {year} {2025}{\natexlab{c}})},\ \Eprint {https://arxiv.org/abs/2508.18081} {arXiv:2508.18081 [gr-qc]} \BibitemShut {NoStop}%
\bibitem [{GWT(2025{\natexlab{a}})}]{GWTC4:Catalog_results}%
  \BibitemOpen
  \bibfield  {title} {\bibinfo {title} {{GWTC-4.0: Updating the Gravitational-Wave Transient Catalog with Observations from the First Part of the Fourth LIGO-Virgo-KAGRA Observing Run}},\ }\href@noop {} {\  (\bibinfo {year} {2025}{\natexlab{a}})},\ \Eprint {https://arxiv.org/abs/2508.18082} {arXiv:2508.18082 [gr-qc]} \BibitemShut {NoStop}%
\bibitem [{\citenamefont {Aasi}\ \emph {et~al.}(2015)\citenamefont {Aasi} \emph {et~al.}}]{LIGOScientific:2014pky}%
  \BibitemOpen
  \bibfield  {author} {\bibinfo {author} {\bibfnamefont {J.}~\bibnamefont {Aasi}} \emph {et~al.} (\bibinfo {collaboration} {LIGO Scientific}),\ }\bibfield  {title} {\bibinfo {title} {{Advanced LIGO}},\ }\href {https://doi.org/10.1088/0264-9381/32/7/074001} {\bibfield  {journal} {\bibinfo  {journal} {Class. Quant. Grav.}\ }\textbf {\bibinfo {volume} {32}},\ \bibinfo {pages} {074001} (\bibinfo {year} {2015})},\ \Eprint {https://arxiv.org/abs/1411.4547} {arXiv:1411.4547 [gr-qc]} \BibitemShut {NoStop}%
\bibitem [{\citenamefont {Acernese}\ \emph {et~al.}(2015)\citenamefont {Acernese} \emph {et~al.}}]{VIRGO:2014yos}%
  \BibitemOpen
  \bibfield  {author} {\bibinfo {author} {\bibfnamefont {F.}~\bibnamefont {Acernese}} \emph {et~al.} (\bibinfo {collaboration} {VIRGO}),\ }\bibfield  {title} {\bibinfo {title} {{Advanced Virgo: a second-generation interferometric gravitational wave detector}},\ }\href {https://doi.org/10.1088/0264-9381/32/2/024001} {\bibfield  {journal} {\bibinfo  {journal} {Class. Quant. Grav.}\ }\textbf {\bibinfo {volume} {32}},\ \bibinfo {pages} {024001} (\bibinfo {year} {2015})},\ \Eprint {https://arxiv.org/abs/1408.3978} {arXiv:1408.3978 [gr-qc]} \BibitemShut {NoStop}%
\bibitem [{\citenamefont {Akutsu}\ \emph {et~al.}(2021)\citenamefont {Akutsu} \emph {et~al.}}]{KAGRA:2020tym}%
  \BibitemOpen
  \bibfield  {author} {\bibinfo {author} {\bibfnamefont {T.}~\bibnamefont {Akutsu}} \emph {et~al.} (\bibinfo {collaboration} {KAGRA}),\ }\bibfield  {title} {\bibinfo {title} {{Overview of KAGRA: Detector design and construction history}},\ }\href {https://doi.org/10.1093/ptep/ptaa125} {\bibfield  {journal} {\bibinfo  {journal} {PTEP}\ }\textbf {\bibinfo {volume} {2021}},\ \bibinfo {pages} {05A101} (\bibinfo {year} {2021})},\ \Eprint {https://arxiv.org/abs/2005.05574} {arXiv:2005.05574 [physics.ins-det]} \BibitemShut {NoStop}%
\bibitem [{GWT(2025{\natexlab{b}})}]{GWTC4:astrodist}%
  \BibitemOpen
  \bibfield  {title} {\bibinfo {title} {{GWTC-4.0: Population Properties of Merging Compact Binaries}},\ }\href@noop {} {\  (\bibinfo {year} {2025}{\natexlab{b}})},\ \Eprint {https://arxiv.org/abs/2508.18083} {arXiv:2508.18083 [astro-ph.HE]} \BibitemShut {NoStop}%
\bibitem [{\citenamefont {Vitale}\ \emph {et~al.}(2017)\citenamefont {Vitale}, \citenamefont {Lynch}, \citenamefont {Sturani},\ and\ \citenamefont {Graff}}]{Vitale:2015tea}%
  \BibitemOpen
  \bibfield  {author} {\bibinfo {author} {\bibfnamefont {S.}~\bibnamefont {Vitale}}, \bibinfo {author} {\bibfnamefont {R.}~\bibnamefont {Lynch}}, \bibinfo {author} {\bibfnamefont {R.}~\bibnamefont {Sturani}},\ and\ \bibinfo {author} {\bibfnamefont {P.}~\bibnamefont {Graff}},\ }\bibfield  {title} {\bibinfo {title} {{Use of gravitational waves to probe the formation channels of compact binaries}},\ }\href {https://doi.org/10.1088/1361-6382/aa552e} {\bibfield  {journal} {\bibinfo  {journal} {Class. Quant. Grav.}\ }\textbf {\bibinfo {volume} {34}},\ \bibinfo {pages} {03LT01} (\bibinfo {year} {2017})},\ \Eprint {https://arxiv.org/abs/1503.04307} {arXiv:1503.04307 [gr-qc]} \BibitemShut {NoStop}%
\bibitem [{\citenamefont {Stevenson}\ \emph {et~al.}(2017)\citenamefont {Stevenson}, \citenamefont {Berry},\ and\ \citenamefont {Mandel}}]{Stevenson:2017dlk}%
  \BibitemOpen
  \bibfield  {author} {\bibinfo {author} {\bibfnamefont {S.}~\bibnamefont {Stevenson}}, \bibinfo {author} {\bibfnamefont {C.~P.~L.}\ \bibnamefont {Berry}},\ and\ \bibinfo {author} {\bibfnamefont {I.}~\bibnamefont {Mandel}},\ }\bibfield  {title} {\bibinfo {title} {{Hierarchical analysis of gravitational-wave measurements of binary black hole spin{\textendash}orbit misalignments}},\ }\href {https://doi.org/10.1093/mnras/stx1764} {\bibfield  {journal} {\bibinfo  {journal} {Mon. Not. Roy. Astron. Soc.}\ }\textbf {\bibinfo {volume} {471}},\ \bibinfo {pages} {2801} (\bibinfo {year} {2017})},\ \Eprint {https://arxiv.org/abs/1703.06873} {arXiv:1703.06873 [astro-ph.HE]} \BibitemShut {NoStop}%
\bibitem [{\citenamefont {Farr}\ \emph {et~al.}(2018)\citenamefont {Farr}, \citenamefont {Holz},\ and\ \citenamefont {Farr}}]{Farr:2017gtv}%
  \BibitemOpen
  \bibfield  {author} {\bibinfo {author} {\bibfnamefont {B.}~\bibnamefont {Farr}}, \bibinfo {author} {\bibfnamefont {D.~E.}\ \bibnamefont {Holz}},\ and\ \bibinfo {author} {\bibfnamefont {W.~M.}\ \bibnamefont {Farr}},\ }\bibfield  {title} {\bibinfo {title} {{Using Spin to Understand the Formation of LIGO and Virgo{\textquoteright}s Black Holes}},\ }\href {https://doi.org/10.3847/2041-8213/aaaa64} {\bibfield  {journal} {\bibinfo  {journal} {Astrophys. J. Lett.}\ }\textbf {\bibinfo {volume} {854}},\ \bibinfo {pages} {L9} (\bibinfo {year} {2018})},\ \Eprint {https://arxiv.org/abs/1709.07896} {arXiv:1709.07896 [astro-ph.HE]} \BibitemShut {NoStop}%
\bibitem [{\citenamefont {Mandel}\ and\ \citenamefont {Broekgaarden}(2022)}]{Mandel:2021smh}%
  \BibitemOpen
  \bibfield  {author} {\bibinfo {author} {\bibfnamefont {I.}~\bibnamefont {Mandel}}\ and\ \bibinfo {author} {\bibfnamefont {F.~S.}\ \bibnamefont {Broekgaarden}},\ }\bibfield  {title} {\bibinfo {title} {{Rates of compact object coalescences}},\ }\href {https://doi.org/10.1007/s41114-021-00034-3} {\bibfield  {journal} {\bibinfo  {journal} {Living Rev. Rel.}\ }\textbf {\bibinfo {volume} {25}},\ \bibinfo {pages} {1} (\bibinfo {year} {2022})},\ \Eprint {https://arxiv.org/abs/2107.14239} {arXiv:2107.14239 [astro-ph.HE]} \BibitemShut {NoStop}%
\bibitem [{\citenamefont {Zevin}\ \emph {et~al.}(2021)\citenamefont {Zevin}, \citenamefont {Bavera}, \citenamefont {Berry}, \citenamefont {Kalogera}, \citenamefont {Fragos}, \citenamefont {Marchant}, \citenamefont {Rodriguez}, \citenamefont {Antonini}, \citenamefont {Holz},\ and\ \citenamefont {Pankow}}]{Zevin:2020gbd}%
  \BibitemOpen
  \bibfield  {author} {\bibinfo {author} {\bibfnamefont {M.}~\bibnamefont {Zevin}}, \bibinfo {author} {\bibfnamefont {S.~S.}\ \bibnamefont {Bavera}}, \bibinfo {author} {\bibfnamefont {C.~P.~L.}\ \bibnamefont {Berry}}, \bibinfo {author} {\bibfnamefont {V.}~\bibnamefont {Kalogera}}, \bibinfo {author} {\bibfnamefont {T.}~\bibnamefont {Fragos}}, \bibinfo {author} {\bibfnamefont {P.}~\bibnamefont {Marchant}}, \bibinfo {author} {\bibfnamefont {C.~L.}\ \bibnamefont {Rodriguez}}, \bibinfo {author} {\bibfnamefont {F.}~\bibnamefont {Antonini}}, \bibinfo {author} {\bibfnamefont {D.~E.}\ \bibnamefont {Holz}},\ and\ \bibinfo {author} {\bibfnamefont {C.}~\bibnamefont {Pankow}},\ }\bibfield  {title} {\bibinfo {title} {{One Channel to Rule Them All? Constraining the Origins of Binary Black Holes Using Multiple Formation Pathways}},\ }\href {https://doi.org/10.3847/1538-4357/abe40e} {\bibfield  {journal} {\bibinfo  {journal} {Astrophys. J.}\ }\textbf {\bibinfo {volume} {910}},\ \bibinfo {pages} {152} (\bibinfo {year}
  {2021})},\ \Eprint {https://arxiv.org/abs/2011.10057} {arXiv:2011.10057 [astro-ph.HE]} \BibitemShut {NoStop}%
\bibitem [{\citenamefont {Afroz}\ and\ \citenamefont {Mukherjee}(2025{\natexlab{a}})}]{Afroz:2024fzp}%
  \BibitemOpen
  \bibfield  {author} {\bibinfo {author} {\bibfnamefont {S.}~\bibnamefont {Afroz}}\ and\ \bibinfo {author} {\bibfnamefont {S.}~\bibnamefont {Mukherjee}},\ }\bibfield  {title} {\bibinfo {title} {{Phase space of binary black holes from gravitational wave observations to unveil its formation history}},\ }\href {https://doi.org/10.1103/7zc2-g9vq} {\bibfield  {journal} {\bibinfo  {journal} {Phys. Rev. D}\ }\textbf {\bibinfo {volume} {112}},\ \bibinfo {pages} {023531} (\bibinfo {year} {2025}{\natexlab{a}})},\ \Eprint {https://arxiv.org/abs/2411.07304} {arXiv:2411.07304 [astro-ph.HE]} \BibitemShut {NoStop}%
\bibitem [{\citenamefont {Banagiri}\ \emph {et~al.}(2025)\citenamefont {Banagiri}, \citenamefont {Callister}, \citenamefont {Doctor},\ and\ \citenamefont {Kalogera}}]{Banagiri:2025dxo}%
  \BibitemOpen
  \bibfield  {author} {\bibinfo {author} {\bibfnamefont {S.}~\bibnamefont {Banagiri}}, \bibinfo {author} {\bibfnamefont {T.~A.}\ \bibnamefont {Callister}}, \bibinfo {author} {\bibfnamefont {Z.}~\bibnamefont {Doctor}},\ and\ \bibinfo {author} {\bibfnamefont {V.}~\bibnamefont {Kalogera}},\ }\bibfield  {title} {\bibinfo {title} {{Structure and Skewness of the Effective Inspiral Spin Distribution of Binary Black Hole Mergers}},\ }\href@noop {} {\  (\bibinfo {year} {2025})},\ \Eprint {https://arxiv.org/abs/2501.06712} {arXiv:2501.06712 [astro-ph.HE]} \BibitemShut {NoStop}%
\bibitem [{\citenamefont {Colloms}\ \emph {et~al.}(2025)\citenamefont {Colloms}, \citenamefont {Berry}, \citenamefont {Veitch},\ and\ \citenamefont {Zevin}}]{Colloms:2025hib}%
  \BibitemOpen
  \bibfield  {author} {\bibinfo {author} {\bibfnamefont {S.}~\bibnamefont {Colloms}}, \bibinfo {author} {\bibfnamefont {C.~P.~L.}\ \bibnamefont {Berry}}, \bibinfo {author} {\bibfnamefont {J.}~\bibnamefont {Veitch}},\ and\ \bibinfo {author} {\bibfnamefont {M.}~\bibnamefont {Zevin}},\ }\bibfield  {title} {\bibinfo {title} {{Exploring the evolution of gravitational-wave emitters with efficient emulation: Constraining the origins of binary black holes using normalising flows}},\ }\href {https://doi.org/10.3847/1538-4357/ade546} {\bibfield  {journal} {\bibinfo  {journal} {Astrophys. J.}\ }\textbf {\bibinfo {volume} {988}},\ \bibinfo {pages} {189} (\bibinfo {year} {2025})},\ \Eprint {https://arxiv.org/abs/2503.03819} {arXiv:2503.03819 [astro-ph.HE]} \BibitemShut {NoStop}%
\bibitem [{\citenamefont {Talbot}\ and\ \citenamefont {Thrane}(2017{\natexlab{a}})}]{Talbot:2017yur}%
  \BibitemOpen
  \bibfield  {author} {\bibinfo {author} {\bibfnamefont {C.}~\bibnamefont {Talbot}}\ and\ \bibinfo {author} {\bibfnamefont {E.}~\bibnamefont {Thrane}},\ }\bibfield  {title} {\bibinfo {title} {{Determining the population properties of spinning black holes}},\ }\href {https://doi.org/10.1103/PhysRevD.96.023012} {\bibfield  {journal} {\bibinfo  {journal} {Phys. Rev. D}\ }\textbf {\bibinfo {volume} {96}},\ \bibinfo {pages} {023012} (\bibinfo {year} {2017}{\natexlab{a}})},\ \Eprint {https://arxiv.org/abs/1704.08370} {arXiv:1704.08370 [astro-ph.HE]} \BibitemShut {NoStop}%
\bibitem [{\citenamefont {Baibhav}\ \emph {et~al.}(2023)\citenamefont {Baibhav}, \citenamefont {Doctor},\ and\ \citenamefont {Kalogera}}]{Baibhav:2022qxm}%
  \BibitemOpen
  \bibfield  {author} {\bibinfo {author} {\bibfnamefont {V.}~\bibnamefont {Baibhav}}, \bibinfo {author} {\bibfnamefont {Z.}~\bibnamefont {Doctor}},\ and\ \bibinfo {author} {\bibfnamefont {V.}~\bibnamefont {Kalogera}},\ }\bibfield  {title} {\bibinfo {title} {{Dropping Anchor: Understanding the Populations of Binary Black Holes with Random and Aligned-spin Orientations}},\ }\href {https://doi.org/10.3847/1538-4357/acbf4c} {\bibfield  {journal} {\bibinfo  {journal} {Astrophys. J.}\ }\textbf {\bibinfo {volume} {946}},\ \bibinfo {pages} {50} (\bibinfo {year} {2023})},\ \Eprint {https://arxiv.org/abs/2212.12113} {arXiv:2212.12113 [astro-ph.HE]} \BibitemShut {NoStop}%
\bibitem [{\citenamefont {Vitale}\ \emph {et~al.}(2022)\citenamefont {Vitale}, \citenamefont {Biscoveanu},\ and\ \citenamefont {Talbot}}]{Vitale:2022dpa}%
  \BibitemOpen
  \bibfield  {author} {\bibinfo {author} {\bibfnamefont {S.}~\bibnamefont {Vitale}}, \bibinfo {author} {\bibfnamefont {S.}~\bibnamefont {Biscoveanu}},\ and\ \bibinfo {author} {\bibfnamefont {C.}~\bibnamefont {Talbot}},\ }\bibfield  {title} {\bibinfo {title} {{Spin it as you like: The (lack of a) measurement of the spin tilt distribution with LIGO-Virgo-KAGRA binary black holes}},\ }\href {https://doi.org/10.1051/0004-6361/202245084} {\bibfield  {journal} {\bibinfo  {journal} {Astron. Astrophys.}\ }\textbf {\bibinfo {volume} {668}},\ \bibinfo {pages} {L2} (\bibinfo {year} {2022})},\ \Eprint {https://arxiv.org/abs/2209.06978} {arXiv:2209.06978 [astro-ph.HE]} \BibitemShut {NoStop}%
\bibitem [{\citenamefont {Pierra}\ \emph {et~al.}(2024)\citenamefont {Pierra}, \citenamefont {Mastrogiovanni},\ and\ \citenamefont {Perri\`es}}]{Pierra:2024fbl}%
  \BibitemOpen
  \bibfield  {author} {\bibinfo {author} {\bibfnamefont {G.}~\bibnamefont {Pierra}}, \bibinfo {author} {\bibfnamefont {S.}~\bibnamefont {Mastrogiovanni}},\ and\ \bibinfo {author} {\bibfnamefont {S.}~\bibnamefont {Perri\`es}},\ }\bibfield  {title} {\bibinfo {title} {{The spin magnitude of stellar-mass black holes evolves with the mass}},\ }\href {https://doi.org/10.1051/0004-6361/202452545} {\bibfield  {journal} {\bibinfo  {journal} {Astron. Astrophys.}\ }\textbf {\bibinfo {volume} {692}},\ \bibinfo {pages} {A80} (\bibinfo {year} {2024})},\ \Eprint {https://arxiv.org/abs/2406.01679} {arXiv:2406.01679 [gr-qc]} \BibitemShut {NoStop}%
\bibitem [{\citenamefont {Li}\ \emph {et~al.}(2024{\natexlab{a}})\citenamefont {Li}, \citenamefont {Wang}, \citenamefont {Tang},\ and\ \citenamefont {Fan}}]{Li:2023yyt}%
  \BibitemOpen
  \bibfield  {author} {\bibinfo {author} {\bibfnamefont {Y.-J.}\ \bibnamefont {Li}}, \bibinfo {author} {\bibfnamefont {Y.-Z.}\ \bibnamefont {Wang}}, \bibinfo {author} {\bibfnamefont {S.-P.}\ \bibnamefont {Tang}},\ and\ \bibinfo {author} {\bibfnamefont {Y.-Z.}\ \bibnamefont {Fan}},\ }\bibfield  {title} {\bibinfo {title} {{Resolving the Stellar-Collapse and Hierarchical-Merger Origins of the Coalescing Black Holes}},\ }\href {https://doi.org/10.1103/PhysRevLett.133.051401} {\bibfield  {journal} {\bibinfo  {journal} {Phys. Rev. Lett.}\ }\textbf {\bibinfo {volume} {133}},\ \bibinfo {pages} {051401} (\bibinfo {year} {2024}{\natexlab{a}})},\ \Eprint {https://arxiv.org/abs/2303.02973} {arXiv:2303.02973 [astro-ph.HE]} \BibitemShut {NoStop}%
\bibitem [{\citenamefont {Kalogera}(2000)}]{Kalogera:1999tq}%
  \BibitemOpen
  \bibfield  {author} {\bibinfo {author} {\bibfnamefont {V.}~\bibnamefont {Kalogera}},\ }\bibfield  {title} {\bibinfo {title} {{Spin orbit misalignment in close binaries with two compact objects}},\ }\href {https://doi.org/10.1086/309400} {\bibfield  {journal} {\bibinfo  {journal} {Astrophys. J.}\ }\textbf {\bibinfo {volume} {541}},\ \bibinfo {pages} {319} (\bibinfo {year} {2000})},\ \Eprint {https://arxiv.org/abs/astro-ph/9911417} {arXiv:astro-ph/9911417} \BibitemShut {NoStop}%
\bibitem [{\citenamefont {Rodriguez}\ \emph {et~al.}(2016)\citenamefont {Rodriguez}, \citenamefont {Zevin}, \citenamefont {Pankow}, \citenamefont {Kalogera},\ and\ \citenamefont {Rasio}}]{Rodriguez:2016vmx}%
  \BibitemOpen
  \bibfield  {author} {\bibinfo {author} {\bibfnamefont {C.~L.}\ \bibnamefont {Rodriguez}}, \bibinfo {author} {\bibfnamefont {M.}~\bibnamefont {Zevin}}, \bibinfo {author} {\bibfnamefont {C.}~\bibnamefont {Pankow}}, \bibinfo {author} {\bibfnamefont {V.}~\bibnamefont {Kalogera}},\ and\ \bibinfo {author} {\bibfnamefont {F.~A.}\ \bibnamefont {Rasio}},\ }\bibfield  {title} {\bibinfo {title} {{Illuminating Black Hole Binary Formation Channels with Spins in Advanced LIGO}},\ }\href {https://doi.org/10.3847/2041-8205/832/1/L2} {\bibfield  {journal} {\bibinfo  {journal} {Astrophys. J. Lett.}\ }\textbf {\bibinfo {volume} {832}},\ \bibinfo {pages} {L2} (\bibinfo {year} {2016})},\ \Eprint {https://arxiv.org/abs/1609.05916} {arXiv:1609.05916 [astro-ph.HE]} \BibitemShut {NoStop}%
\bibitem [{\citenamefont {Farr}\ \emph {et~al.}(2017)\citenamefont {Farr}, \citenamefont {Stevenson}, \citenamefont {Coleman~Miller}, \citenamefont {Mandel}, \citenamefont {Farr},\ and\ \citenamefont {Vecchio}}]{Farr:2017uvj}%
  \BibitemOpen
  \bibfield  {author} {\bibinfo {author} {\bibfnamefont {W.~M.}\ \bibnamefont {Farr}}, \bibinfo {author} {\bibfnamefont {S.}~\bibnamefont {Stevenson}}, \bibinfo {author} {\bibfnamefont {M.}~\bibnamefont {Coleman~Miller}}, \bibinfo {author} {\bibfnamefont {I.}~\bibnamefont {Mandel}}, \bibinfo {author} {\bibfnamefont {B.}~\bibnamefont {Farr}},\ and\ \bibinfo {author} {\bibfnamefont {A.}~\bibnamefont {Vecchio}},\ }\bibfield  {title} {\bibinfo {title} {{Distinguishing Spin-Aligned and Isotropic Black Hole Populations With Gravitational Waves}},\ }\href {https://doi.org/10.1038/nature23453} {\bibfield  {journal} {\bibinfo  {journal} {Nature}\ }\textbf {\bibinfo {volume} {548}},\ \bibinfo {pages} {426} (\bibinfo {year} {2017})},\ \Eprint {https://arxiv.org/abs/1706.01385} {arXiv:1706.01385 [astro-ph.HE]} \BibitemShut {NoStop}%
\bibitem [{\citenamefont {Gerosa}\ \emph {et~al.}(2018)\citenamefont {Gerosa}, \citenamefont {Berti}, \citenamefont {O'Shaughnessy}, \citenamefont {Belczynski}, \citenamefont {Kesden}, \citenamefont {Wysocki},\ and\ \citenamefont {Gladysz}}]{Gerosa:2018wbw}%
  \BibitemOpen
  \bibfield  {author} {\bibinfo {author} {\bibfnamefont {D.}~\bibnamefont {Gerosa}}, \bibinfo {author} {\bibfnamefont {E.}~\bibnamefont {Berti}}, \bibinfo {author} {\bibfnamefont {R.}~\bibnamefont {O'Shaughnessy}}, \bibinfo {author} {\bibfnamefont {K.}~\bibnamefont {Belczynski}}, \bibinfo {author} {\bibfnamefont {M.}~\bibnamefont {Kesden}}, \bibinfo {author} {\bibfnamefont {D.}~\bibnamefont {Wysocki}},\ and\ \bibinfo {author} {\bibfnamefont {W.}~\bibnamefont {Gladysz}},\ }\bibfield  {title} {\bibinfo {title} {{Spin orientations of merging black holes formed from the evolution of stellar binaries}},\ }\href {https://doi.org/10.1103/PhysRevD.98.084036} {\bibfield  {journal} {\bibinfo  {journal} {Phys. Rev. D}\ }\textbf {\bibinfo {volume} {98}},\ \bibinfo {pages} {084036} (\bibinfo {year} {2018})},\ \Eprint {https://arxiv.org/abs/1808.02491} {arXiv:1808.02491 [astro-ph.HE]} \BibitemShut {NoStop}%
\bibitem [{\citenamefont {Wang}\ \emph {et~al.}(2021)\citenamefont {Wang}, \citenamefont {McKernan}, \citenamefont {Ford}, \citenamefont {Perna}, \citenamefont {Leigh},\ and\ \citenamefont {Mac~Low}}]{Wang:2021yjf}%
  \BibitemOpen
  \bibfield  {author} {\bibinfo {author} {\bibfnamefont {Y.-H.}\ \bibnamefont {Wang}}, \bibinfo {author} {\bibfnamefont {B.}~\bibnamefont {McKernan}}, \bibinfo {author} {\bibfnamefont {S.}~\bibnamefont {Ford}}, \bibinfo {author} {\bibfnamefont {R.}~\bibnamefont {Perna}}, \bibinfo {author} {\bibfnamefont {N.~W.~C.}\ \bibnamefont {Leigh}},\ and\ \bibinfo {author} {\bibfnamefont {M.-M.}\ \bibnamefont {Mac~Low}},\ }\bibfield  {title} {\bibinfo {title} {{Symmetry Breaking in Dynamical Encounters in the Disks of Active Galactic Nuclei}},\ }\href {https://doi.org/10.3847/2041-8213/ac400a} {\bibfield  {journal} {\bibinfo  {journal} {Astrophys. J. Lett.}\ }\textbf {\bibinfo {volume} {923}},\ \bibinfo {pages} {L23} (\bibinfo {year} {2021})},\ \Eprint {https://arxiv.org/abs/2110.03698} {arXiv:2110.03698 [astro-ph.HE]} \BibitemShut {NoStop}%
\bibitem [{\citenamefont {Banerjee}\ \emph {et~al.}(2023)\citenamefont {Banerjee}, \citenamefont {Olejak},\ and\ \citenamefont {Belczynski}}]{Banerjee:2023ycw}%
  \BibitemOpen
  \bibfield  {author} {\bibinfo {author} {\bibfnamefont {S.}~\bibnamefont {Banerjee}}, \bibinfo {author} {\bibfnamefont {A.}~\bibnamefont {Olejak}},\ and\ \bibinfo {author} {\bibfnamefont {K.}~\bibnamefont {Belczynski}},\ }\bibfield  {title} {\bibinfo {title} {{Symmetry Breaking in Merging Binary Black Holes from Young Massive Clusters and Isolated Binaries}},\ }\href {https://doi.org/10.3847/1538-4357/acdd59} {\bibfield  {journal} {\bibinfo  {journal} {Astrophys. J.}\ }\textbf {\bibinfo {volume} {953}},\ \bibinfo {pages} {80} (\bibinfo {year} {2023})},\ \Eprint {https://arxiv.org/abs/2302.10851} {arXiv:2302.10851 [astro-ph.HE]} \BibitemShut {NoStop}%
\bibitem [{\citenamefont {Baibhav}\ and\ \citenamefont {Kalogera}(2024)}]{Baibhav:2024rkn}%
  \BibitemOpen
  \bibfield  {author} {\bibinfo {author} {\bibfnamefont {V.}~\bibnamefont {Baibhav}}\ and\ \bibinfo {author} {\bibfnamefont {V.}~\bibnamefont {Kalogera}},\ }\bibfield  {title} {\bibinfo {title} {{Revising the Spin and Kick Connection in Isolated Binary Black Holes}},\ }\href@noop {} {\  (\bibinfo {year} {2024})},\ \Eprint {https://arxiv.org/abs/2412.03461} {arXiv:2412.03461 [astro-ph.HE]} \BibitemShut {NoStop}%
\bibitem [{\citenamefont {K\i{}ro\u{g}lu}\ \emph {et~al.}(2025)\citenamefont {K\i{}ro\u{g}lu}, \citenamefont {Lombardi}, \citenamefont {Kremer}, \citenamefont {Vanderzyden},\ and\ \citenamefont {Rasio}}]{Kiroglu:2025bbp}%
  \BibitemOpen
  \bibfield  {author} {\bibinfo {author} {\bibfnamefont {F.}~\bibnamefont {K\i{}ro\u{g}lu}}, \bibinfo {author} {\bibfnamefont {J.~C.}\ \bibnamefont {Lombardi}}, \bibinfo {author} {\bibfnamefont {K.}~\bibnamefont {Kremer}}, \bibinfo {author} {\bibfnamefont {H.~D.}\ \bibnamefont {Vanderzyden}},\ and\ \bibinfo {author} {\bibfnamefont {F.~A.}\ \bibnamefont {Rasio}},\ }\bibfield  {title} {\bibinfo {title} {{Spin-Orbit Alignment in Merging Binary Black Holes Following Collisions with Massive Stars}},\ }\href@noop {} {\  (\bibinfo {year} {2025})},\ \Eprint {https://arxiv.org/abs/2501.09068} {arXiv:2501.09068 [astro-ph.HE]} \BibitemShut {NoStop}%
\bibitem [{\citenamefont {Callister}\ \emph {et~al.}(2021)\citenamefont {Callister}, \citenamefont {Haster}, \citenamefont {Ng}, \citenamefont {Vitale},\ and\ \citenamefont {Farr}}]{Callister:2021fpo}%
  \BibitemOpen
  \bibfield  {author} {\bibinfo {author} {\bibfnamefont {T.~A.}\ \bibnamefont {Callister}}, \bibinfo {author} {\bibfnamefont {C.-J.}\ \bibnamefont {Haster}}, \bibinfo {author} {\bibfnamefont {K.~K.~Y.}\ \bibnamefont {Ng}}, \bibinfo {author} {\bibfnamefont {S.}~\bibnamefont {Vitale}},\ and\ \bibinfo {author} {\bibfnamefont {W.~M.}\ \bibnamefont {Farr}},\ }\bibfield  {title} {\bibinfo {title} {{Who Ordered That? Unequal-mass Binary Black Hole Mergers Have Larger Effective Spins}},\ }\href {https://doi.org/10.3847/2041-8213/ac2ccc} {\bibfield  {journal} {\bibinfo  {journal} {Astrophys. J. Lett.}\ }\textbf {\bibinfo {volume} {922}},\ \bibinfo {pages} {L5} (\bibinfo {year} {2021})},\ \Eprint {https://arxiv.org/abs/2106.00521} {arXiv:2106.00521 [astro-ph.HE]} \BibitemShut {NoStop}%
\bibitem [{\citenamefont {Biscoveanu}\ \emph {et~al.}(2022)\citenamefont {Biscoveanu}, \citenamefont {Callister}, \citenamefont {Haster}, \citenamefont {Ng}, \citenamefont {Vitale},\ and\ \citenamefont {Farr}}]{Biscoveanu:2022qac}%
  \BibitemOpen
  \bibfield  {author} {\bibinfo {author} {\bibfnamefont {S.}~\bibnamefont {Biscoveanu}}, \bibinfo {author} {\bibfnamefont {T.~A.}\ \bibnamefont {Callister}}, \bibinfo {author} {\bibfnamefont {C.-J.}\ \bibnamefont {Haster}}, \bibinfo {author} {\bibfnamefont {K.~K.~Y.}\ \bibnamefont {Ng}}, \bibinfo {author} {\bibfnamefont {S.}~\bibnamefont {Vitale}},\ and\ \bibinfo {author} {\bibfnamefont {W.~M.}\ \bibnamefont {Farr}},\ }\bibfield  {title} {\bibinfo {title} {{The Binary Black Hole Spin Distribution Likely Broadens with Redshift}},\ }\href {https://doi.org/10.3847/2041-8213/ac71a8} {\bibfield  {journal} {\bibinfo  {journal} {Astrophys. J. Lett.}\ }\textbf {\bibinfo {volume} {932}},\ \bibinfo {pages} {L19} (\bibinfo {year} {2022})},\ \Eprint {https://arxiv.org/abs/2204.01578} {arXiv:2204.01578 [astro-ph.HE]} \BibitemShut {NoStop}%
\bibitem [{\citenamefont {Adamcewicz}\ and\ \citenamefont {Thrane}(2022)}]{Adamcewicz:2022hce}%
  \BibitemOpen
  \bibfield  {author} {\bibinfo {author} {\bibfnamefont {C.}~\bibnamefont {Adamcewicz}}\ and\ \bibinfo {author} {\bibfnamefont {E.}~\bibnamefont {Thrane}},\ }\bibfield  {title} {\bibinfo {title} {{Do unequal-mass binary black hole systems have larger \ensuremath{\chi}eff? Probing correlations with copulas in gravitational-wave astronomy}},\ }\href {https://doi.org/10.1093/mnras/stac2961} {\bibfield  {journal} {\bibinfo  {journal} {Mon. Not. Roy. Astron. Soc.}\ }\textbf {\bibinfo {volume} {517}},\ \bibinfo {pages} {3928} (\bibinfo {year} {2022})},\ \Eprint {https://arxiv.org/abs/2208.03405} {arXiv:2208.03405 [astro-ph.HE]} \BibitemShut {NoStop}%
\bibitem [{\citenamefont {Adamcewicz}\ \emph {et~al.}(2023)\citenamefont {Adamcewicz}, \citenamefont {Lasky},\ and\ \citenamefont {Thrane}}]{Adamcewicz:2023mov}%
  \BibitemOpen
  \bibfield  {author} {\bibinfo {author} {\bibfnamefont {C.}~\bibnamefont {Adamcewicz}}, \bibinfo {author} {\bibfnamefont {P.~D.}\ \bibnamefont {Lasky}},\ and\ \bibinfo {author} {\bibfnamefont {E.}~\bibnamefont {Thrane}},\ }\bibfield  {title} {\bibinfo {title} {{Evidence for a Correlation between Binary Black Hole Mass Ratio and Black Hole Spins}},\ }\href {https://doi.org/10.3847/1538-4357/acf763} {\bibfield  {journal} {\bibinfo  {journal} {Astrophys. J.}\ }\textbf {\bibinfo {volume} {958}},\ \bibinfo {pages} {13} (\bibinfo {year} {2023})},\ \Eprint {https://arxiv.org/abs/2307.15278} {arXiv:2307.15278 [astro-ph.HE]} \BibitemShut {NoStop}%
\bibitem [{\citenamefont {Li}\ \emph {et~al.}(2022)\citenamefont {Li}, \citenamefont {Wang}, \citenamefont {Tang}, \citenamefont {Yuan}, \citenamefont {Fan},\ and\ \citenamefont {Wei}}]{Li:2022jge}%
  \BibitemOpen
  \bibfield  {author} {\bibinfo {author} {\bibfnamefont {Y.-J.}\ \bibnamefont {Li}}, \bibinfo {author} {\bibfnamefont {Y.-Z.}\ \bibnamefont {Wang}}, \bibinfo {author} {\bibfnamefont {S.-P.}\ \bibnamefont {Tang}}, \bibinfo {author} {\bibfnamefont {Q.}~\bibnamefont {Yuan}}, \bibinfo {author} {\bibfnamefont {Y.-Z.}\ \bibnamefont {Fan}},\ and\ \bibinfo {author} {\bibfnamefont {D.-M.}\ \bibnamefont {Wei}},\ }\bibfield  {title} {\bibinfo {title} {{Divergence in Mass Ratio Distributions between Low-mass and High-mass Coalescing Binary Black Holes}},\ }\href {https://doi.org/10.3847/2041-8213/ac78dd} {\bibfield  {journal} {\bibinfo  {journal} {Astrophys. J. Lett.}\ }\textbf {\bibinfo {volume} {933}},\ \bibinfo {pages} {L14} (\bibinfo {year} {2022})},\ \Eprint {https://arxiv.org/abs/2201.01905} {arXiv:2201.01905 [astro-ph.HE]} \BibitemShut {NoStop}%
\bibitem [{\citenamefont {Godfrey}\ \emph {et~al.}(2023)\citenamefont {Godfrey}, \citenamefont {Edelman},\ and\ \citenamefont {Farr}}]{Godfrey:2023oxb}%
  \BibitemOpen
  \bibfield  {author} {\bibinfo {author} {\bibfnamefont {J.}~\bibnamefont {Godfrey}}, \bibinfo {author} {\bibfnamefont {B.}~\bibnamefont {Edelman}},\ and\ \bibinfo {author} {\bibfnamefont {B.}~\bibnamefont {Farr}},\ }\bibfield  {title} {\bibinfo {title} {{Cosmic Cousins: Identification of a Subpopulation of Binary Black Holes Consistent with Isolated Binary Evolution}},\ }\href@noop {} {\  (\bibinfo {year} {2023})},\ \Eprint {https://arxiv.org/abs/2304.01288} {arXiv:2304.01288 [astro-ph.HE]} \BibitemShut {NoStop}%
\bibitem [{\citenamefont {Hussain}\ \emph {et~al.}(2024)\citenamefont {Hussain}, \citenamefont {Isi},\ and\ \citenamefont {Zimmerman}}]{Hussain:2024qzl}%
  \BibitemOpen
  \bibfield  {author} {\bibinfo {author} {\bibfnamefont {A.}~\bibnamefont {Hussain}}, \bibinfo {author} {\bibfnamefont {M.}~\bibnamefont {Isi}},\ and\ \bibinfo {author} {\bibfnamefont {A.}~\bibnamefont {Zimmerman}},\ }\bibfield  {title} {\bibinfo {title} {{Hints of spin-magnitude correlations and a rapidly spinning subpopulation of binary black holes}},\ }\href@noop {} {\  (\bibinfo {year} {2024})},\ \Eprint {https://arxiv.org/abs/2411.02252} {arXiv:2411.02252 [astro-ph.HE]} \BibitemShut {NoStop}%
\bibitem [{\citenamefont {Ray}\ \emph {et~al.}(2024)\citenamefont {Ray}, \citenamefont {Maga{\~n}a~Hernandez}, \citenamefont {Breivik},\ and\ \citenamefont {Creighton}}]{Ray:2024hos}%
  \BibitemOpen
  \bibfield  {author} {\bibinfo {author} {\bibfnamefont {A.}~\bibnamefont {Ray}}, \bibinfo {author} {\bibfnamefont {I.}~\bibnamefont {Maga{\~n}a~Hernandez}}, \bibinfo {author} {\bibfnamefont {K.}~\bibnamefont {Breivik}},\ and\ \bibinfo {author} {\bibfnamefont {J.}~\bibnamefont {Creighton}},\ }\bibfield  {title} {\bibinfo {title} {{Searching for binary black hole sub-populations in gravitational wave data using binned Gaussian processes}},\ }\href@noop {} {\  (\bibinfo {year} {2024})},\ \Eprint {https://arxiv.org/abs/2404.03166} {arXiv:2404.03166 [astro-ph.HE]} \BibitemShut {NoStop}%
\bibitem [{\citenamefont {Antonini}\ \emph {et~al.}(2025{\natexlab{a}})\citenamefont {Antonini}, \citenamefont {Romero-Shaw},\ and\ \citenamefont {Callister}}]{Antonini:2024het}%
  \BibitemOpen
  \bibfield  {author} {\bibinfo {author} {\bibfnamefont {F.}~\bibnamefont {Antonini}}, \bibinfo {author} {\bibfnamefont {I.~M.}\ \bibnamefont {Romero-Shaw}},\ and\ \bibinfo {author} {\bibfnamefont {T.}~\bibnamefont {Callister}},\ }\bibfield  {title} {\bibinfo {title} {{Star Cluster Population of High Mass Black Hole Mergers in Gravitational Wave Data}},\ }\href {https://doi.org/10.1103/PhysRevLett.134.011401} {\bibfield  {journal} {\bibinfo  {journal} {Phys. Rev. Lett.}\ }\textbf {\bibinfo {volume} {134}},\ \bibinfo {pages} {011401} (\bibinfo {year} {2025}{\natexlab{a}})},\ \Eprint {https://arxiv.org/abs/2406.19044} {arXiv:2406.19044 [astro-ph.HE]} \BibitemShut {NoStop}%
\bibitem [{\citenamefont {Antonini}\ \emph {et~al.}(2025{\natexlab{b}})\citenamefont {Antonini}, \citenamefont {Callister}, \citenamefont {Dosopoulou}, \citenamefont {Romero-Shaw},\ and\ \citenamefont {Chattopadhyay}}]{Antonini:2025zzw}%
  \BibitemOpen
  \bibfield  {author} {\bibinfo {author} {\bibfnamefont {F.}~\bibnamefont {Antonini}}, \bibinfo {author} {\bibfnamefont {T.}~\bibnamefont {Callister}}, \bibinfo {author} {\bibfnamefont {F.}~\bibnamefont {Dosopoulou}}, \bibinfo {author} {\bibfnamefont {I.}~\bibnamefont {Romero-Shaw}},\ and\ \bibinfo {author} {\bibfnamefont {D.}~\bibnamefont {Chattopadhyay}},\ }\bibfield  {title} {\bibinfo {title} {{Inferring the pair-instability mass gap from gravitational wave data using flexible models}},\ }\href@noop {} {\  (\bibinfo {year} {2025}{\natexlab{b}})},\ \Eprint {https://arxiv.org/abs/2506.09154} {arXiv:2506.09154 [astro-ph.HE]} \BibitemShut {NoStop}%
\bibitem [{\citenamefont {Galaudage}\ and\ \citenamefont {Lamberts}(2025)}]{Galaudage:2024meo}%
  \BibitemOpen
  \bibfield  {author} {\bibinfo {author} {\bibfnamefont {S.}~\bibnamefont {Galaudage}}\ and\ \bibinfo {author} {\bibfnamefont {A.}~\bibnamefont {Lamberts}},\ }\bibfield  {title} {\bibinfo {title} {{Compactness peaks: An astrophysical interpretation of the mass distribution of merging binary black holes}},\ }\href {https://doi.org/10.1051/0004-6361/202451654} {\bibfield  {journal} {\bibinfo  {journal} {Astron. Astrophys.}\ }\textbf {\bibinfo {volume} {694}},\ \bibinfo {pages} {A186} (\bibinfo {year} {2025})},\ \Eprint {https://arxiv.org/abs/2407.17561} {arXiv:2407.17561 [astro-ph.HE]} \BibitemShut {NoStop}%
\bibitem [{\citenamefont {Tong}\ \emph {et~al.}(2025)\citenamefont {Tong} \emph {et~al.}}]{Tong2025}%
  \BibitemOpen
  \bibfield  {author} {\bibinfo {author} {\bibfnamefont {H.}~\bibnamefont {Tong}} \emph {et~al.},\ }\bibfield  {title} {\bibinfo {title} {{Evidence of the pair instability gap in the distribution of black hole masses}},\ }\href@noop {} {\  (\bibinfo {year} {2025})},\ \Eprint {https://arxiv.org/abs/2509.04151} {arXiv:2509.04151 [astro-ph.HE]} \BibitemShut {NoStop}%
\bibitem [{\citenamefont {Antonini}\ \emph {et~al.}(2025{\natexlab{c}})\citenamefont {Antonini}, \citenamefont {Romero-Shaw}, \citenamefont {Callister}, \citenamefont {Dosopoulou}, \citenamefont {Chattopadhyay}, \citenamefont {Gieles},\ and\ \citenamefont {Mapelli}}]{Antonini:2025ilj}%
  \BibitemOpen
  \bibfield  {author} {\bibinfo {author} {\bibfnamefont {F.}~\bibnamefont {Antonini}}, \bibinfo {author} {\bibfnamefont {I.}~\bibnamefont {Romero-Shaw}}, \bibinfo {author} {\bibfnamefont {T.}~\bibnamefont {Callister}}, \bibinfo {author} {\bibfnamefont {F.}~\bibnamefont {Dosopoulou}}, \bibinfo {author} {\bibfnamefont {D.}~\bibnamefont {Chattopadhyay}}, \bibinfo {author} {\bibfnamefont {M.}~\bibnamefont {Gieles}},\ and\ \bibinfo {author} {\bibfnamefont {M.}~\bibnamefont {Mapelli}},\ }\bibfield  {title} {\bibinfo {title} {{Gravitational waves reveal the pair-instability mass gap and constrain nuclear burning in massive stars}},\ }\href@noop {} {\  (\bibinfo {year} {2025}{\natexlab{c}})},\ \Eprint {https://arxiv.org/abs/2509.04637} {arXiv:2509.04637 [astro-ph.HE]} \BibitemShut {NoStop}%
\bibitem [{\citenamefont {Li}\ \emph {et~al.}(2024{\natexlab{b}})\citenamefont {Li}, \citenamefont {Tang}, \citenamefont {Gao}, \citenamefont {Wu},\ and\ \citenamefont {Wang}}]{Li:2024jzi}%
  \BibitemOpen
  \bibfield  {author} {\bibinfo {author} {\bibfnamefont {Y.-J.}\ \bibnamefont {Li}}, \bibinfo {author} {\bibfnamefont {S.-P.}\ \bibnamefont {Tang}}, \bibinfo {author} {\bibfnamefont {S.-J.}\ \bibnamefont {Gao}}, \bibinfo {author} {\bibfnamefont {D.-C.}\ \bibnamefont {Wu}},\ and\ \bibinfo {author} {\bibfnamefont {Y.-Z.}\ \bibnamefont {Wang}},\ }\bibfield  {title} {\bibinfo {title} {{Exploring Field-evolution and Dynamical-capture Coalescing Binary Black Holes in GWTC-3}},\ }\href {https://doi.org/10.3847/1538-4357/ad83b5} {\bibfield  {journal} {\bibinfo  {journal} {Astrophys. J.}\ }\textbf {\bibinfo {volume} {977}},\ \bibinfo {pages} {67} (\bibinfo {year} {2024}{\natexlab{b}})},\ \Eprint {https://arxiv.org/abs/2404.09668} {arXiv:2404.09668 [astro-ph.HE]} \BibitemShut {NoStop}%
\bibitem [{\citenamefont {Wang}\ \emph {et~al.}(2022)\citenamefont {Wang}, \citenamefont {Li}, \citenamefont {Vink}, \citenamefont {Fan}, \citenamefont {Tang}, \citenamefont {Qin},\ and\ \citenamefont {Wei}}]{Wang:2022gnx}%
  \BibitemOpen
  \bibfield  {author} {\bibinfo {author} {\bibfnamefont {Y.-Z.}\ \bibnamefont {Wang}}, \bibinfo {author} {\bibfnamefont {Y.-J.}\ \bibnamefont {Li}}, \bibinfo {author} {\bibfnamefont {J.~S.}\ \bibnamefont {Vink}}, \bibinfo {author} {\bibfnamefont {Y.-Z.}\ \bibnamefont {Fan}}, \bibinfo {author} {\bibfnamefont {S.-P.}\ \bibnamefont {Tang}}, \bibinfo {author} {\bibfnamefont {Y.}~\bibnamefont {Qin}},\ and\ \bibinfo {author} {\bibfnamefont {D.-M.}\ \bibnamefont {Wei}},\ }\bibfield  {title} {\bibinfo {title} {{Potential Subpopulations and Assembling Tendency of the Merging Black Holes}},\ }\href {https://doi.org/10.3847/2041-8213/aca89f} {\bibfield  {journal} {\bibinfo  {journal} {Astrophys. J. Lett.}\ }\textbf {\bibinfo {volume} {941}},\ \bibinfo {pages} {L39} (\bibinfo {year} {2022})},\ \Eprint {https://arxiv.org/abs/2208.11871} {arXiv:2208.11871 [astro-ph.HE]} \BibitemShut {NoStop}%
\bibitem [{\citenamefont {Talbot}\ and\ \citenamefont {Thrane}(2017{\natexlab{b}})}]{spin}%
  \BibitemOpen
  \bibfield  {author} {\bibinfo {author} {\bibfnamefont {C.}~\bibnamefont {Talbot}}\ and\ \bibinfo {author} {\bibfnamefont {E.}~\bibnamefont {Thrane}},\ }\bibfield  {title} {\bibinfo {title} {Determining the population properties of spinning black holes},\ }\href@noop {} {\bibfield  {journal} {\bibinfo  {journal} {Phys. Rev. D}\ }\textbf {\bibinfo {volume} {96}},\ \bibinfo {pages} {023012} (\bibinfo {year} {2017}{\natexlab{b}})}\BibitemShut {NoStop}%
\bibitem [{\citenamefont {Fishbach}\ \emph {et~al.}(2018)\citenamefont {Fishbach}, \citenamefont {Holz},\ and\ \citenamefont {Farr}}]{Fishbach:2018edt}%
  \BibitemOpen
  \bibfield  {author} {\bibinfo {author} {\bibfnamefont {M.}~\bibnamefont {Fishbach}}, \bibinfo {author} {\bibfnamefont {D.~E.}\ \bibnamefont {Holz}},\ and\ \bibinfo {author} {\bibfnamefont {W.~M.}\ \bibnamefont {Farr}},\ }\bibfield  {title} {\bibinfo {title} {{Does the Black Hole Merger Rate Evolve with Redshift?}},\ }\href {https://doi.org/10.3847/2041-8213/aad800} {\bibfield  {journal} {\bibinfo  {journal} {Astrophys. J. Lett.}\ }\textbf {\bibinfo {volume} {863}},\ \bibinfo {pages} {L41} (\bibinfo {year} {2018})},\ \Eprint {https://arxiv.org/abs/1805.10270} {arXiv:1805.10270 [astro-ph.HE]} \BibitemShut {NoStop}%
\bibitem [{\citenamefont {Mandel}\ \emph {et~al.}(2019)\citenamefont {Mandel}, \citenamefont {Farr},\ and\ \citenamefont {Gair}}]{Mandel:2018mve}%
  \BibitemOpen
  \bibfield  {author} {\bibinfo {author} {\bibfnamefont {I.}~\bibnamefont {Mandel}}, \bibinfo {author} {\bibfnamefont {W.~M.}\ \bibnamefont {Farr}},\ and\ \bibinfo {author} {\bibfnamefont {J.~R.}\ \bibnamefont {Gair}},\ }\bibfield  {title} {\bibinfo {title} {{Extracting distribution parameters from multiple uncertain observations with selection biases}},\ }\href {https://doi.org/10.1093/mnras/stz896} {\bibfield  {journal} {\bibinfo  {journal} {Mon. Not. Roy. Astron. Soc.}\ }\textbf {\bibinfo {volume} {486}},\ \bibinfo {pages} {1086} (\bibinfo {year} {2019})},\ \Eprint {https://arxiv.org/abs/1809.02063} {arXiv:1809.02063 [physics.data-an]} \BibitemShut {NoStop}%
\bibitem [{\citenamefont {Thrane}\ and\ \citenamefont {Talbot}(2019)}]{Thrane:2019}%
  \BibitemOpen
  \bibfield  {author} {\bibinfo {author} {\bibfnamefont {E.}~\bibnamefont {Thrane}}\ and\ \bibinfo {author} {\bibfnamefont {C.}~\bibnamefont {Talbot}},\ }\bibfield  {title} {\bibinfo {title} {An introduction to bayesian inference in gravitational-wave astronomy: Parameter estimation, model selection, and hierarchical models},\ }\bibfield  {journal} {\bibinfo  {journal} {Publications of the Astronomical Society of Australia}\ }\textbf {\bibinfo {volume} {36}},\ \href {https://doi.org/10.1017/pasa.2019.2} {10.1017/pasa.2019.2} (\bibinfo {year} {2019})\BibitemShut {NoStop}%
\bibitem [{\citenamefont {Vitale}\ \emph {et~al.}(2020)\citenamefont {Vitale}, \citenamefont {Gerosa}, \citenamefont {Farr},\ and\ \citenamefont {Taylor}}]{Vitale:2020aaz}%
  \BibitemOpen
  \bibfield  {author} {\bibinfo {author} {\bibfnamefont {S.}~\bibnamefont {Vitale}}, \bibinfo {author} {\bibfnamefont {D.}~\bibnamefont {Gerosa}}, \bibinfo {author} {\bibfnamefont {W.~M.}\ \bibnamefont {Farr}},\ and\ \bibinfo {author} {\bibfnamefont {S.~R.}\ \bibnamefont {Taylor}},\ }\bibfield  {title} {\bibinfo {title} {{Inferring the properties of a population of compact binaries in presence of selection effects}},\ }\href@noop {} {\  (\bibinfo {year} {2020})},\ \Eprint {https://arxiv.org/abs/2007.05579} {arXiv:2007.05579 [astro-ph.IM]} \BibitemShut {NoStop}%
\bibitem [{LIG(2025)}]{LIGOScientific:2025snk}%
  \BibitemOpen
  \bibfield  {title} {\bibinfo {title} {{Open Data from LIGO, Virgo, and KAGRA through the First Part of the Fourth Observing Run}},\ }\href@noop {} {\  (\bibinfo {year} {2025})},\ \Eprint {https://arxiv.org/abs/2508.18079} {arXiv:2508.18079 [gr-qc]} \BibitemShut {NoStop}%
\bibitem [{\citenamefont {Collaboration}\ \emph {et~al.}(2025{\natexlab{a}})\citenamefont {Collaboration}, \citenamefont {Collaboration},\ and\ \citenamefont {Collaboration}}]{LIGOScientific:2025gwtc4zenodo}%
  \BibitemOpen
  \bibfield  {author} {\bibinfo {author} {\bibfnamefont {L.~S.}\ \bibnamefont {Collaboration}}, \bibinfo {author} {\bibfnamefont {V.}~\bibnamefont {Collaboration}},\ and\ \bibinfo {author} {\bibfnamefont {K.}~\bibnamefont {Collaboration}},\ }\bibfield  {title} {\bibinfo {title} {Gwtc-4.0: Parameter estimation data release},\ }\href {https://doi.org/10.5281/zenodo.16053484} {10.5281/zenodo.16053484} (\bibinfo {year} {2025}{\natexlab{a}})\BibitemShut {NoStop}%
\bibitem [{\citenamefont {Collaboration}\ \emph {et~al.}(2021)\citenamefont {Collaboration}, \citenamefont {Collaboration},\ and\ \citenamefont {Collaboration}}]{gwtc3_pe:zenodo}%
  \BibitemOpen
  \bibfield  {author} {\bibinfo {author} {\bibfnamefont {L.~S.}\ \bibnamefont {Collaboration}}, \bibinfo {author} {\bibfnamefont {V.}~\bibnamefont {Collaboration}},\ and\ \bibinfo {author} {\bibfnamefont {K.}~\bibnamefont {Collaboration}},\ }\bibfield  {title} {\bibinfo {title} {Gwtc-3: Compact binary coalescences observed by ligo and virgo during the second part of the third observing run — parameter estimation data release},\ }\href {https://doi.org/10.5281/zenodo.5546663} {10.5281/zenodo.5546663} (\bibinfo {year} {2021})\BibitemShut {NoStop}%
\bibitem [{\citenamefont {Collaboration}\ and\ \citenamefont {Collaboration}(2022)}]{gwtc2_pe:zenodo}%
  \BibitemOpen
  \bibfield  {author} {\bibinfo {author} {\bibfnamefont {L.~S.}\ \bibnamefont {Collaboration}}\ and\ \bibinfo {author} {\bibfnamefont {V.}~\bibnamefont {Collaboration}},\ }\bibfield  {title} {\bibinfo {title} {Gwtc-2.1: Deep extended catalog of compact binary coalescences observed by ligo and virgo during the first half of the third observing run - parameter estimation data release},\ }\href {https://doi.org/10.5281/zenodo.6513631} {10.5281/zenodo.6513631} (\bibinfo {year} {2022})\BibitemShut {NoStop}%
\bibitem [{\citenamefont {Abbott}\ \emph {et~al.}(2020{\natexlab{a}})\citenamefont {Abbott} \emph {et~al.}}]{LIGOScientific:2020zkf}%
  \BibitemOpen
  \bibfield  {author} {\bibinfo {author} {\bibfnamefont {R.}~\bibnamefont {Abbott}} \emph {et~al.} (\bibinfo {collaboration} {LIGO Scientific, Virgo}),\ }\bibfield  {title} {\bibinfo {title} {{GW190814: Gravitational Waves from the Coalescence of a 23 Solar Mass Black Hole with a 2.6 Solar Mass Compact Object}},\ }\href {https://doi.org/10.3847/2041-8213/ab960f} {\bibfield  {journal} {\bibinfo  {journal} {Astrophys. J. Lett.}\ }\textbf {\bibinfo {volume} {896}},\ \bibinfo {pages} {L44} (\bibinfo {year} {2020}{\natexlab{a}})},\ \Eprint {https://arxiv.org/abs/2006.12611} {arXiv:2006.12611 [astro-ph.HE]} \BibitemShut {NoStop}%
\bibitem [{\citenamefont {Abac}\ \emph {et~al.}(2025{\natexlab{d}})\citenamefont {Abac} \emph {et~al.}}]{LIGOScientific:2025rsn}%
  \BibitemOpen
  \bibfield  {author} {\bibinfo {author} {\bibfnamefont {A.~G.}\ \bibnamefont {Abac}} \emph {et~al.} (\bibinfo {collaboration} {LIGO Scientific, VIRGO, KAGRA}),\ }\bibfield  {title} {\bibinfo {title} {{GW231123: a Binary Black Hole Merger with Total Mass 190-265 $M_{\odot}$}},\ }\href@noop {} {\  (\bibinfo {year} {2025}{\natexlab{d}})},\ \Eprint {https://arxiv.org/abs/2507.08219} {arXiv:2507.08219 [astro-ph.HE]} \BibitemShut {NoStop}%
\bibitem [{\citenamefont {Adamcewicz}\ \emph {et~al.}(2025)\citenamefont {Adamcewicz}, \citenamefont {Guttman}, \citenamefont {Lasky},\ and\ \citenamefont {Thrane}}]{Adamcewicz:2025phm}%
  \BibitemOpen
  \bibfield  {author} {\bibinfo {author} {\bibfnamefont {C.}~\bibnamefont {Adamcewicz}}, \bibinfo {author} {\bibfnamefont {N.}~\bibnamefont {Guttman}}, \bibinfo {author} {\bibfnamefont {P.~D.}\ \bibnamefont {Lasky}},\ and\ \bibinfo {author} {\bibfnamefont {E.}~\bibnamefont {Thrane}},\ }\bibfield  {title} {\bibinfo {title} {{Do Both Black Holes Spin in Merging Binaries? Evidence from GWTC-4 and Astrophysical Implications}},\ }\href {https://doi.org/10.3847/1538-4357/ae1370} {\bibfield  {journal} {\bibinfo  {journal} {Astrophys. J.}\ }\textbf {\bibinfo {volume} {994}},\ \bibinfo {pages} {261} (\bibinfo {year} {2025})},\ \Eprint {https://arxiv.org/abs/2509.04706} {arXiv:2509.04706 [astro-ph.HE]} \BibitemShut {NoStop}%
\bibitem [{\citenamefont {Plunkett}\ \emph {et~al.}(2026)\citenamefont {Plunkett}, \citenamefont {Callister}, \citenamefont {Zevin},\ and\ \citenamefont {Vitale}}]{Plunkett:2026pxt}%
  \BibitemOpen
  \bibfield  {author} {\bibinfo {author} {\bibfnamefont {C.}~\bibnamefont {Plunkett}}, \bibinfo {author} {\bibfnamefont {T.}~\bibnamefont {Callister}}, \bibinfo {author} {\bibfnamefont {M.}~\bibnamefont {Zevin}},\ and\ \bibinfo {author} {\bibfnamefont {S.}~\bibnamefont {Vitale}},\ }\bibfield  {title} {\bibinfo {title} {{Signatures of a subpopulation of hierarchical mergers in the GWTC-4 gravitational-wave dataset}},\ }\href@noop {} {\  (\bibinfo {year} {2026})},\ \Eprint {https://arxiv.org/abs/2601.07908} {arXiv:2601.07908 [gr-qc]} \BibitemShut {NoStop}%
\bibitem [{\citenamefont {Wang}\ \emph {et~al.}(2025)\citenamefont {Wang}, \citenamefont {Li}, \citenamefont {Gao}, \citenamefont {Tang},\ and\ \citenamefont {Fan}}]{Wang:2025nhf}%
  \BibitemOpen
  \bibfield  {author} {\bibinfo {author} {\bibfnamefont {Y.-Z.}\ \bibnamefont {Wang}}, \bibinfo {author} {\bibfnamefont {Y.-J.}\ \bibnamefont {Li}}, \bibinfo {author} {\bibfnamefont {S.-J.}\ \bibnamefont {Gao}}, \bibinfo {author} {\bibfnamefont {S.-P.}\ \bibnamefont {Tang}},\ and\ \bibinfo {author} {\bibfnamefont {Y.-Z.}\ \bibnamefont {Fan}},\ }\bibfield  {title} {\bibinfo {title} {{A new group of low-spin $50-70M_\odot$ Black Holes and the high pair-instability mass cutoff}},\ }\href@noop {} {\  (\bibinfo {year} {2025})},\ \Eprint {https://arxiv.org/abs/2510.22698} {arXiv:2510.22698 [astro-ph.HE]} \BibitemShut {NoStop}%
\bibitem [{\citenamefont {Ray}\ and\ \citenamefont {Kalogera}(2026)}]{Ray:2025xti}%
  \BibitemOpen
  \bibfield  {author} {\bibinfo {author} {\bibfnamefont {A.}~\bibnamefont {Ray}}\ and\ \bibinfo {author} {\bibfnamefont {V.}~\bibnamefont {Kalogera}},\ }\bibfield  {title} {\bibinfo {title} {{Reexamining Evidence of a Pair-instability Mass Gap in the Binary Black Hole Population}},\ }\href {https://doi.org/10.3847/2041-8213/ae374d} {\bibfield  {journal} {\bibinfo  {journal} {Astrophys. J. Lett.}\ }\textbf {\bibinfo {volume} {998}},\ \bibinfo {pages} {L20} (\bibinfo {year} {2026})},\ \Eprint {https://arxiv.org/abs/2510.18867} {arXiv:2510.18867 [astro-ph.HE]} \BibitemShut {NoStop}%
\bibitem [{\citenamefont {Fuller}\ \emph {et~al.}(2019)\citenamefont {Fuller}, \citenamefont {Piro},\ and\ \citenamefont {Jermyn}}]{Fuller:2019abc}%
  \BibitemOpen
  \bibfield  {author} {\bibinfo {author} {\bibfnamefont {J.}~\bibnamefont {Fuller}}, \bibinfo {author} {\bibfnamefont {A.~L.}\ \bibnamefont {Piro}},\ and\ \bibinfo {author} {\bibfnamefont {A.~S.}\ \bibnamefont {Jermyn}},\ }\bibfield  {title} {\bibinfo {title} {Slowing the spins of stellar cores},\ }\href {https://doi.org/10.1093/mnras/stz514} {\bibfield  {journal} {\bibinfo  {journal} {Monthly Notices of the Royal Astronomical Society}\ }\textbf {\bibinfo {volume} {485}},\ \bibinfo {pages} {3661–3680} (\bibinfo {year} {2019})}\BibitemShut {NoStop}%
\bibitem [{\citenamefont {Ma}\ and\ \citenamefont {Fuller}(2019)}]{Ma:2019cpr}%
  \BibitemOpen
  \bibfield  {author} {\bibinfo {author} {\bibfnamefont {L.}~\bibnamefont {Ma}}\ and\ \bibinfo {author} {\bibfnamefont {J.}~\bibnamefont {Fuller}},\ }\bibfield  {title} {\bibinfo {title} {{Angular momentum transport in massive stars and natal neutron star rotation rates}},\ }\href {https://doi.org/10.1093/mnras/stz2009} {\bibfield  {journal} {\bibinfo  {journal} {Mon. Not. Roy. Astron. Soc.}\ }\textbf {\bibinfo {volume} {488}},\ \bibinfo {pages} {4338} (\bibinfo {year} {2019})},\ \Eprint {https://arxiv.org/abs/1907.03713} {arXiv:1907.03713 [astro-ph.SR]} \BibitemShut {NoStop}%
\bibitem [{\citenamefont {Fuller}\ and\ \citenamefont {Ma}(2019)}]{Fuller:2019sxi}%
  \BibitemOpen
  \bibfield  {author} {\bibinfo {author} {\bibfnamefont {J.}~\bibnamefont {Fuller}}\ and\ \bibinfo {author} {\bibfnamefont {L.}~\bibnamefont {Ma}},\ }\bibfield  {title} {\bibinfo {title} {{Most Black Holes are Born Very Slowly Rotating}},\ }\href {https://doi.org/10.3847/2041-8213/ab339b} {\bibfield  {journal} {\bibinfo  {journal} {Astrophys. J. Lett.}\ }\textbf {\bibinfo {volume} {881}},\ \bibinfo {pages} {L1} (\bibinfo {year} {2019})},\ \Eprint {https://arxiv.org/abs/1907.03714} {arXiv:1907.03714 [astro-ph.SR]} \BibitemShut {NoStop}%
\bibitem [{\citenamefont {Spruit}(2002)}]{Spruit:2001tz}%
  \BibitemOpen
  \bibfield  {author} {\bibinfo {author} {\bibfnamefont {H.~C.}\ \bibnamefont {Spruit}},\ }\bibfield  {title} {\bibinfo {title} {{Dynamo action by differential rotation in a stably stratified stellar interior}},\ }\href {https://doi.org/10.1051/0004-6361:20011465} {\bibfield  {journal} {\bibinfo  {journal} {Astron. Astrophys.}\ }\textbf {\bibinfo {volume} {381}},\ \bibinfo {pages} {923} (\bibinfo {year} {2002})},\ \Eprint {https://arxiv.org/abs/astro-ph/0108207} {arXiv:astro-ph/0108207} \BibitemShut {NoStop}%
\bibitem [{\citenamefont {Heger}\ \emph {et~al.}(2005)\citenamefont {Heger}, \citenamefont {Woosley},\ and\ \citenamefont {Spruit}}]{Heger:2004qp}%
  \BibitemOpen
  \bibfield  {author} {\bibinfo {author} {\bibfnamefont {A.}~\bibnamefont {Heger}}, \bibinfo {author} {\bibfnamefont {S.~E.}\ \bibnamefont {Woosley}},\ and\ \bibinfo {author} {\bibfnamefont {H.~C.}\ \bibnamefont {Spruit}},\ }\bibfield  {title} {\bibinfo {title} {{Presupernova evolution of differentially rotating massive stars including magnetic fields}},\ }\href {https://doi.org/10.1086/429868} {\bibfield  {journal} {\bibinfo  {journal} {Astrophys. J.}\ }\textbf {\bibinfo {volume} {626}},\ \bibinfo {pages} {350} (\bibinfo {year} {2005})},\ \Eprint {https://arxiv.org/abs/astro-ph/0409422} {arXiv:astro-ph/0409422} \BibitemShut {NoStop}%
\bibitem [{\citenamefont {Qin}\ \emph {et~al.}(2018)\citenamefont {Qin}, \citenamefont {Fragos}, \citenamefont {Meynet}, \citenamefont {Andrews}, \citenamefont {S{\o}rensen},\ and\ \citenamefont {Song}}]{Qin:2018vaa}%
  \BibitemOpen
  \bibfield  {author} {\bibinfo {author} {\bibfnamefont {Y.}~\bibnamefont {Qin}}, \bibinfo {author} {\bibfnamefont {T.}~\bibnamefont {Fragos}}, \bibinfo {author} {\bibfnamefont {G.}~\bibnamefont {Meynet}}, \bibinfo {author} {\bibfnamefont {J.}~\bibnamefont {Andrews}}, \bibinfo {author} {\bibfnamefont {M.}~\bibnamefont {S{\o}rensen}},\ and\ \bibinfo {author} {\bibfnamefont {H.~F.}\ \bibnamefont {Song}},\ }\bibfield  {title} {\bibinfo {title} {{The spin of the second-born black hole in coalescing binary black holes}},\ }\href {https://doi.org/10.1051/0004-6361/201832839} {\bibfield  {journal} {\bibinfo  {journal} {Astron. Astrophys.}\ }\textbf {\bibinfo {volume} {616}},\ \bibinfo {pages} {A28} (\bibinfo {year} {2018})},\ \Eprint {https://arxiv.org/abs/1802.05738} {arXiv:1802.05738 [astro-ph.SR]} \BibitemShut {NoStop}%
\bibitem [{\citenamefont {Belczynski}\ \emph {et~al.}(2020)\citenamefont {Belczynski} \emph {et~al.}}]{Belczynski:2017gds}%
  \BibitemOpen
  \bibfield  {author} {\bibinfo {author} {\bibfnamefont {K.}~\bibnamefont {Belczynski}} \emph {et~al.},\ }\bibfield  {title} {\bibinfo {title} {{Evolutionary roads leading to low effective spins, high black hole masses, and O1/O2 rates for LIGO/Virgo binary black holes}},\ }\href {https://doi.org/10.1051/0004-6361/201936528} {\bibfield  {journal} {\bibinfo  {journal} {Astron. Astrophys.}\ }\textbf {\bibinfo {volume} {636}},\ \bibinfo {pages} {A104} (\bibinfo {year} {2020})},\ \Eprint {https://arxiv.org/abs/1706.07053} {arXiv:1706.07053 [astro-ph.HE]} \BibitemShut {NoStop}%
\bibitem [{\citenamefont {Tiwari}\ and\ \citenamefont {Fairhurst}(2021)}]{Tiwari:2020otp}%
  \BibitemOpen
  \bibfield  {author} {\bibinfo {author} {\bibfnamefont {V.}~\bibnamefont {Tiwari}}\ and\ \bibinfo {author} {\bibfnamefont {S.}~\bibnamefont {Fairhurst}},\ }\bibfield  {title} {\bibinfo {title} {{The Emergence of Structure in the Binary Black Hole Mass Distribution}},\ }\href {https://doi.org/10.3847/2041-8213/abfbe7} {\bibfield  {journal} {\bibinfo  {journal} {Astrophys. J. Lett.}\ }\textbf {\bibinfo {volume} {913}},\ \bibinfo {pages} {L19} (\bibinfo {year} {2021})},\ \Eprint {https://arxiv.org/abs/2011.04502} {arXiv:2011.04502 [astro-ph.HE]} \BibitemShut {NoStop}%
\bibitem [{\citenamefont {Edelman}\ \emph {et~al.}(2022)\citenamefont {Edelman}, \citenamefont {Doctor}, \citenamefont {Godfrey},\ and\ \citenamefont {Farr}}]{Edelman:2021zkw}%
  \BibitemOpen
  \bibfield  {author} {\bibinfo {author} {\bibfnamefont {B.}~\bibnamefont {Edelman}}, \bibinfo {author} {\bibfnamefont {Z.}~\bibnamefont {Doctor}}, \bibinfo {author} {\bibfnamefont {J.}~\bibnamefont {Godfrey}},\ and\ \bibinfo {author} {\bibfnamefont {B.}~\bibnamefont {Farr}},\ }\bibfield  {title} {\bibinfo {title} {{Ain{\textquoteright}t No Mountain High Enough: Semiparametric Modeling of LIGO{\textendash}Virgo{\textquoteright}s Binary Black Hole Mass Distribution}},\ }\href {https://doi.org/10.3847/1538-4357/ac3667} {\bibfield  {journal} {\bibinfo  {journal} {Astrophys. J.}\ }\textbf {\bibinfo {volume} {924}},\ \bibinfo {pages} {101} (\bibinfo {year} {2022})},\ \Eprint {https://arxiv.org/abs/2109.06137} {arXiv:2109.06137 [astro-ph.HE]} \BibitemShut {NoStop}%
\bibitem [{\citenamefont {Abbott}\ \emph {et~al.}(2021)\citenamefont {Abbott} \emph {et~al.}}]{o3a_pop}%
  \BibitemOpen
  \bibfield  {author} {\bibinfo {author} {\bibfnamefont {R.}~\bibnamefont {Abbott}} \emph {et~al.},\ }\bibfield  {title} {\bibinfo {title} {{Population properties of compact objects from the second LIGO-Virgo Gravitational-Wave Transient Catalog}},\ }\href@noop {} {\ \textbf {\bibinfo {volume} {913}},\ \bibinfo {pages} {L7} (\bibinfo {year} {2021})}\BibitemShut {NoStop}%
\bibitem [{\citenamefont {Farah}\ \emph {et~al.}(2024)\citenamefont {Farah}, \citenamefont {Fishbach},\ and\ \citenamefont {Holz}}]{Farah:2023swu}%
  \BibitemOpen
  \bibfield  {author} {\bibinfo {author} {\bibfnamefont {A.~M.}\ \bibnamefont {Farah}}, \bibinfo {author} {\bibfnamefont {M.}~\bibnamefont {Fishbach}},\ and\ \bibinfo {author} {\bibfnamefont {D.~E.}\ \bibnamefont {Holz}},\ }\bibfield  {title} {\bibinfo {title} {{Two of a Kind: Comparing Big and Small Black Holes in Binaries with Gravitational Waves}},\ }\href {https://doi.org/10.3847/1538-4357/ad0558} {\bibfield  {journal} {\bibinfo  {journal} {Astrophys. J.}\ }\textbf {\bibinfo {volume} {962}},\ \bibinfo {pages} {69} (\bibinfo {year} {2024})},\ \Eprint {https://arxiv.org/abs/2308.05102} {arXiv:2308.05102 [astro-ph.HE]} \BibitemShut {NoStop}%
\bibitem [{\citenamefont {Romero-Shaw}\ \emph {et~al.}(2022)\citenamefont {Romero-Shaw}, \citenamefont {Thrane},\ and\ \citenamefont {Lasky}}]{wmf}%
  \BibitemOpen
  \bibfield  {author} {\bibinfo {author} {\bibfnamefont {I.~M.}\ \bibnamefont {Romero-Shaw}}, \bibinfo {author} {\bibfnamefont {E.}~\bibnamefont {Thrane}},\ and\ \bibinfo {author} {\bibfnamefont {P.~D.}\ \bibnamefont {Lasky}},\ }\bibfield  {title} {\bibinfo {title} {When models fail: an introduction to posterior predictive checks and model misspecification in gravitational-wave astronomy},\ }\href@noop {} {\ \textbf {\bibinfo {volume} {39}},\ \bibinfo {pages} {E025} (\bibinfo {year} {2022})}\BibitemShut {NoStop}%
\bibitem [{\citenamefont {Roy}\ \emph {et~al.}(2025)\citenamefont {Roy}, \citenamefont {van Son},\ and\ \citenamefont {Farr}}]{Roy:2025ktr}%
  \BibitemOpen
  \bibfield  {author} {\bibinfo {author} {\bibfnamefont {S.~K.}\ \bibnamefont {Roy}}, \bibinfo {author} {\bibfnamefont {L.~A.~C.}\ \bibnamefont {van Son}},\ and\ \bibinfo {author} {\bibfnamefont {W.~M.}\ \bibnamefont {Farr}},\ }\bibfield  {title} {\bibinfo {title} {{A Mid-Thirties Crisis: Dissecting the Properties of Gravitational Wave Sources Near the 35 Solar Mass Peak}},\ }\href@noop {} {\  (\bibinfo {year} {2025})},\ \Eprint {https://arxiv.org/abs/2507.01086} {arXiv:2507.01086 [astro-ph.HE]} \BibitemShut {NoStop}%
\bibitem [{\citenamefont {{de Mink}}\ \emph {et~al.}(2010)\citenamefont {{de Mink}}, \citenamefont {{Cantiello}}, \citenamefont {{Langer}}, \citenamefont {{Pols}},\ and\ \citenamefont {{Yoon}}}]{deMink:2009}%
  \BibitemOpen
  \bibfield  {author} {\bibinfo {author} {\bibfnamefont {S.~E.}\ \bibnamefont {{de Mink}}}, \bibinfo {author} {\bibfnamefont {M.}~\bibnamefont {{Cantiello}}}, \bibinfo {author} {\bibfnamefont {N.}~\bibnamefont {{Langer}}}, \bibinfo {author} {\bibfnamefont {O.~R.}\ \bibnamefont {{Pols}}},\ and\ \bibinfo {author} {\bibfnamefont {S.~C.}\ \bibnamefont {{Yoon}}},\ }\bibfield  {title} {\bibinfo {title} {{A New Evolutionary Scenario for the Formation of Massive Black-Hole Binaries such as M33 X-7 and IC 10 X-1}},\ }in\ \href {https://doi.org/10.48550/arXiv.0910.3694} {\emph {\bibinfo {booktitle} {Binaries - Key to Comprehension of the Universe}}},\ \bibinfo {series} {Astronomical Society of the Pacific Conference Series}, Vol.\ \bibinfo {volume} {435},\ \bibinfo {editor} {edited by\ \bibinfo {editor} {\bibfnamefont {A.}~\bibnamefont {{Pr{\v{s}}a}}}\ and\ \bibinfo {editor} {\bibfnamefont {M.}~\bibnamefont {{Zejda}}}}\ (\bibinfo {year} {2010})\ p.\ \bibinfo {pages} {179},\ \Eprint {https://arxiv.org/abs/0910.3694}
  {arXiv:0910.3694 [astro-ph.SR]} \BibitemShut {NoStop}%
\bibitem [{\citenamefont {{de Mink}}\ \emph {et~al.}(2009)\citenamefont {{de Mink}}, \citenamefont {{Cantiello}}, \citenamefont {{Langer}}, \citenamefont {{Pols}}, \citenamefont {{Brott}},\ and\ \citenamefont {{Yoon}}}]{deMink20092}%
  \BibitemOpen
  \bibfield  {author} {\bibinfo {author} {\bibfnamefont {S.~E.}\ \bibnamefont {{de Mink}}}, \bibinfo {author} {\bibfnamefont {M.}~\bibnamefont {{Cantiello}}}, \bibinfo {author} {\bibfnamefont {N.}~\bibnamefont {{Langer}}}, \bibinfo {author} {\bibfnamefont {O.~R.}\ \bibnamefont {{Pols}}}, \bibinfo {author} {\bibfnamefont {I.}~\bibnamefont {{Brott}}},\ and\ \bibinfo {author} {\bibfnamefont {S.~C.}\ \bibnamefont {{Yoon}}},\ }\bibfield  {title} {\bibinfo {title} {{Rotational mixing in massive binaries. Detached short-period systems}},\ }\href {https://doi.org/10.1051/0004-6361/200811439} {\bibfield  {journal} {\bibinfo  {journal} {Astronomy and Astrophysics}\ }\textbf {\bibinfo {volume} {497}},\ \bibinfo {pages} {243} (\bibinfo {year} {2009})},\ \Eprint {https://arxiv.org/abs/0902.1751} {arXiv:0902.1751 [astro-ph.SR]} \BibitemShut {NoStop}%
\bibitem [{\citenamefont {de~Mink}\ and\ \citenamefont {Mandel}(2016)}]{deMink:2016vkw}%
  \BibitemOpen
  \bibfield  {author} {\bibinfo {author} {\bibfnamefont {S.~E.}\ \bibnamefont {de~Mink}}\ and\ \bibinfo {author} {\bibfnamefont {I.}~\bibnamefont {Mandel}},\ }\bibfield  {title} {\bibinfo {title} {{The chemically homogeneous evolutionary channel for binary black hole mergers: rates and properties of gravitational-wave events detectable by advanced LIGO}},\ }\href {https://doi.org/10.1093/mnras/stw1219} {\bibfield  {journal} {\bibinfo  {journal} {Mon. Not. Roy. Astron. Soc.}\ }\textbf {\bibinfo {volume} {460}},\ \bibinfo {pages} {3545} (\bibinfo {year} {2016})},\ \Eprint {https://arxiv.org/abs/1603.02291} {arXiv:1603.02291 [astro-ph.HE]} \BibitemShut {NoStop}%
\bibitem [{\citenamefont {Mandel}\ and\ \citenamefont {de~Mink}(2016)}]{Mandel:2015qlu}%
  \BibitemOpen
  \bibfield  {author} {\bibinfo {author} {\bibfnamefont {I.}~\bibnamefont {Mandel}}\ and\ \bibinfo {author} {\bibfnamefont {S.~E.}\ \bibnamefont {de~Mink}},\ }\bibfield  {title} {\bibinfo {title} {{Merging binary black holes formed through chemically homogeneous evolution in short-period stellar binaries}},\ }\href {https://doi.org/10.1093/mnras/stw379} {\bibfield  {journal} {\bibinfo  {journal} {Mon. Not. Roy. Astron. Soc.}\ }\textbf {\bibinfo {volume} {458}},\ \bibinfo {pages} {2634} (\bibinfo {year} {2016})},\ \Eprint {https://arxiv.org/abs/1601.00007} {arXiv:1601.00007 [astro-ph.HE]} \BibitemShut {NoStop}%
\bibitem [{\citenamefont {Marchant}\ \emph {et~al.}(2016)\citenamefont {Marchant}, \citenamefont {Langer}, \citenamefont {Podsiadlowski}, \citenamefont {Tauris},\ and\ \citenamefont {Moriya}}]{Marchant:2016wow}%
  \BibitemOpen
  \bibfield  {author} {\bibinfo {author} {\bibfnamefont {P.}~\bibnamefont {Marchant}}, \bibinfo {author} {\bibfnamefont {N.}~\bibnamefont {Langer}}, \bibinfo {author} {\bibfnamefont {P.}~\bibnamefont {Podsiadlowski}}, \bibinfo {author} {\bibfnamefont {T.~M.}\ \bibnamefont {Tauris}},\ and\ \bibinfo {author} {\bibfnamefont {T.~J.}\ \bibnamefont {Moriya}},\ }\bibfield  {title} {\bibinfo {title} {{A new route towards merging massive black holes}},\ }\href {https://doi.org/10.1051/0004-6361/201628133} {\bibfield  {journal} {\bibinfo  {journal} {Astron. Astrophys.}\ }\textbf {\bibinfo {volume} {588}},\ \bibinfo {pages} {A50} (\bibinfo {year} {2016})},\ \Eprint {https://arxiv.org/abs/1601.03718} {arXiv:1601.03718 [astro-ph.SR]} \BibitemShut {NoStop}%
\bibitem [{\citenamefont {Hastings}\ \emph {et~al.}(2020)\citenamefont {Hastings}, \citenamefont {Langer},\ and\ \citenamefont {Koenigsberger}}]{Hastings:2020}%
  \BibitemOpen
  \bibfield  {author} {\bibinfo {author} {\bibfnamefont {B.}~\bibnamefont {Hastings}}, \bibinfo {author} {\bibfnamefont {N.}~\bibnamefont {Langer}},\ and\ \bibinfo {author} {\bibfnamefont {G.}~\bibnamefont {Koenigsberger}},\ }\bibfield  {title} {\bibinfo {title} {Internal circulation in tidally locked massive binary stars: Consequences for double black hole formation},\ }\href {https://doi.org/10.1051/0004-6361/202038499} {\bibfield  {journal} {\bibinfo  {journal} {Astronomy \&; Astrophysics}\ }\textbf {\bibinfo {volume} {641}},\ \bibinfo {pages} {A86} (\bibinfo {year} {2020})}\BibitemShut {NoStop}%
\bibitem [{\citenamefont {Riley}\ \emph {et~al.}(2021)\citenamefont {Riley}, \citenamefont {Mandel}, \citenamefont {Marchant}, \citenamefont {Butler}, \citenamefont {Nathaniel}, \citenamefont {Neijssel}, \citenamefont {Shortt},\ and\ \citenamefont {Vigna-Gomez}}]{Riley:2020btf}%
  \BibitemOpen
  \bibfield  {author} {\bibinfo {author} {\bibfnamefont {J.}~\bibnamefont {Riley}}, \bibinfo {author} {\bibfnamefont {I.}~\bibnamefont {Mandel}}, \bibinfo {author} {\bibfnamefont {P.}~\bibnamefont {Marchant}}, \bibinfo {author} {\bibfnamefont {E.}~\bibnamefont {Butler}}, \bibinfo {author} {\bibfnamefont {K.}~\bibnamefont {Nathaniel}}, \bibinfo {author} {\bibfnamefont {C.}~\bibnamefont {Neijssel}}, \bibinfo {author} {\bibfnamefont {S.}~\bibnamefont {Shortt}},\ and\ \bibinfo {author} {\bibfnamefont {A.}~\bibnamefont {Vigna-Gomez}},\ }\bibfield  {title} {\bibinfo {title} {{Chemically homogeneous evolution: a rapid population synthesis approach}},\ }\href {https://doi.org/10.1093/mnras/stab1291} {\bibfield  {journal} {\bibinfo  {journal} {Mon. Not. Roy. Astron. Soc.}\ }\textbf {\bibinfo {volume} {505}},\ \bibinfo {pages} {663} (\bibinfo {year} {2021})},\ \Eprint {https://arxiv.org/abs/2010.00002} {arXiv:2010.00002 [astro-ph.SR]} \BibitemShut {NoStop}%
\bibitem [{\citenamefont {Marchant}\ \emph {et~al.}(2024)\citenamefont {Marchant}, \citenamefont {Podsiadlowski},\ and\ \citenamefont {Mandel}}]{Marchant:2023ncp}%
  \BibitemOpen
  \bibfield  {author} {\bibinfo {author} {\bibfnamefont {P.}~\bibnamefont {Marchant}}, \bibinfo {author} {\bibfnamefont {P.}~\bibnamefont {Podsiadlowski}},\ and\ \bibinfo {author} {\bibfnamefont {I.}~\bibnamefont {Mandel}},\ }\bibfield  {title} {\bibinfo {title} {{An upper limit on the spins of merging binary black holes formed through isolated binary evolution}},\ }\href {https://doi.org/10.1051/0004-6361/202348190} {\bibfield  {journal} {\bibinfo  {journal} {Astron. Astrophys.}\ }\textbf {\bibinfo {volume} {691}},\ \bibinfo {pages} {A339} (\bibinfo {year} {2024})},\ \Eprint {https://arxiv.org/abs/2311.14041} {arXiv:2311.14041 [astro-ph.HE]} \BibitemShut {NoStop}%
\bibitem [{\citenamefont {Kinugawa}\ \emph {et~al.}(2014)\citenamefont {Kinugawa}, \citenamefont {Inayoshi}, \citenamefont {Hotokezaka}, \citenamefont {Nakauchi},\ and\ \citenamefont {Nakamura}}]{Kinugawa:2014zha}%
  \BibitemOpen
  \bibfield  {author} {\bibinfo {author} {\bibfnamefont {T.}~\bibnamefont {Kinugawa}}, \bibinfo {author} {\bibfnamefont {K.}~\bibnamefont {Inayoshi}}, \bibinfo {author} {\bibfnamefont {K.}~\bibnamefont {Hotokezaka}}, \bibinfo {author} {\bibfnamefont {D.}~\bibnamefont {Nakauchi}},\ and\ \bibinfo {author} {\bibfnamefont {T.}~\bibnamefont {Nakamura}},\ }\bibfield  {title} {\bibinfo {title} {{Possible Indirect Confirmation of the Existence of Pop III Massive Stars by Gravitational Wave}},\ }\href {https://doi.org/10.1093/mnras/stu1022} {\bibfield  {journal} {\bibinfo  {journal} {Mon. Not. Roy. Astron. Soc.}\ }\textbf {\bibinfo {volume} {442}},\ \bibinfo {pages} {2963} (\bibinfo {year} {2014})},\ \Eprint {https://arxiv.org/abs/1402.6672} {arXiv:1402.6672 [astro-ph.HE]} \BibitemShut {NoStop}%
\bibitem [{\citenamefont {Kinugawa}\ \emph {et~al.}(2020)\citenamefont {Kinugawa}, \citenamefont {Nakamura},\ and\ \citenamefont {Nakano}}]{Kinugawa:2020ego}%
  \BibitemOpen
  \bibfield  {author} {\bibinfo {author} {\bibfnamefont {T.}~\bibnamefont {Kinugawa}}, \bibinfo {author} {\bibfnamefont {T.}~\bibnamefont {Nakamura}},\ and\ \bibinfo {author} {\bibfnamefont {H.}~\bibnamefont {Nakano}},\ }\bibfield  {title} {\bibinfo {title} {{Chirp Mass and Spin of Binary Black Holes from First Star Remnants}},\ }\href {https://doi.org/10.1093/mnras/staa2511} {\bibfield  {journal} {\bibinfo  {journal} {Mon. Not. Roy. Astron. Soc.}\ }\textbf {\bibinfo {volume} {498}},\ \bibinfo {pages} {3946} (\bibinfo {year} {2020})},\ \Eprint {https://arxiv.org/abs/2005.09795} {arXiv:2005.09795 [astro-ph.HE]} \BibitemShut {NoStop}%
\bibitem [{\citenamefont {Iwaya}\ \emph {et~al.}(2023)\citenamefont {Iwaya}, \citenamefont {Kinugawa},\ and\ \citenamefont {Tagoshi}}]{Iwaya:2023mse}%
  \BibitemOpen
  \bibfield  {author} {\bibinfo {author} {\bibfnamefont {M.}~\bibnamefont {Iwaya}}, \bibinfo {author} {\bibfnamefont {T.}~\bibnamefont {Kinugawa}},\ and\ \bibinfo {author} {\bibfnamefont {H.}~\bibnamefont {Tagoshi}},\ }\bibfield  {title} {\bibinfo {title} {{Constraint on the progenitor of binary black hole merger using Population III star formation channel}},\ }\href@noop {} {\  (\bibinfo {year} {2023})},\ \Eprint {https://arxiv.org/abs/2312.17491} {arXiv:2312.17491 [astro-ph.HE]} \BibitemShut {NoStop}%
\bibitem [{\citenamefont {Santoliquido}\ \emph {et~al.}(2023)\citenamefont {Santoliquido}, \citenamefont {Mapelli}, \citenamefont {Iorio}, \citenamefont {Costa}, \citenamefont {Glover}, \citenamefont {Hartwig}, \citenamefont {Klessen},\ and\ \citenamefont {Merli}}]{Santoliquido:2023wzn}%
  \BibitemOpen
  \bibfield  {author} {\bibinfo {author} {\bibfnamefont {F.}~\bibnamefont {Santoliquido}}, \bibinfo {author} {\bibfnamefont {M.}~\bibnamefont {Mapelli}}, \bibinfo {author} {\bibfnamefont {G.}~\bibnamefont {Iorio}}, \bibinfo {author} {\bibfnamefont {G.}~\bibnamefont {Costa}}, \bibinfo {author} {\bibfnamefont {S.~C.~O.}\ \bibnamefont {Glover}}, \bibinfo {author} {\bibfnamefont {T.}~\bibnamefont {Hartwig}}, \bibinfo {author} {\bibfnamefont {R.~S.}\ \bibnamefont {Klessen}},\ and\ \bibinfo {author} {\bibfnamefont {L.}~\bibnamefont {Merli}},\ }\bibfield  {title} {\bibinfo {title} {{Binary black hole mergers from population III stars: uncertainties from star formation and binary star properties}},\ }\href {https://doi.org/10.1093/mnras/stad1860} {\bibfield  {journal} {\bibinfo  {journal} {Mon. Not. Roy. Astron. Soc.}\ }\textbf {\bibinfo {volume} {524}},\ \bibinfo {pages} {307} (\bibinfo {year} {2023})},\ \bibinfo {note} {[Erratum: Mon.Not.Roy.Astron.Soc. 528, 954--962 (2024)]},\ \Eprint
  {https://arxiv.org/abs/2303.15515} {arXiv:2303.15515 [astro-ph.GA]} \BibitemShut {NoStop}%
\bibitem [{\citenamefont {Fishbach}\ and\ \citenamefont {Holz}(2017)}]{Fishbach:2017zga}%
  \BibitemOpen
  \bibfield  {author} {\bibinfo {author} {\bibfnamefont {M.}~\bibnamefont {Fishbach}}\ and\ \bibinfo {author} {\bibfnamefont {D.~E.}\ \bibnamefont {Holz}},\ }\bibfield  {title} {\bibinfo {title} {{Where Are LIGO{\textquoteright}s Big Black Holes?}},\ }\href {https://doi.org/10.3847/2041-8213/aa9bf6} {\bibfield  {journal} {\bibinfo  {journal} {Astrophys. J. Lett.}\ }\textbf {\bibinfo {volume} {851}},\ \bibinfo {pages} {L25} (\bibinfo {year} {2017})},\ \Eprint {https://arxiv.org/abs/1709.08584} {arXiv:1709.08584 [astro-ph.HE]} \BibitemShut {NoStop}%
\bibitem [{\citenamefont {Talbot}\ and\ \citenamefont {Thrane}(2018)}]{Talbot:2018cva}%
  \BibitemOpen
  \bibfield  {author} {\bibinfo {author} {\bibfnamefont {C.}~\bibnamefont {Talbot}}\ and\ \bibinfo {author} {\bibfnamefont {E.}~\bibnamefont {Thrane}},\ }\bibfield  {title} {\bibinfo {title} {{Measuring the binary black hole mass spectrum with an astrophysically motivated parameterization}},\ }\href {https://doi.org/10.3847/1538-4357/aab34c} {\bibfield  {journal} {\bibinfo  {journal} {Astrophys. J.}\ }\textbf {\bibinfo {volume} {856}},\ \bibinfo {pages} {173} (\bibinfo {year} {2018})},\ \Eprint {https://arxiv.org/abs/1801.02699} {arXiv:1801.02699 [astro-ph.HE]} \BibitemShut {NoStop}%
\bibitem [{\citenamefont {Stevenson}\ \emph {et~al.}(2019)\citenamefont {Stevenson}, \citenamefont {Sampson}, \citenamefont {Powell}, \citenamefont {Vigna-Gómez}, \citenamefont {Neijssel}, \citenamefont {Sz{\'e}csi},\ and\ \citenamefont {Mandel}}]{stevenson}%
  \BibitemOpen
  \bibfield  {author} {\bibinfo {author} {\bibfnamefont {S.}~\bibnamefont {Stevenson}}, \bibinfo {author} {\bibfnamefont {M.}~\bibnamefont {Sampson}}, \bibinfo {author} {\bibfnamefont {J.}~\bibnamefont {Powell}}, \bibinfo {author} {\bibfnamefont {A.}~\bibnamefont {Vigna-Gómez}}, \bibinfo {author} {\bibfnamefont {C.~J.}\ \bibnamefont {Neijssel}}, \bibinfo {author} {\bibfnamefont {D.}~\bibnamefont {Sz{\'e}csi}},\ and\ \bibinfo {author} {\bibfnamefont {I.}~\bibnamefont {Mandel}},\ }\bibfield  {title} {\bibinfo {title} {The impact of pair-instability mass loss on the binary black hole mass distribution},\ }\href@noop {} {\bibfield  {journal} {\bibinfo  {journal} {Astrophys. J.}\ }\textbf {\bibinfo {volume} {882}},\ \bibinfo {pages} {121} (\bibinfo {year} {2019})}\BibitemShut {NoStop}%
\bibitem [{\citenamefont {Farmer}\ \emph {et~al.}(2019)\citenamefont {Farmer}, \citenamefont {Renzo}, \citenamefont {de~Mink}, \citenamefont {Marchant},\ and\ \citenamefont {Justham}}]{Farmer:2019jed}%
  \BibitemOpen
  \bibfield  {author} {\bibinfo {author} {\bibfnamefont {R.}~\bibnamefont {Farmer}}, \bibinfo {author} {\bibfnamefont {M.}~\bibnamefont {Renzo}}, \bibinfo {author} {\bibfnamefont {S.~E.}\ \bibnamefont {de~Mink}}, \bibinfo {author} {\bibfnamefont {P.}~\bibnamefont {Marchant}},\ and\ \bibinfo {author} {\bibfnamefont {S.}~\bibnamefont {Justham}},\ }\bibfield  {title} {\bibinfo {title} {{Mind the gap: The location of the lower edge of the pair instability supernovae black hole mass gap}}\ }\href {https://doi.org/10.3847/1538-4357/ab518b} {10.3847/1538-4357/ab518b} (\bibinfo {year} {2019}),\ \Eprint {https://arxiv.org/abs/1910.12874} {arXiv:1910.12874 [astro-ph.SR]} \BibitemShut {NoStop}%
\bibitem [{\citenamefont {Farmer}\ \emph {et~al.}(2020)\citenamefont {Farmer}, \citenamefont {Renzo}, \citenamefont {de~Mink}, \citenamefont {Fishbach},\ and\ \citenamefont {Justham}}]{Farmer:2020xne}%
  \BibitemOpen
  \bibfield  {author} {\bibinfo {author} {\bibfnamefont {R.}~\bibnamefont {Farmer}}, \bibinfo {author} {\bibfnamefont {M.}~\bibnamefont {Renzo}}, \bibinfo {author} {\bibfnamefont {S.}~\bibnamefont {de~Mink}}, \bibinfo {author} {\bibfnamefont {M.}~\bibnamefont {Fishbach}},\ and\ \bibinfo {author} {\bibfnamefont {S.}~\bibnamefont {Justham}},\ }\bibfield  {title} {\bibinfo {title} {{Constraints from gravitational wave detections of binary black hole mergers on the $^{12}\rm{C}\left(\alpha,\gamma\right)^{16}\!\rm{O}$ rate}},\ }\href {https://doi.org/10.3847/2041-8213/abbadd} {\bibfield  {journal} {\bibinfo  {journal} {Astrophys. J. Lett.}\ }\textbf {\bibinfo {volume} {902}},\ \bibinfo {pages} {L36} (\bibinfo {year} {2020})},\ \Eprint {https://arxiv.org/abs/2006.06678} {arXiv:2006.06678 [astro-ph.HE]} \BibitemShut {NoStop}%
\bibitem [{\citenamefont {Hendriks}\ \emph {et~al.}(2023)\citenamefont {Hendriks}, \citenamefont {van Son}, \citenamefont {Renzo}, \citenamefont {Izzard},\ and\ \citenamefont {Farmer}}]{Hendriks:2023yrw}%
  \BibitemOpen
  \bibfield  {author} {\bibinfo {author} {\bibfnamefont {D.~D.}\ \bibnamefont {Hendriks}}, \bibinfo {author} {\bibfnamefont {L.~A.~C.}\ \bibnamefont {van Son}}, \bibinfo {author} {\bibfnamefont {M.}~\bibnamefont {Renzo}}, \bibinfo {author} {\bibfnamefont {R.~G.}\ \bibnamefont {Izzard}},\ and\ \bibinfo {author} {\bibfnamefont {R.}~\bibnamefont {Farmer}},\ }\bibfield  {title} {\bibinfo {title} {{Pulsational pair-instability supernovae in gravitational-wave and electromagnetic transients}},\ }\href {https://doi.org/10.1093/mnras/stad2857} {\bibfield  {journal} {\bibinfo  {journal} {Mon. Not. Roy. Astron. Soc.}\ }\textbf {\bibinfo {volume} {526}},\ \bibinfo {pages} {4130} (\bibinfo {year} {2023})},\ \Eprint {https://arxiv.org/abs/2309.09339} {arXiv:2309.09339 [astro-ph.HE]} \BibitemShut {NoStop}%
\bibitem [{\citenamefont {Renzo}\ \emph {et~al.}(2020)\citenamefont {Renzo}, \citenamefont {Farmer}, \citenamefont {Justham}, \citenamefont {de~Mink}, \citenamefont {G{\"o}tberg},\ and\ \citenamefont {Marchant}}]{Renzo:2020rzx}%
  \BibitemOpen
  \bibfield  {author} {\bibinfo {author} {\bibfnamefont {M.}~\bibnamefont {Renzo}}, \bibinfo {author} {\bibfnamefont {R.~J.}\ \bibnamefont {Farmer}}, \bibinfo {author} {\bibfnamefont {S.}~\bibnamefont {Justham}}, \bibinfo {author} {\bibfnamefont {S.~E.}\ \bibnamefont {de~Mink}}, \bibinfo {author} {\bibfnamefont {Y.}~\bibnamefont {G{\"o}tberg}},\ and\ \bibinfo {author} {\bibfnamefont {P.}~\bibnamefont {Marchant}},\ }\bibfield  {title} {\bibinfo {title} {{Sensitivity of the lower-edge of the pair instability black hole mass gap to the treatment of time dependent convection}},\ }\href {https://doi.org/10.1093/mnras/staa549} {\bibfield  {journal} {\bibinfo  {journal} {Mon. Not. Roy. Astron. Soc.}\ }\textbf {\bibinfo {volume} {493}},\ \bibinfo {pages} {4333} (\bibinfo {year} {2020})},\ \Eprint {https://arxiv.org/abs/2002.08200} {arXiv:2002.08200 [astro-ph.SR]} \BibitemShut {NoStop}%
\bibitem [{\citenamefont {Croon}\ and\ \citenamefont {Sakstein}(2023)}]{Croon}%
  \BibitemOpen
  \bibfield  {author} {\bibinfo {author} {\bibfnamefont {D.}~\bibnamefont {Croon}}\ and\ \bibinfo {author} {\bibfnamefont {J.}~\bibnamefont {Sakstein}},\ }\bibfield  {title} {\bibinfo {title} {{Prediction of Multiple Features in the Black Hole Mass Function due to Pulsational Pair-Instability Supernovae}},\ }\href@noop {} {\  (\bibinfo {year} {2023})},\ \Eprint {https://arxiv.org/abs/2312.13459} {arXiv:2312.13459} \BibitemShut {NoStop}%
\bibitem [{\citenamefont {Winch}\ \emph {et~al.}(2025)\citenamefont {Winch}, \citenamefont {Sabhahit}, \citenamefont {Vink},\ and\ \citenamefont {Higgins}}]{Winch}%
  \BibitemOpen
  \bibfield  {author} {\bibinfo {author} {\bibfnamefont {E.~R.~J.}\ \bibnamefont {Winch}}, \bibinfo {author} {\bibfnamefont {G.~N.}\ \bibnamefont {Sabhahit}}, \bibinfo {author} {\bibfnamefont {J.~S.}\ \bibnamefont {Vink}},\ and\ \bibinfo {author} {\bibfnamefont {E.~R.}\ \bibnamefont {Higgins}},\ }\bibfield  {title} {\bibinfo {title} {{The black hole - pair instability boundary for high stellar rotation}},\ }\href@noop {} {\ \textbf {\bibinfo {volume} {540}},\ \bibinfo {pages} {90} (\bibinfo {year} {2025})}\BibitemShut {NoStop}%
\bibitem [{\citenamefont {Abbott}\ \emph {et~al.}(2020{\natexlab{b}})\citenamefont {Abbott} \emph {et~al.}}]{LIGOScientific:2020iuh}%
  \BibitemOpen
  \bibfield  {author} {\bibinfo {author} {\bibfnamefont {R.}~\bibnamefont {Abbott}} \emph {et~al.} (\bibinfo {collaboration} {LIGO Scientific, Virgo}),\ }\bibfield  {title} {\bibinfo {title} {{GW190521: A Binary Black Hole Merger with a Total Mass of $150 M_{\odot}$}},\ }\href {https://doi.org/10.1103/PhysRevLett.125.101102} {\bibfield  {journal} {\bibinfo  {journal} {Phys. Rev. Lett.}\ }\textbf {\bibinfo {volume} {125}},\ \bibinfo {pages} {101102} (\bibinfo {year} {2020}{\natexlab{b}})},\ \Eprint {https://arxiv.org/abs/2009.01075} {arXiv:2009.01075 [gr-qc]} \BibitemShut {NoStop}%
\bibitem [{\citenamefont {Kimball}\ \emph {et~al.}(2020{\natexlab{a}})\citenamefont {Kimball}, \citenamefont {Talbot}, \citenamefont {Berry}, \citenamefont {Carney}, \citenamefont {Zevin}, \citenamefont {Thrane},\ and\ \citenamefont {Kalogera}}]{hierarchical}%
  \BibitemOpen
  \bibfield  {author} {\bibinfo {author} {\bibfnamefont {C.}~\bibnamefont {Kimball}}, \bibinfo {author} {\bibfnamefont {C.}~\bibnamefont {Talbot}}, \bibinfo {author} {\bibfnamefont {C.~P.~L.}\ \bibnamefont {Berry}}, \bibinfo {author} {\bibfnamefont {M.}~\bibnamefont {Carney}}, \bibinfo {author} {\bibfnamefont {M.}~\bibnamefont {Zevin}}, \bibinfo {author} {\bibfnamefont {E.}~\bibnamefont {Thrane}},\ and\ \bibinfo {author} {\bibfnamefont {V.}~\bibnamefont {Kalogera}},\ }\bibfield  {title} {\bibinfo {title} {Black hole genealogy: Identifying hierarchical mergers with gravitational waves},\ }\href@noop {} {\bibfield  {journal} {\bibinfo  {journal} {Astrophys. J.}\ }\textbf {\bibinfo {volume} {900}},\ \bibinfo {pages} {177} (\bibinfo {year} {2020}{\natexlab{a}})}\BibitemShut {NoStop}%
\bibitem [{\citenamefont {Kimball}\ \emph {et~al.}(2021)\citenamefont {Kimball}, \citenamefont {Talbot}, \citenamefont {Berry}, \citenamefont {Zevin}, \citenamefont {Thrane} \emph {et~al.}}]{gwtc2_hierarchical}%
  \BibitemOpen
  \bibfield  {author} {\bibinfo {author} {\bibfnamefont {C.}~\bibnamefont {Kimball}}, \bibinfo {author} {\bibfnamefont {C.}~\bibnamefont {Talbot}}, \bibinfo {author} {\bibfnamefont {C.~P.}\ \bibnamefont {Berry}}, \bibinfo {author} {\bibfnamefont {M.}~\bibnamefont {Zevin}}, \bibinfo {author} {\bibfnamefont {E.}~\bibnamefont {Thrane}}, \emph {et~al.},\ }\bibfield  {title} {\bibinfo {title} {{Evidence for hierarchical black hole mergers in the second LIGO--Virgo gravitational-wave catalog}},\ }\href@noop {} {\ \textbf {\bibinfo {volume} {915}},\ \bibinfo {pages} {L35} (\bibinfo {year} {2021})}\BibitemShut {NoStop}%
\bibitem [{\citenamefont {Fishbach}\ \emph {et~al.}(2017)\citenamefont {Fishbach}, \citenamefont {Holz},\ and\ \citenamefont {Farr}}]{fishbach2017}%
  \BibitemOpen
  \bibfield  {author} {\bibinfo {author} {\bibfnamefont {M.}~\bibnamefont {Fishbach}}, \bibinfo {author} {\bibfnamefont {D.~E.}\ \bibnamefont {Holz}},\ and\ \bibinfo {author} {\bibfnamefont {B.}~\bibnamefont {Farr}},\ }\bibfield  {title} {\bibinfo {title} {{Are LIGO's Black Holes Made from Smaller Black Holes?}},\ }\href@noop {} {\ \textbf {\bibinfo {volume} {840}},\ \bibinfo {pages} {L24} (\bibinfo {year} {2017})}\BibitemShut {NoStop}%
\bibitem [{\citenamefont {Chatziioannou}\ \emph {et~al.}(2019)\citenamefont {Chatziioannou} \emph {et~al.}}]{Chatziioannou:2019dsz}%
  \BibitemOpen
  \bibfield  {author} {\bibinfo {author} {\bibfnamefont {K.}~\bibnamefont {Chatziioannou}} \emph {et~al.},\ }\bibfield  {title} {\bibinfo {title} {{On the properties of the massive binary black hole merger GW170729}},\ }\href {https://doi.org/10.1103/PhysRevD.100.104015} {\bibfield  {journal} {\bibinfo  {journal} {Phys. Rev. D}\ }\textbf {\bibinfo {volume} {100}},\ \bibinfo {pages} {104015} (\bibinfo {year} {2019})},\ \Eprint {https://arxiv.org/abs/1903.06742} {arXiv:1903.06742 [gr-qc]} \BibitemShut {NoStop}%
\bibitem [{\citenamefont {Doctor}\ \emph {et~al.}(2019)\citenamefont {Doctor}, \citenamefont {Wysocki}, \citenamefont {O'Shaughnessy}, \citenamefont {Holz},\ and\ \citenamefont {Farr}}]{Doctor:2019ruh}%
  \BibitemOpen
  \bibfield  {author} {\bibinfo {author} {\bibfnamefont {Z.}~\bibnamefont {Doctor}}, \bibinfo {author} {\bibfnamefont {D.}~\bibnamefont {Wysocki}}, \bibinfo {author} {\bibfnamefont {R.}~\bibnamefont {O'Shaughnessy}}, \bibinfo {author} {\bibfnamefont {D.~E.}\ \bibnamefont {Holz}},\ and\ \bibinfo {author} {\bibfnamefont {B.}~\bibnamefont {Farr}},\ }\bibfield  {title} {\bibinfo {title} {{Black Hole Coagulation: Modeling Hierarchical Mergers in Black Hole Populations}}\ }\href {https://doi.org/10.3847/1538-4357/ab7fac} {10.3847/1538-4357/ab7fac} (\bibinfo {year} {2019}),\ \Eprint {https://arxiv.org/abs/1911.04424} {arXiv:1911.04424 [astro-ph.HE]} \BibitemShut {NoStop}%
\bibitem [{\citenamefont {Kimball}\ \emph {et~al.}(2020{\natexlab{b}})\citenamefont {Kimball}, \citenamefont {Berry},\ and\ \citenamefont {Kalogera}}]{Kimball:2019mfs}%
  \BibitemOpen
  \bibfield  {author} {\bibinfo {author} {\bibfnamefont {C.}~\bibnamefont {Kimball}}, \bibinfo {author} {\bibfnamefont {C.~P.~L.}\ \bibnamefont {Berry}},\ and\ \bibinfo {author} {\bibfnamefont {V.}~\bibnamefont {Kalogera}},\ }\bibfield  {title} {\bibinfo {title} {{What GW170729's exceptional mass and spin tells us about its family tree}},\ }\href {https://doi.org/10.3847/2515-5172/ab66be} {\bibfield  {journal} {\bibinfo  {journal} {Res. Notes AAS}\ }\textbf {\bibinfo {volume} {4}},\ \bibinfo {pages} {2} (\bibinfo {year} {2020}{\natexlab{b}})},\ \Eprint {https://arxiv.org/abs/1903.07813} {arXiv:1903.07813 [astro-ph.HE]} \BibitemShut {NoStop}%
\bibitem [{\citenamefont {Bartos}\ and\ \citenamefont {Haiman}(2025)}]{Bartos:2025pkv}%
  \BibitemOpen
  \bibfield  {author} {\bibinfo {author} {\bibfnamefont {I.}~\bibnamefont {Bartos}}\ and\ \bibinfo {author} {\bibfnamefont {Z.}~\bibnamefont {Haiman}},\ }\bibfield  {title} {\bibinfo {title} {{Accretion is All You Need: Black Hole Spin Alignment in Merger GW231123 Indicates Accretion Pathway}},\ }\href@noop {} {\  (\bibinfo {year} {2025})},\ \Eprint {https://arxiv.org/abs/2508.08558} {arXiv:2508.08558 [astro-ph.HE]} \BibitemShut {NoStop}%
\bibitem [{\citenamefont {Safarzadeh}\ and\ \citenamefont {Haiman}(2020)}]{Safarzadeh:2020vbv}%
  \BibitemOpen
  \bibfield  {author} {\bibinfo {author} {\bibfnamefont {M.}~\bibnamefont {Safarzadeh}}\ and\ \bibinfo {author} {\bibfnamefont {Z.}~\bibnamefont {Haiman}},\ }\bibfield  {title} {\bibinfo {title} {{Formation of GW190521 via gas accretion onto Population III stellar black hole remnants born in high-redshift minihalos}},\ }\href {https://doi.org/10.3847/2041-8213/abc253} {\bibfield  {journal} {\bibinfo  {journal} {Astrophys. J. Lett.}\ }\textbf {\bibinfo {volume} {903}},\ \bibinfo {pages} {L21} (\bibinfo {year} {2020})},\ \Eprint {https://arxiv.org/abs/2009.09320} {arXiv:2009.09320 [astro-ph.HE]} \BibitemShut {NoStop}%
\bibitem [{\citenamefont {van Son}\ \emph {et~al.}(2020)\citenamefont {van Son}, \citenamefont {de~Mink}, \citenamefont {Broekgaarden}, \citenamefont {Renzo}, \citenamefont {Justham}, \citenamefont {Laplace}, \citenamefont {Moran-Fraile}, \citenamefont {Hendriks},\ and\ \citenamefont {Farmer}}]{vanSon:2020zbk}%
  \BibitemOpen
  \bibfield  {author} {\bibinfo {author} {\bibfnamefont {L.~A.~C.}\ \bibnamefont {van Son}}, \bibinfo {author} {\bibfnamefont {S.~E.}\ \bibnamefont {de~Mink}}, \bibinfo {author} {\bibfnamefont {F.~S.}\ \bibnamefont {Broekgaarden}}, \bibinfo {author} {\bibfnamefont {M.}~\bibnamefont {Renzo}}, \bibinfo {author} {\bibfnamefont {S.}~\bibnamefont {Justham}}, \bibinfo {author} {\bibfnamefont {E.}~\bibnamefont {Laplace}}, \bibinfo {author} {\bibfnamefont {J.}~\bibnamefont {Moran-Fraile}}, \bibinfo {author} {\bibfnamefont {D.~D.}\ \bibnamefont {Hendriks}},\ and\ \bibinfo {author} {\bibfnamefont {R.}~\bibnamefont {Farmer}},\ }\bibfield  {title} {\bibinfo {title} {{Polluting the pair-instability mass gap for binary black holes through super-Eddington accretion in isolated binaries}},\ }\href {https://doi.org/10.3847/1538-4357/ab9809} {\bibfield  {journal} {\bibinfo  {journal} {Astrophys. J.}\ }\textbf {\bibinfo {volume} {897}},\ \bibinfo {pages} {100} (\bibinfo {year} {2020})},\ \Eprint
  {https://arxiv.org/abs/2004.05187} {arXiv:2004.05187 [astro-ph.HE]} \BibitemShut {NoStop}%
\bibitem [{\citenamefont {Tiwari}(2022)}]{Tiwari:2021yvr}%
  \BibitemOpen
  \bibfield  {author} {\bibinfo {author} {\bibfnamefont {V.}~\bibnamefont {Tiwari}},\ }\bibfield  {title} {\bibinfo {title} {{Exploring Features in the Binary Black Hole Population}},\ }\href {https://doi.org/10.3847/1538-4357/ac589a} {\bibfield  {journal} {\bibinfo  {journal} {Astrophys. J.}\ }\textbf {\bibinfo {volume} {928}},\ \bibinfo {pages} {155} (\bibinfo {year} {2022})},\ \Eprint {https://arxiv.org/abs/2111.13991} {arXiv:2111.13991 [astro-ph.HE]} \BibitemShut {NoStop}%
\bibitem [{\citenamefont {Afroz}\ and\ \citenamefont {Mukherjee}(2025{\natexlab{b}})}]{Afroz:2025ikg}%
  \BibitemOpen
  \bibfield  {author} {\bibinfo {author} {\bibfnamefont {S.}~\bibnamefont {Afroz}}\ and\ \bibinfo {author} {\bibfnamefont {S.}~\bibnamefont {Mukherjee}},\ }\bibfield  {title} {\bibinfo {title} {{Binary Black Hole Phase Space Discovers the Signature of Pair Instability Supernovae Mass Gap}},\ }\href@noop {} {\  (\bibinfo {year} {2025}{\natexlab{b}})},\ \Eprint {https://arxiv.org/abs/2509.09123} {arXiv:2509.09123 [astro-ph.HE]} \BibitemShut {NoStop}%
\bibitem [{\citenamefont {Banerjee}(2022)}]{Banerjee:2021xzp}%
  \BibitemOpen
  \bibfield  {author} {\bibinfo {author} {\bibfnamefont {S.}~\bibnamefont {Banerjee}},\ }\bibfield  {title} {\bibinfo {title} {{Merger rate density of stellar-mass binary black holes from young massive clusters, open clusters, and isolated binaries: Comparisons with LIGO-Virgo-KAGRA results}},\ }\href {https://doi.org/10.1103/PhysRevD.105.023004} {\bibfield  {journal} {\bibinfo  {journal} {Phys. Rev. D}\ }\textbf {\bibinfo {volume} {105}},\ \bibinfo {pages} {023004} (\bibinfo {year} {2022})},\ \Eprint {https://arxiv.org/abs/2108.04250} {arXiv:2108.04250 [astro-ph.HE]} \BibitemShut {NoStop}%
\bibitem [{\citenamefont {Giacobbo}\ and\ \citenamefont {Mapelli}(2018)}]{Giacobbo:2018etu}%
  \BibitemOpen
  \bibfield  {author} {\bibinfo {author} {\bibfnamefont {N.}~\bibnamefont {Giacobbo}}\ and\ \bibinfo {author} {\bibfnamefont {M.}~\bibnamefont {Mapelli}},\ }\bibfield  {title} {\bibinfo {title} {{The progenitors of compact-object binaries: impact of metallicity, common envelope and natal kicks}},\ }\href {https://doi.org/10.1093/mnras/sty1999} {\bibfield  {journal} {\bibinfo  {journal} {Mon. Not. Roy. Astron. Soc.}\ }\textbf {\bibinfo {volume} {480}},\ \bibinfo {pages} {2011} (\bibinfo {year} {2018})},\ \Eprint {https://arxiv.org/abs/1806.00001} {arXiv:1806.00001 [astro-ph.HE]} \BibitemShut {NoStop}%
\bibitem [{\citenamefont {Neijssel}\ \emph {et~al.}(2019)\citenamefont {Neijssel}, \citenamefont {Vigna-Gómez}, \citenamefont {Stevenson}, \citenamefont {Barrett}, \citenamefont {Gaebel}, \citenamefont {Broekgaarden}, \citenamefont {de Mink}, \citenamefont {Szécsi}, \citenamefont {Vinciguerra},\ and\ \citenamefont {Mandel}}]{Neijssel:2019}%
  \BibitemOpen
  \bibfield  {author} {\bibinfo {author} {\bibfnamefont {C.~J.}\ \bibnamefont {Neijssel}}, \bibinfo {author} {\bibfnamefont {A.}~\bibnamefont {Vigna-Gómez}}, \bibinfo {author} {\bibfnamefont {S.}~\bibnamefont {Stevenson}}, \bibinfo {author} {\bibfnamefont {J.~W.}\ \bibnamefont {Barrett}}, \bibinfo {author} {\bibfnamefont {S.~M.}\ \bibnamefont {Gaebel}}, \bibinfo {author} {\bibfnamefont {F.~S.}\ \bibnamefont {Broekgaarden}}, \bibinfo {author} {\bibfnamefont {S.~E.}\ \bibnamefont {de Mink}}, \bibinfo {author} {\bibfnamefont {D.}~\bibnamefont {Szécsi}}, \bibinfo {author} {\bibfnamefont {S.}~\bibnamefont {Vinciguerra}},\ and\ \bibinfo {author} {\bibfnamefont {I.}~\bibnamefont {Mandel}},\ }\bibfield  {title} {\bibinfo {title} {The effect of the metallicity-specific star formation history on double compact object mergers},\ }\href {https://doi.org/10.1093/mnras/stz2840} {\bibfield  {journal} {\bibinfo  {journal} {Monthly Notices of the Royal Astronomical Society}\ }\textbf {\bibinfo {volume} {490}},\ \bibinfo
  {pages} {3740–3759} (\bibinfo {year} {2019})}\BibitemShut {NoStop}%
\bibitem [{\citenamefont {Tanikawa}\ \emph {et~al.}(2022)\citenamefont {Tanikawa}, \citenamefont {Yoshida}, \citenamefont {Kinugawa}, \citenamefont {Trani}, \citenamefont {Hosokawa}, \citenamefont {Susa},\ and\ \citenamefont {Omukai}}]{Tanikawa:2021qqi}%
  \BibitemOpen
  \bibfield  {author} {\bibinfo {author} {\bibfnamefont {A.}~\bibnamefont {Tanikawa}}, \bibinfo {author} {\bibfnamefont {T.}~\bibnamefont {Yoshida}}, \bibinfo {author} {\bibfnamefont {T.}~\bibnamefont {Kinugawa}}, \bibinfo {author} {\bibfnamefont {A.~A.}\ \bibnamefont {Trani}}, \bibinfo {author} {\bibfnamefont {T.}~\bibnamefont {Hosokawa}}, \bibinfo {author} {\bibfnamefont {H.}~\bibnamefont {Susa}},\ and\ \bibinfo {author} {\bibfnamefont {K.}~\bibnamefont {Omukai}},\ }\bibfield  {title} {\bibinfo {title} {{Merger Rate Density of Binary Black Holes through Isolated Population I, II, III and Extremely Metal-poor Binary Star Evolution}},\ }\href {https://doi.org/10.3847/1538-4357/ac4247} {\bibfield  {journal} {\bibinfo  {journal} {Astrophys. J.}\ }\textbf {\bibinfo {volume} {926}},\ \bibinfo {pages} {83} (\bibinfo {year} {2022})},\ \Eprint {https://arxiv.org/abs/2110.10846} {arXiv:2110.10846 [astro-ph.HE]} \BibitemShut {NoStop}%
\bibitem [{\citenamefont {van Son}\ \emph {et~al.}(2022{\natexlab{a}})\citenamefont {van Son}, \citenamefont {de~Mink}, \citenamefont {Chruslinska}, \citenamefont {Conroy}, \citenamefont {Pakmor},\ and\ \citenamefont {Hernquist}}]{vanSon:2022ylf}%
  \BibitemOpen
  \bibfield  {author} {\bibinfo {author} {\bibfnamefont {L.~A.~C.}\ \bibnamefont {van Son}}, \bibinfo {author} {\bibfnamefont {S.~E.}\ \bibnamefont {de~Mink}}, \bibinfo {author} {\bibfnamefont {M.}~\bibnamefont {Chruslinska}}, \bibinfo {author} {\bibfnamefont {C.}~\bibnamefont {Conroy}}, \bibinfo {author} {\bibfnamefont {R.}~\bibnamefont {Pakmor}},\ and\ \bibinfo {author} {\bibfnamefont {L.}~\bibnamefont {Hernquist}},\ }\bibfield  {title} {\bibinfo {title} {{The locations of features in the mass distribution of merging binary black holes are robust against uncertainties in the metallicity-dependent cosmic star formation history}}\ }\href {https://doi.org/10.3847/1538-4357/acbf51} {10.3847/1538-4357/acbf51} (\bibinfo {year} {2022}{\natexlab{a}}),\ \Eprint {https://arxiv.org/abs/2209.03385} {arXiv:2209.03385 [astro-ph.GA]} \BibitemShut {NoStop}%
\bibitem [{\citenamefont {van Son}\ \emph {et~al.}(2022{\natexlab{b}})\citenamefont {van Son}, \citenamefont {de~Mink}, \citenamefont {Renzo}, \citenamefont {Justham}, \citenamefont {Zapartas}, \citenamefont {Breivik}, \citenamefont {Callister}, \citenamefont {Farr},\ and\ \citenamefont {Conroy}}]{vanSon:2022myr}%
  \BibitemOpen
  \bibfield  {author} {\bibinfo {author} {\bibfnamefont {L.~A.~C.}\ \bibnamefont {van Son}}, \bibinfo {author} {\bibfnamefont {S.~E.}\ \bibnamefont {de~Mink}}, \bibinfo {author} {\bibfnamefont {M.}~\bibnamefont {Renzo}}, \bibinfo {author} {\bibfnamefont {S.}~\bibnamefont {Justham}}, \bibinfo {author} {\bibfnamefont {E.}~\bibnamefont {Zapartas}}, \bibinfo {author} {\bibfnamefont {K.}~\bibnamefont {Breivik}}, \bibinfo {author} {\bibfnamefont {T.}~\bibnamefont {Callister}}, \bibinfo {author} {\bibfnamefont {W.~M.}\ \bibnamefont {Farr}},\ and\ \bibinfo {author} {\bibfnamefont {C.}~\bibnamefont {Conroy}},\ }\bibfield  {title} {\bibinfo {title} {{No Peaks without Valleys: The Stable Mass Transfer Channel for Gravitational-wave Sources in Light of the Neutron Star{\textendash}Black Hole Mass Gap}},\ }\href {https://doi.org/10.3847/1538-4357/ac9b0a} {\bibfield  {journal} {\bibinfo  {journal} {Astrophys. J.}\ }\textbf {\bibinfo {volume} {940}},\ \bibinfo {pages} {184} (\bibinfo {year} {2022}{\natexlab{b}})},\ \Eprint
  {https://arxiv.org/abs/2209.13609} {arXiv:2209.13609 [astro-ph.HE]} \BibitemShut {NoStop}%
\bibitem [{\citenamefont {Ye}\ and\ \citenamefont {Fishbach}(2024)}]{Ye:2024ypm}%
  \BibitemOpen
  \bibfield  {author} {\bibinfo {author} {\bibfnamefont {C.~S.}\ \bibnamefont {Ye}}\ and\ \bibinfo {author} {\bibfnamefont {M.}~\bibnamefont {Fishbach}},\ }\bibfield  {title} {\bibinfo {title} {{The Redshift Evolution of the Binary Black Hole Mass Distribution from Dense Star Clusters}},\ }\href {https://doi.org/10.3847/1538-4357/ad3ba8} {\bibfield  {journal} {\bibinfo  {journal} {Astrophys. J.}\ }\textbf {\bibinfo {volume} {967}},\ \bibinfo {pages} {62} (\bibinfo {year} {2024})},\ \Eprint {https://arxiv.org/abs/2402.12444} {arXiv:2402.12444 [astro-ph.HE]} \BibitemShut {NoStop}%
\bibitem [{\citenamefont {Ye}\ \emph {et~al.}(2025)\citenamefont {Ye}, \citenamefont {Fishbach}, \citenamefont {Kremer},\ and\ \citenamefont {Reina-Campos}}]{Ye:2025ano}%
  \BibitemOpen
  \bibfield  {author} {\bibinfo {author} {\bibfnamefont {C.~S.}\ \bibnamefont {Ye}}, \bibinfo {author} {\bibfnamefont {M.}~\bibnamefont {Fishbach}}, \bibinfo {author} {\bibfnamefont {K.}~\bibnamefont {Kremer}},\ and\ \bibinfo {author} {\bibfnamefont {M.}~\bibnamefont {Reina-Campos}},\ }\bibfield  {title} {\bibinfo {title} {{Mass Distribution of Binary Black Hole Mergers from Young and Old Dense Star Clusters}},\ }\href@noop {} {\  (\bibinfo {year} {2025})},\ \Eprint {https://arxiv.org/abs/2507.07183} {arXiv:2507.07183 [astro-ph.HE]} \BibitemShut {NoStop}%
\bibitem [{\citenamefont {Antonini}\ \emph {et~al.}(2023)\citenamefont {Antonini}, \citenamefont {Gieles}, \citenamefont {Dosopoulou},\ and\ \citenamefont {Chattopadhyay}}]{Antonini:2022vib}%
  \BibitemOpen
  \bibfield  {author} {\bibinfo {author} {\bibfnamefont {F.}~\bibnamefont {Antonini}}, \bibinfo {author} {\bibfnamefont {M.}~\bibnamefont {Gieles}}, \bibinfo {author} {\bibfnamefont {F.}~\bibnamefont {Dosopoulou}},\ and\ \bibinfo {author} {\bibfnamefont {D.}~\bibnamefont {Chattopadhyay}},\ }\bibfield  {title} {\bibinfo {title} {{Coalescing black hole binaries from globular clusters: mass distributions and comparison to gravitational wave data from GWTC-3}},\ }\href {https://doi.org/10.1093/mnras/stad972} {\bibfield  {journal} {\bibinfo  {journal} {Mon. Not. Roy. Astron. Soc.}\ }\textbf {\bibinfo {volume} {522}},\ \bibinfo {pages} {466} (\bibinfo {year} {2023})},\ \Eprint {https://arxiv.org/abs/2208.01081} {arXiv:2208.01081 [astro-ph.HE]} \BibitemShut {NoStop}%
\bibitem [{\citenamefont {Banagiri}\ \emph {et~al.}(2026)\citenamefont {Banagiri}, \citenamefont {Thrane},\ and\ \citenamefont {Lasky}}]{banagiri_2026_20320700}%
  \BibitemOpen
  \bibfield  {author} {\bibinfo {author} {\bibfnamefont {S.}~\bibnamefont {Banagiri}}, \bibinfo {author} {\bibfnamefont {E.}~\bibnamefont {Thrane}},\ and\ \bibinfo {author} {\bibfnamefont {P.}~\bibnamefont {Lasky}},\ }\bibfield  {title} {\bibinfo {title} {Evidence for three subpopulations of merging binary black holes at different primary masses},\ }\href {https://doi.org/10.5281/zenodo.20320700} {10.5281/zenodo.20320700} (\bibinfo {year} {2026})\BibitemShut {NoStop}%
\bibitem [{\citenamefont {Essick}\ and\ \citenamefont {Fishbach}(2024)}]{Essick:2023upv}%
  \BibitemOpen
  \bibfield  {author} {\bibinfo {author} {\bibfnamefont {R.}~\bibnamefont {Essick}}\ and\ \bibinfo {author} {\bibfnamefont {M.}~\bibnamefont {Fishbach}},\ }\bibfield  {title} {\bibinfo {title} {{Ensuring Consistency between Noise and Detection in Hierarchical Bayesian Inference}},\ }\href {https://doi.org/10.3847/1538-4357/ad1604} {\bibfield  {journal} {\bibinfo  {journal} {Astrophys. J.}\ }\textbf {\bibinfo {volume} {962}},\ \bibinfo {pages} {169} (\bibinfo {year} {2024})},\ \Eprint {https://arxiv.org/abs/2310.02017} {arXiv:2310.02017 [gr-qc]} \BibitemShut {NoStop}%
\bibitem [{\citenamefont {Callister}\ and\ \citenamefont {Farr}(2024)}]{Callister:2023tgi}%
  \BibitemOpen
  \bibfield  {author} {\bibinfo {author} {\bibfnamefont {T.~A.}\ \bibnamefont {Callister}}\ and\ \bibinfo {author} {\bibfnamefont {W.~M.}\ \bibnamefont {Farr}},\ }\bibfield  {title} {\bibinfo {title} {{Parameter-Free Tour of the Binary Black Hole Population}},\ }\href {https://doi.org/10.1103/PhysRevX.14.021005} {\bibfield  {journal} {\bibinfo  {journal} {Phys. Rev. X}\ }\textbf {\bibinfo {volume} {14}},\ \bibinfo {pages} {021005} (\bibinfo {year} {2024})},\ \Eprint {https://arxiv.org/abs/2302.07289} {arXiv:2302.07289 [astro-ph.HE]} \BibitemShut {NoStop}%
\bibitem [{\citenamefont {Essick}\ \emph {et~al.}(2025)\citenamefont {Essick} \emph {et~al.}}]{Essick:2025zed}%
  \BibitemOpen
  \bibfield  {author} {\bibinfo {author} {\bibfnamefont {R.}~\bibnamefont {Essick}} \emph {et~al.},\ }\bibfield  {title} {\bibinfo {title} {{Compact Binary Coalescence Sensitivity Estimates with Injection Campaigns during the LIGO-Virgo-KAGRA Collaborations' Fourth Observing Run}},\ }\href@noop {} {\  (\bibinfo {year} {2025})},\ \Eprint {https://arxiv.org/abs/2508.10638} {arXiv:2508.10638 [gr-qc]} \BibitemShut {NoStop}%
\bibitem [{\citenamefont {Collaboration}\ \emph {et~al.}(2025{\natexlab{b}})\citenamefont {Collaboration}, \citenamefont {Collaboration},\ and\ \citenamefont {Collaboration}}]{LVK:2025gwtc4sensitivityzenodo}%
  \BibitemOpen
  \bibfield  {author} {\bibinfo {author} {\bibfnamefont {L.~S.}\ \bibnamefont {Collaboration}}, \bibinfo {author} {\bibfnamefont {V.}~\bibnamefont {Collaboration}},\ and\ \bibinfo {author} {\bibfnamefont {K.}~\bibnamefont {Collaboration}},\ }\bibfield  {title} {\bibinfo {title} {Gwtc-4.0 cumulative search sensitivity estimates},\ }\href {https://doi.org/10.5281/zenodo.16740128} {10.5281/zenodo.16740128} (\bibinfo {year} {2025}{\natexlab{b}})\BibitemShut {NoStop}%
\bibitem [{\citenamefont {Essick}\ and\ \citenamefont {Farr}(2022)}]{Essick:2022ojx}%
  \BibitemOpen
  \bibfield  {author} {\bibinfo {author} {\bibfnamefont {R.}~\bibnamefont {Essick}}\ and\ \bibinfo {author} {\bibfnamefont {W.}~\bibnamefont {Farr}},\ }\bibfield  {title} {\bibinfo {title} {{Precision Requirements for Monte Carlo Sums within Hierarchical Bayesian Inference}},\ }\href@noop {} {\  (\bibinfo {year} {2022})},\ \Eprint {https://arxiv.org/abs/2204.00461} {arXiv:2204.00461 [astro-ph.IM]} \BibitemShut {NoStop}%
\bibitem [{\citenamefont {Talbot}\ and\ \citenamefont {Golomb}(2023)}]{Talbot:2023pex}%
  \BibitemOpen
  \bibfield  {author} {\bibinfo {author} {\bibfnamefont {C.}~\bibnamefont {Talbot}}\ and\ \bibinfo {author} {\bibfnamefont {J.}~\bibnamefont {Golomb}},\ }\bibfield  {title} {\bibinfo {title} {{Growing pains: understanding the impact of likelihood uncertainty on hierarchical Bayesian inference for gravitational-wave astronomy}},\ }\href {https://doi.org/10.1093/mnras/stad2968} {\bibfield  {journal} {\bibinfo  {journal} {Mon. Not. Roy. Astron. Soc.}\ }\textbf {\bibinfo {volume} {526}},\ \bibinfo {pages} {3495} (\bibinfo {year} {2023})},\ \Eprint {https://arxiv.org/abs/2304.06138} {arXiv:2304.06138 [astro-ph.IM]} \BibitemShut {NoStop}%
\bibitem [{\citenamefont {{Talbot}}\ \emph {et~al.}(2019)\citenamefont {{Talbot}}, \citenamefont {{Smith}}, \citenamefont {{Thrane}},\ and\ \citenamefont {{Poole}}}]{Talbot:2019}%
  \BibitemOpen
  \bibfield  {author} {\bibinfo {author} {\bibfnamefont {C.}~\bibnamefont {{Talbot}}}, \bibinfo {author} {\bibfnamefont {R.}~\bibnamefont {{Smith}}}, \bibinfo {author} {\bibfnamefont {E.}~\bibnamefont {{Thrane}}},\ and\ \bibinfo {author} {\bibfnamefont {G.~B.}\ \bibnamefont {{Poole}}},\ }\bibfield  {title} {\bibinfo {title} {{Parallelized inference for gravitational-wave astronomy}},\ }\href {https://doi.org/10.1103/PhysRevD.100.043030} {\bibfield  {journal} {\bibinfo  {journal} {\prd}\ }\textbf {\bibinfo {volume} {100}},\ \bibinfo {eid} {043030} (\bibinfo {year} {2019})},\ \Eprint {https://arxiv.org/abs/1904.02863} {arXiv:1904.02863 [astro-ph.IM]} \BibitemShut {NoStop}%
\bibitem [{\citenamefont {Ashton}\ \emph {et~al.}(2019)\citenamefont {Ashton} \emph {et~al.}}]{Ashton:2019}%
  \BibitemOpen
  \bibfield  {author} {\bibinfo {author} {\bibfnamefont {G.}~\bibnamefont {Ashton}} \emph {et~al.},\ }\bibfield  {title} {\bibinfo {title} {{BILBY: A user-friendly Bayesian inference library for gravitational-wave astronomy}},\ }\href {https://doi.org/10.3847/1538-4365/ab06fc} {\bibfield  {journal} {\bibinfo  {journal} {Astrophys. J. Suppl.}\ }\textbf {\bibinfo {volume} {241}},\ \bibinfo {pages} {27} (\bibinfo {year} {2019})},\ \Eprint {https://arxiv.org/abs/1811.02042} {arXiv:1811.02042 [astro-ph.IM]} \BibitemShut {NoStop}%
\bibitem [{\citenamefont {{Speagle}}(2020)}]{Speagle:2020}%
  \BibitemOpen
  \bibfield  {author} {\bibinfo {author} {\bibfnamefont {J.~S.}\ \bibnamefont {{Speagle}}},\ }\bibfield  {title} {\bibinfo {title} {{DYNESTY: a dynamic nested sampling package for estimating Bayesian posteriors and evidences}},\ }\href {https://doi.org/10.1093/mnras/staa278} {\bibfield  {journal} {\bibinfo  {journal} {Mon. Not. Roy. Astron. Soc.}\ }\textbf {\bibinfo {volume} {493}},\ \bibinfo {pages} {3132} (\bibinfo {year} {2020})},\ \Eprint {https://arxiv.org/abs/1904.02180} {arXiv:1904.02180 [astro-ph.IM]} \BibitemShut {NoStop}%
\end{thebibliography}%
